\documentclass[
 pra,
 amsmath,amssymb,
 aps,
 twocolumn,
 notitlepage,
 superscriptaddress
]{revtex4-2}

\usepackage{float}
\makeatletter
\let\newfloat\newfloat@ltx
\makeatother

\newcommand{\av}[1]{{\textcolor{cyan}{#1}}}   
\newcommand{\ar}[1]{{\textcolor{blue}{#1}}}   

\usepackage{algorithm}
\usepackage{algpseudocode}
\usepackage[colorlinks=true]{hyperref}

\usepackage{ragged2e}
\usepackage{subcaption}
\DeclareCaptionJustification{justified}{\justifying}
\usepackage{cleveref}
\usepackage{styles}
\usepackage{fancyhdr}
\usepackage{txfonts}
\usepackage{xurl} 
\usepackage{bm} 
\usepackage{physics} 
\usepackage{tikz} 
\usetikzlibrary{quantikz} 
\usepackage{tikzscale}

\newcommand{\vv}[1]{\boldsymbol{\mathbf{#1}}} 
\newcommand{\emin}{E_{\textrm{min}}}

\begin{document}

\title{Optimization Applications as Quantum Performance Benchmarks}

\author{Thomas Lubinski}
\affiliation{Quantum Circuits Inc, 25 Science Park, New Haven, CT 06511}
\affiliation{QED-C Technical Advisory Committee on Standards and Performance Benchmarks}

\author{Carleton Coffrin}
\affiliation{Advanced Network Science Initiative, Los Alamos National Laboratory, Los Alamos, New Mexico 87545, USA}

\author{Catherine McGeoch}
\affiliation{D-Wave Systems, Burnaby, British Columbia, Canada, V5G 4M9, Canada}

\author{\\ Pratik Sathe}
\affiliation{Department of Physics and Astronomy, University of California at Los Angeles, USA}
\affiliation{Research Institute of Advanced Computer Science, Universities Space Research Association, Mountain View, CA, USA}
\affiliation{Theoretical Division (T4), Los Alamos National Laboratory, Los Alamos, New Mexico 87545, USA}

\author{Joshua Apanavicius}
\affiliation{Indiana University Department of Physics, Bloomington, Indiana 47405, USA}
\affiliation{Indiana University Quantum Science and Engineering Center, Bloomington, Indiana 47405, USA}

\author{David E. Bernal Neira}
\affiliation{Research Institute of Advanced Computer Science, Universities Space Research Association, Mountain View, CA, USA}
\affiliation{Quantum Artificial Intelligence Laboratory, NASA Ames Research Center, Mountain View, CA, USA}
\affiliation{Davidson School of Chemical Engineering, Purdue University, West Lafayette, IN, USA}

\collaboration{Quantum Economic Development Consortium (QED-C) collaboration} 

\thanks{This work was sponsored by the Quantum Economic Development Consortium (QED-C) (quantumconsortium.org), managed by SRI International, and was performed under the auspices of the QED-C Technical Advisory Committee on Standards and Performance Benchmarks. The authors acknowledge many committee members for their input and feedback on the project and this manuscript.}

\date{\rule[15pt]{0pt}{0pt}\today}

\begin{abstract}

\vspace{0.0cm}
Combinatorial optimization is anticipated to be one of the primary use cases for quantum computation in the coming years. The Quantum Approximate Optimization Algorithm (QAOA) and Quantum Annealing (QA) can potentially demonstrate significant run-time performance benefits over current state-of-the-art solutions.
Inspired by existing methods to characterize classical optimization algorithms, we analyze the solution quality obtained by solving Max-Cut problems using gate-model quantum devices and a quantum annealing device. 
This is used to guide the development of an advanced benchmarking framework for quantum computers designed to evaluate the trade-off between run-time execution performance and the solution quality for iterative hybrid quantum-classical applications.
The framework generates performance profiles through compelling visualizations that show performance progression as a function of time for various problem sizes and illustrates algorithm limitations uncovered by the benchmarking approach.
As an illustration, we explore the factors that influence quantum computing system throughput, using results obtained through execution on various quantum simulators and quantum hardware systems.

\end{abstract}

\keywords{Quantum Computing \and Benchmarks \and Benchmarking \and Algorithms \and Application Benchmarks \and QAOA \and Quantum Approximate Optimization Algorithm \and Max-Cut }

\maketitle
\tableofcontents 

\pagestyle{fancy}

\renewcommand{\headrulewidth}{0.0pt}
\lhead{}
\rhead{\thepage}

\renewcommand{\footrulewidth}{0.4pt}
\cfoot{}
\lfoot{Optimization Applications as Quantum Performance Benchmarks}
\rfoot{\today}

\vspace{0.2cm}


\section{Introduction}
\label{sec:introduction}

In many application domains, it is of utmost importance to efficiently find near-optimal solutions to problems that involve many variables that affect the cost of some operation or function.
For example, in a large power grid, rapidly determining the best allocation of power distribution could prevent a major blackout.
These are known as combinatorial optimization problems and are often cited as a potential use case for quantum computing~\cite{Zahedinejad2017, Egger2020_9222275, mcgeoch2013}. 

\begin{figure}[t]
\includegraphics[width=0.84\columnwidth]
{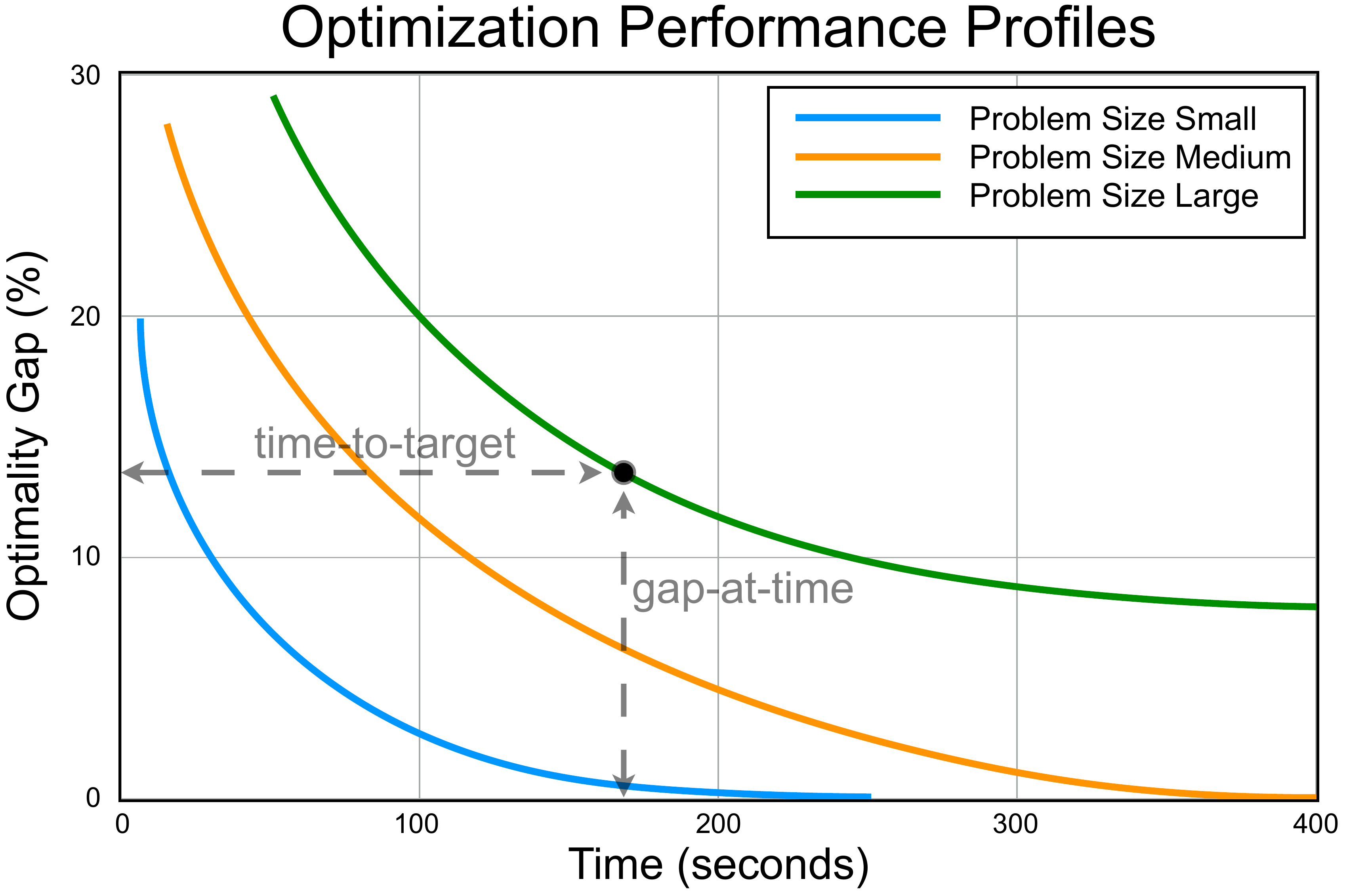}
\caption{
\textbf{An Illustration of a Performance Profile for Benchmarking Optimization Methods.}
\emph{Performance Profile} plots, like the one shown here, are widely used by the Operations Research community to understand, communicate, and compare the performance of optimization methods.
The quality of the solution, as the relative difference from optimal (optimality gap), can evolve over time during the execution of an optimization algorithm.
This permits the user to gauge the time required to obtain a solution of a desired quality (time-to-target) or the solution quality achieved after a specified amount of time (gap-at-time).
The gap-at-time metric is the \emph{de facto} standard used in Operations Research, reflecting use cases for most industrial optimization applications.
Performance profiles tend to change with problem size. It is common for problems with a small number of decision variables to converge to an optimal solution reasonably quickly. In contrast, with larger problems, achieving solutions above a quality threshold can be difficult, which is expected due to the \(\text{NP-HARD}\) nature of challenging optimization tasks.
}
\label{fig:opt_intro_examples_1}
\end{figure}

Classical computer algorithms for addressing such problems are substantially advanced and are implemented across industry, government, and academia.
They perform critical functions in optimizing resource utilization and minimizing cost.
Combinatorial optimization applications are often executed under tight resource constraints (e.g., time, memory, energy, or money), and there is particular emphasis on quantifying the quality of results that could be obtained within a limited budget.
Standard techniques for measuring and comparing the performance of alternative solution methods have matured and are in widespread use~\cite{Zahedinejad2017}.
An illustrative example of a performance profile is shown in~\autoref{fig:opt_intro_examples_1}. 

Quantum computing introduces new techniques for finding solutions to such combinatorial challenges, such as Quantum Annealing (QA)~\cite{Fin1994, PhysRevE.58.5355} and the Quantum Approximate Optimization Algorithm (QAOA)~\cite{farhiQuantumApproximateOptimization2014} that may demonstrate some benefit over classical approaches. 
Theory and classical simulations indicate that, for some problems, QAOA has the potential to outperform classical algorithms~\cite{CrooksPerformance2018, WurtzMaxCut2021}, and some empirical tests of QA systems have demonstrated superior performance over classical alternatives in limited scenarios~\cite{mcgeoch2013, Trummer2015, Pang2021, Tasseff2022Emerging}. 

Numerous efforts have emerged to characterize the performance of quantum computers for applications in optimization (see~\autoref{sec:background}).
However, we find that for such benchmarks to be accessible to users outside the quantum research community, they must both incorporate emerging methods for quantum computing benchmarking and present results meaningfully to experts in domains such as classical optimization and Operations Research (OR).

In this paper, we demonstrate how a properly constructed benchmark program that monitors and characterizes the execution of a combinatorial optimization application on a quantum computing system can provide valuable and critical insights into options for improving its performance and overall throughput.
Additionally, analysis and presentation methodology can be structured in ways familiar to quantum computing specialists but are informed by how Operations Research views the quality of results from a solver in addressing optimization problems.
These enhanced analysis and visualization techniques can provide useful information about the throughput a quantum computing solution can offer and the factors that can be adjusted to improve performance on these systems.
While component and simple application-level benchmarks provide useful information about general performance characteristics, the optimization application supplements this with a detailed understanding of a quantum optimization application's total cost of ownership.
While these techniques have long been used in Operations Research, their effective application to quantum computing is still in the early stages.

Concretely, we introduce a methodology and versatile framework for characterizing the performance of combinatorial optimization solvers executed on quantum computing systems based on different underlying technologies.
We demonstrate this framework's features and highlight its benefits using the QAOA algorithm for execution on gate model systems. We also demonstrate its adaptability to other types of solvers by using QA on annealing hardware.
In future work, we plan to extend this to include other technologies, such as cold atoms.
We demonstrate the capabilities of our framework using the widely studied Max-Cut~\cite{Garey1976, PAPADIMITRIOU1991425} problem, in which the goal is to find the maximum cut size of an undirected graph.
The Max-Cut problem offers a simple early-stage target for evaluating the effectiveness of quantum computing solutions. These solutions could also scale to larger applications and incorporate constraints and other problem features that escalate the challenge.

\begin{figure*}[t]
\centering
\begin{subfigure}{0.32\textwidth}
        \includegraphics[width=\textwidth]{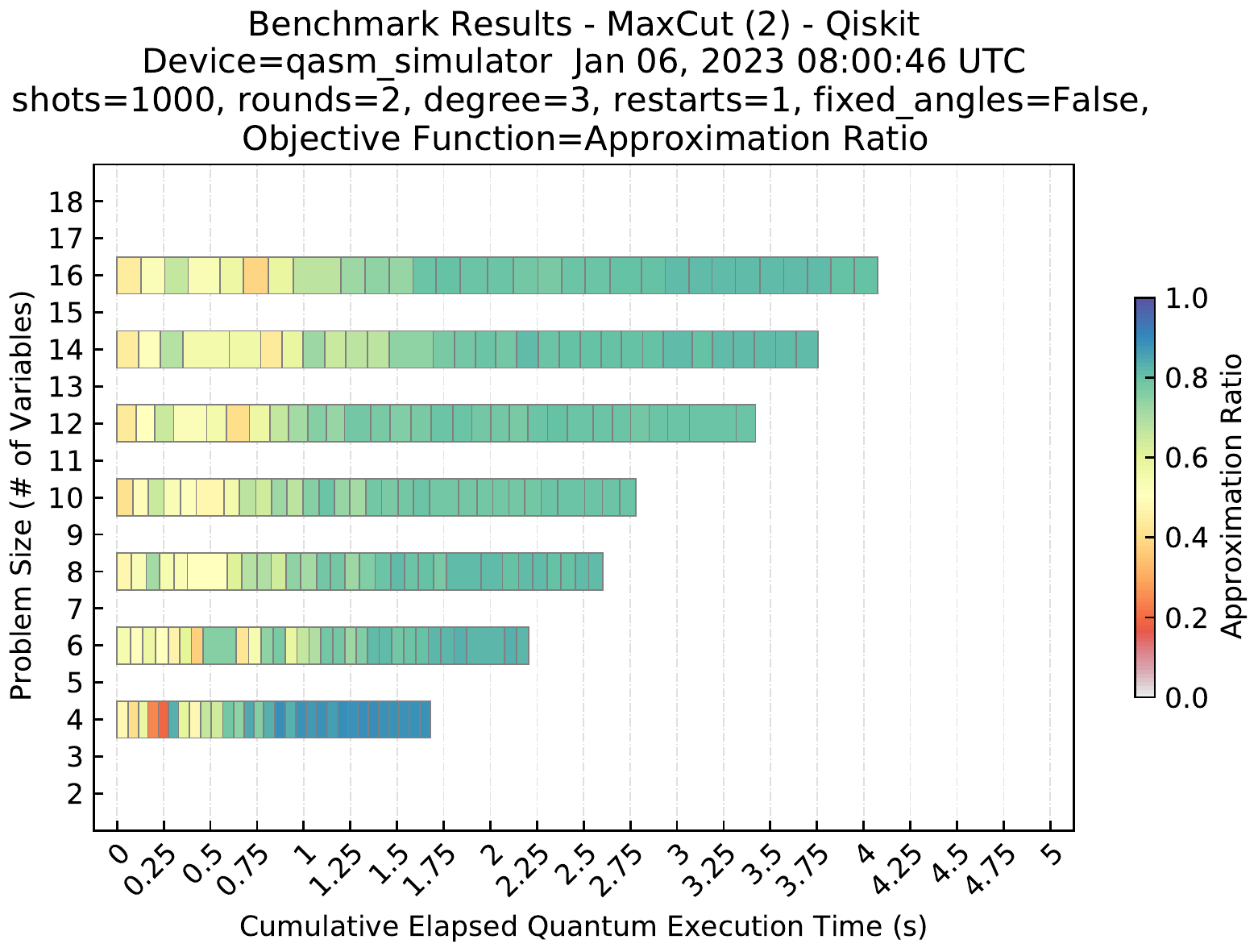}
        \caption{}
        \label{fig:qaoa_area_plot_intro}
    \end{subfigure} \hfill
    \begin{subfigure}{0.32\textwidth}
        \includegraphics[width=\textwidth]{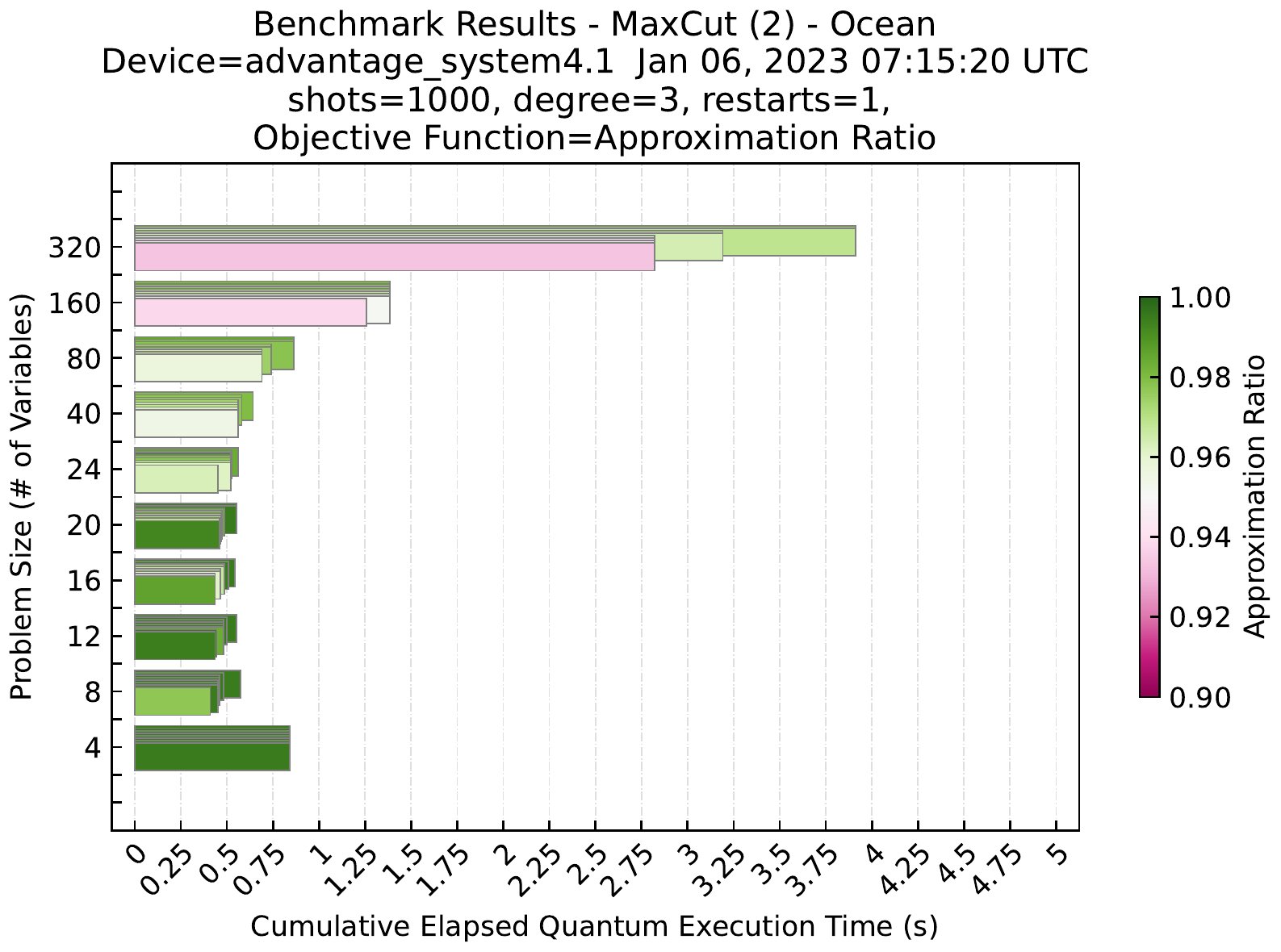}
        \caption{}
        \label{fig:qa_area_plot_intro}
    \end{subfigure}\hfill
    \begin{subfigure}{0.32\textwidth}
        \includegraphics[width=\textwidth]{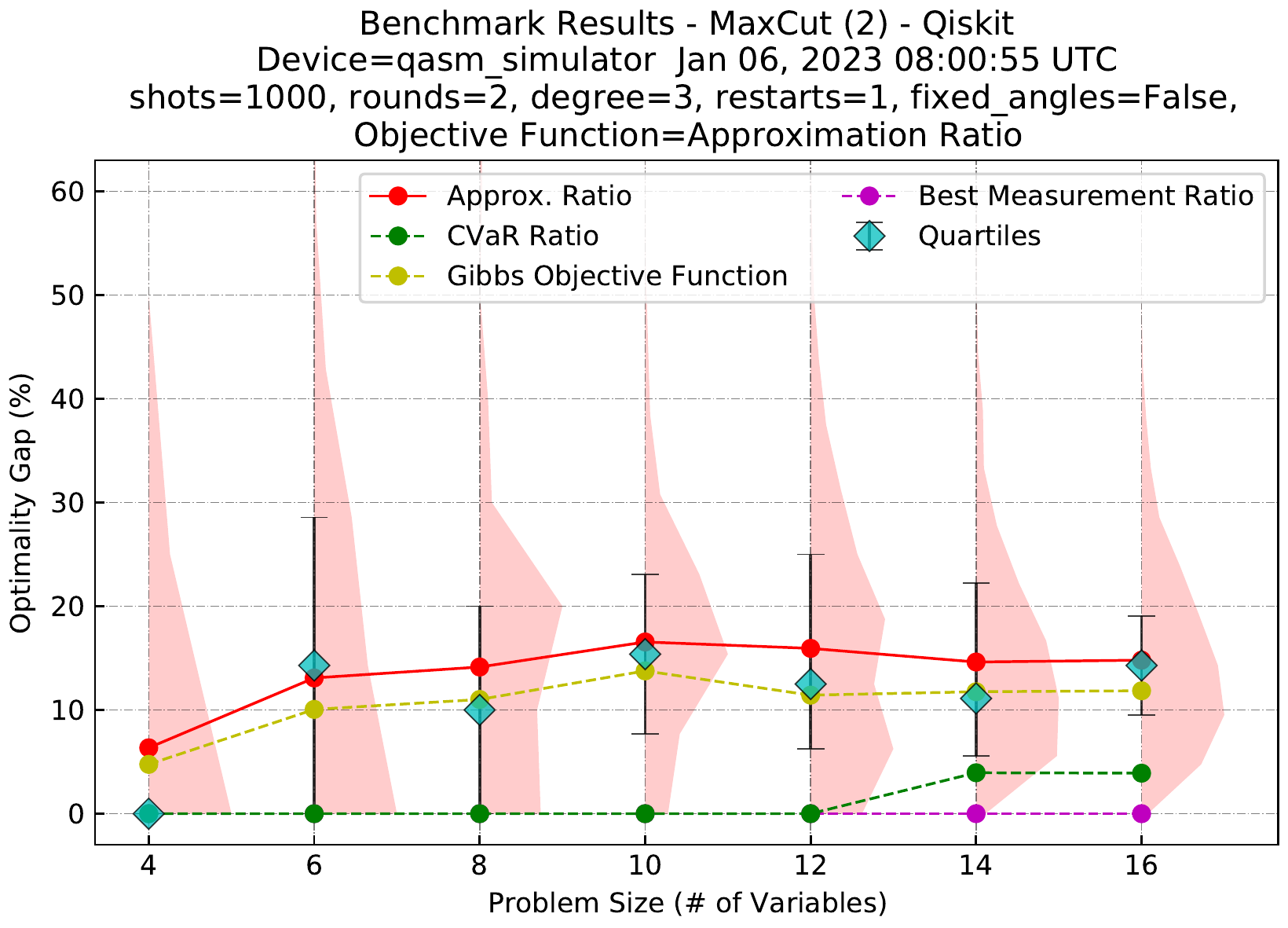}
        \caption{}
        \label{fig:optgaps_intro}
    \end{subfigure}
    
\caption{\textbf{Characterizing Performance of Quantum Computing Solutions.}
These new performance profiles depict the trade-off between result quality and execution time for two quantum computing solutions to the unweighted Max-Cut problem: (a) the Quantum Approximate Optimization Algorithm and (b) Quantum Annealing.
Each row shows a different problem size. The X-axis displays the cumulative execution time, and the rectangle color measures the solution quality defined by the approximation ratio ($= 1-opt\_gap/100$).
For QAOA, successive rectangles depict its iterative execution, tracking the search for appropriate parameters to converge to an optimal solution.
For QA, each stacked rectangle represents a distinct execution at increasing anneal times.
Shown in (c) are several measures of algorithm success, variants of the approximation ratio, plotted over the distribution of final measurements at each problem size.
}
\label{fig:opt_intro_examples_2}
\end{figure*}

\vspace{0.3cm}

The new optimization application benchmark is provided as an enhancement to the existing open-source QED-C Application-Oriented Benchmark suite~\cite{qedc-app-benchmarks,lubinski2023_10061574}.
This is a diverse collection of algorithmic benchmarks for evaluating the performance of (gate-model) quantum computers on problems not currently related to optimization, with support for execution on multiple systems and for collecting, analyzing, and uniformly presenting performance metrics.
Basing our work on this existing framework enabled us to readily extend it with new functionality and make it easy to use and accessible to a broad audience.

The new benchmark exercises multiple components of the integrated hybrid quantum-classical computer systems on which quantum optimization applications run and mimics their real-world use. Combining existing metrics visualization tools with new techniques specific to optimization problems enables the deep exploration of algorithm performance across target systems.
To illustrate these analytics features, we present in~\autoref{fig:opt_intro_examples_2} several visualizations generated by this new benchmark.
While many prior benchmarking studies use Max-Cut as an example of an optimization application, our approach provides a unique level of flexibility, generality, and customization.

\vspace{0.3cm}

We note that the work described in this paper does not include a full-scale comparison between quantum computing systems of different types.  Nor does it address benchmarking of classical solutions to optimization problems, as numerous in-depth studies exist in this area (see~\autoref{sec:optimization_benchmarks}).
The performance results in this paper are intended {\em primarily} to illustrate features and benefits of our benchmarking framework and not to meet methodological expectations for heuristic performance studies from Operations Research. 

The inclusion of benchmarking in the QA algorithm has an essential purpose. QA has been extensively studied over decades, and its performance characteristics on annealing hardware are well understood.  
Including QA illustrates how our benchmark framework readily adapts to quantum computing technologies other than the gate model.
We use QA as a proxy for other solvers that may use quantum technologies in which a large part of the algorithm is executed within the hardware of the remote computing service.
We demonstrate how the framework manages the execution of a series of benchmark problems and collects and analyzes metrics consistently across different technologies.

Our work has identified many variables that impact how well a quantum computer will solve an optimization problem.
However, we did not perform an exhaustive study of these, nor could we tune vendor-specific hardware settings in all cases to achieve optimal results.
As a result, the performance outcomes presented should not be considered generalizable to other test scenarios. A full-scale study of all the factors contributing to quantum performance to tease out the separate contributions of algorithms and hardware is beyond the scope of this paper.

\vspace{0.3cm}

Our contributions to quantum benchmarking are threefold:
\begin{itemize}
    \item Developing a methodology for evaluating the performance of quantum computers running on heterogeneous quantum platforms inspired by standard procedures for assessing classical optimization heuristics.
    \item Implementing and demonstrating an open-source benchmarking procedure for optimization applications that integrates smoothly with the evolving QED-C benchmarking framework and allows users to implement their performance studies easily.
    \item Illustrating the capabilities of this framework and the types of performance analysis that it can support using a familiar \(\text{NP-HARD}\) problem of interest to applications. As an example, we focus on throughput analysis of the application executed on several quantum hardware backend systems as a factor contributing to the total cost of using quantum solutions.
\end{itemize}

We hope this work sheds light on the practical considerations associated with implementing combinatorial optimization solvers on quantum computing systems and will encourage and enable others to measure and record progress in developing quantum algorithms and computing systems.
We propose that the framework could be used to explore many of the recent innovations in quantum algorithms for optimization problems~\cite{Lykov_2023, Zhu_2023,majumdar2021optimizing,herrman2021multiangle,herrman2022relating, Shi_2022,chalupnik2022augmenting,vijendran2023expressive,sciorilli2024largescale,chalupnik2022augmenting}.

\vspace{0.3cm}

The remainder of this paper is structured as follows.
Background on fundamentals of benchmarking the performance of quantum computers and their applications is provided in~\autoref{sec:background}.  
Enhancements to the QED-C Application-Oriented Benchmarks suite are described in \autoref{sec:qedc_benchmarks_iterative} where we describe the benchmark algorithms.
This is followed by a discussion on how we analyze and present the metrics collected by the benchmarks in~\autoref{sec:analysis_benchmark_metrics}.
Results from execution on classically implemented quantum simulators validate that results match expectations and highlight the insights that can be gleaned from these benchmarks. 

In \autoref{sec:execution_on_hardware}, we analyze results obtained from executing these benchmarks on two-gate model quantum hardware systems and a quantum annealing processor using the methods described in the previous section. 
Several appendices are provided at the end of the manuscript to provide detailed information about quantum solutions to combinatorial optimization and to highlight factors that impact the quality of the result, trade-offs in parameter selection, and challenges in scalability inherent in quantum algorithms.


\section{Background}
\label{sec:background}

The benchmarking framework measures performance characteristics of the two leading quantum heuristics for solving combinatorial optimization problems: the Quantum Approximate Optimization Algorithm, which uses a gate-model quantum computer, and Quantum Annealing, which uses an analog quantum computer. 

This paper presents a benchmark of these algorithms in the context of their application to solving the Max-Cut problem.
This section provides an overview of the problem's characteristics and then reviews the quantum solutions we propose to benchmark. We also offer a quick review of existing methodologies for benchmarking and discuss the similarities and differences of our approach with them.


\subsection{The MaxCut Optimization Problem} 
\label{sec:maxcut_problem}

The Max-Cut problem has emerged as a popular benchmark for quantum optimization~\cite{Garey1976, Beaulieu2021MaxCut, Amaro_2022, Zhu_2020} for two reasons:
(1) it is among the most challenging combinatorial optimization tasks, even to obtain an approximate solution, i.e., APX-Hard~\cite{PAPADIMITRIOU1991425, Hastadoptimal2001},
(2) as an unconstrained discrete optimization task, it has a natural encoding as a Quadratic Unconstrained Binary Optimization (QUBO)~\cite{Glover2018QUBO, Zhou_2020} or an Ising model~\cite{Lucas_2014, PhysRevE.58.5355}, ideally fitting current quantum optimization algorithms (QAOA, QA).
These problems often arise when mapping practical applications~\cite{Barahona1988, ALIDAEE1994} to computing hardware and can appear as subroutines in composite algorithms.

The input for a Max-Cut problem is an undirected graph consisting of nodes or vertices ($V$) and edges ($E$). (Each edge of the graph can be accompanied by a `weight', but we only consider unweighted 3-regular graphs in this paper.) A cut is a partition of the graph nodes into two sets. The size of a cut is defined as the number of graph edges that connect nodes belonging to different sets. The Max-Cut problem is identifying a cut with the largest size out of all possible cuts.
(Here, we consider the unweighted version of the MaxCut problem, so that each edge of the graph has the same weight.)

\begin{figure}[t!]
    \includegraphics[width=0.40\textwidth]{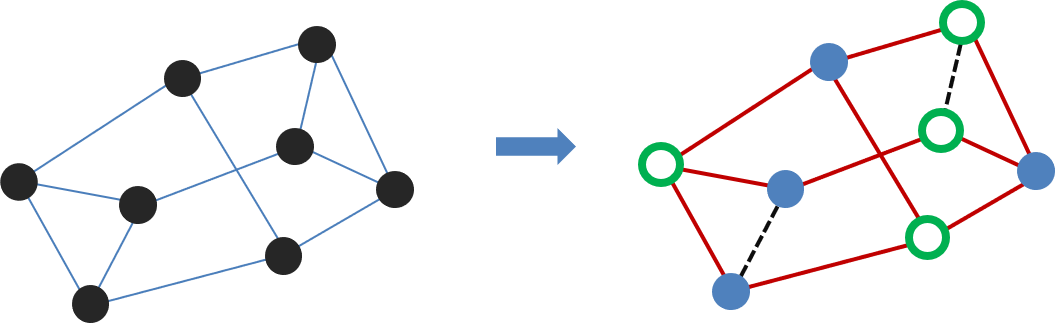}
    \caption{\textbf{The MaxCut Problem.} For an undirected graph consisting of nodes, or vertices, ($V$) and edges ($E$), partition the vertices into complementary sets such that the number of edges between the sets is the greatest.
    This graph shows one solution to one instance of the MaxCut problem for a graph with eight nodes, using colored nodes and edges. Nodes with different colors belong to the two sets of the solution cut. The number of solid red edges that connect nodes from different sets is the MaxCut of that graph.}
    \label{fig:maxcut_problem}
\end{figure}

\autoref{fig:maxcut_problem} presents a representative 8-node graph in which each node is connected to others by precisely three edges.
This type of graph is known as 3-regular or of degree 3.
Graphs of degree 3 are considered the most difficult to solve~\cite{BAZGAN2008510}.
This figure shows one solution to the Max-Cut problem, using colored nodes and edges. 
The number of red edges that connect the nodes is the maximum cut of the graph.

\vspace{0.3cm}

Due to the challenges associated with finding even approximate solutions to the MaxCut problem at a larger scale (for both classical and quantum algorithms), a quantity called the approximation ratio is often computed to characterize the quality of the solution obtained.
For example, in the problem depicted in~\autoref{fig:maxcut_problem}, the Max-Cut size is 10 (out of 12 total edges).
A naive optimization algorithm that randomly tested various cuts but ran out of time to test them all might conclude that the largest cut size was 9.  
In this case, the approximation ratio would be a number smaller than $9/10$ or $0.90$, as it is a statistical function of the distribution of all solutions found.

It is common to report the quality of a result in terms of its distance from an optimal solution when benchmarking classical algorithms for optimization problems.
The optimality gap is related to the approximation ratio by~\autoref{eq:optimality_gap}.
\begin{equation}
    \label{eq:optimality_gap}
    optimality\_gap = (1.0 - approximation\_ratio) \times 100
\end{equation}

Both of these measures of quantum system performance are relevant in distinct contexts. In our work, we incorporate both metrics to facilitate discussions on the outcomes attained with our benchmarking implementation.


\subsection{Quantum Algorithms for Optimization Problems} 
\label{sec:quantum_algorithms}

Quantum annealing and circuit-based quantum computers solve a combinatorial optimization problem using fundamentally different strategies.
To provide context for our work, we briefly outline how these quantum algorithms function to find solutions to these problems.
Additional detail about these algorithms is provided in~\autoref{apdx:heuristic_solutions}.

Optimization problems, such as Max-Cut, can be described by a Hamiltonian $H_P$ that is unique to the problem and represents its variables and constraints.
The optimal problem solution then corresponds to this Hamiltonian's ground state(s).
The QAOA and QA algorithms use quantum state evolution to compute the energy expectation value for $H_P$ and identify values for variables $\vv \beta$ and $\vv \gamma$ that yield the lowest energy eigenstate(s) for
\begin{equation}
    F_{\vv \beta, \vv \gamma}\coloneqq \bra{\vv \beta, \vv \gamma} H_P \ket{\vv \beta, \vv \gamma}. \label{eq:problem_hamiltonian}
\end{equation}

The quality of solution, or approximation ratio ($AR$), can then be defined as the ratio of the computed energy state $F_{\vv \beta, \vv \gamma}$ and the true ground state energy $\emin$ (assuming it is known). In our case, the Hamiltonian energies are always negative or zero.
\begin{equation}
    AR = F_{\vv \beta, \vv \gamma} / \emin
    \label{eq:problem_hamiltonian_2}
\end{equation}

\vspace{0.3cm}
\paragraph{Quantum Approximate Optimization Algorithm}

QAOA is arguably the leading candidate for solving combinatorial optimization problems using gate-model quantum processors.
QAOA belongs to the class of Variational Quantum Algorithms (VQA)~\cite{Cerezo_2021} and is usually implemented iteratively wherein a classical optimizer `trains' a parameterized quantum circuit.
QAOA is a heuristic that attempts to solve combinatorial optimization problems, such as QUBO problems. Specifically, the problem is encoded in the form of a specified quadratic function of binary variables, and the objective is to find an assignment for those variables that minimizes the function. 

At the core of QAOA is an `ansatz circuit', a parameterized quantum circuit. 
Measurements in the computational basis at the end of the circuit correspond to sampling from a probability distribution over possible answers to the problem. 
A classical optimizer is used to obtain parameter values likely to produce optimal or near-optimal solutions by repeatedly taking circuit measurements while varying parameter values.

\vspace{0.3cm}
\paragraph{Quantum Annealing}

Quantum annealing effectively addresses optimization problems by using a versatile approach to identifying the global minimum of a function using a systematic process.
The algorithmic approach of QA is inspired by the adiabatic theorem from quantum mechanics to transform an easy-to-prepare ground state of an initial Hamiltonian into the ground state of the `target' Hamiltonian that encodes the combinatorial optimization problem.

At a high level, the protocol strives to identify the low-energy states of a user-specified $H_{\text{Target}}$ model by conducting an analog interpolation process of the following Hamiltonian, arriving at minimum energy states at the end of the evolution:
\begin{equation}
    H(s) = (1-s)H_{\text{Init}} + (s) H_{\text{Target}}.
    \label{eq:qah_bg}
\end{equation}

With quantum annealing, convergence to a solution is performed entirely within the quantum system from the user's view. 
The problem is mapped to an initial state (equal superposition with respect to the problem basis) on the hardware, and the system is set to anneal towards a solution.
Longer annealing times are associated with higher solution quality.


\subsection{Benchmarking Quantum Computers}
\label{sec:benchmarking_quantum_computers}

This section reviews concepts and definitions from prior benchmarking work that we reference throughout this manuscript.
We focus primarily on the application-oriented level of performance evaluation in our benchmarking of combinatorial optimization applications.

\vspace{0.2cm}

\paragraph{System-Level Benchmarks}
\label{sec:system_level_benchmarks}

A large body of reference material exists for gate model computing systems on component and system-level benchmarks~\cite{PhysRevA.77.012307, PhysRevLett.106.180504, Blume-Kohout2017-no,gambetta2012characterization,sarovar2019detecting, Proctor_2022,mckay2023benchmarking}.
We use two well-known system-level performance benchmarks, Quantum Volume (QV)~\cite{Cross_2019,qiskit_measuring_quantum_volume} and Volumetric Benchmarking (VB)~\cite{BlumeKohout2020volumetricframework,proctor2020measuring} as a backdrop in several of our benchmark plots.
While these two methods characterize quantum circuit execution quality and scale, neither provides information about the time it takes a program to run, which is a critical factor in evaluating the total cost of any computing solution.

Quantum annealing systems have been available for empirical study since 2011~\cite{Johnson2011ManuQA, Perdomo_Ortiz_2019, Chakrabarti2022}.
Examples of component- and system-level approaches to evaluating quantum annealing processors may be found in early papers
~\cite{Johnson2011ManuQA,Dickson2013ThermallyAQ, Lanting_2014} and in recent proposals for benchmarking of large-scale quantum annealing hardware~\cite{9319535,9465651,PRXQuantum.3.020317}.


\vspace{0.2cm}
\paragraph{Application-Level Benchmarks}
\label{sec:application_level_benchmarks}

\begin{figure}[t!]
\includegraphics[width=0.92\columnwidth]{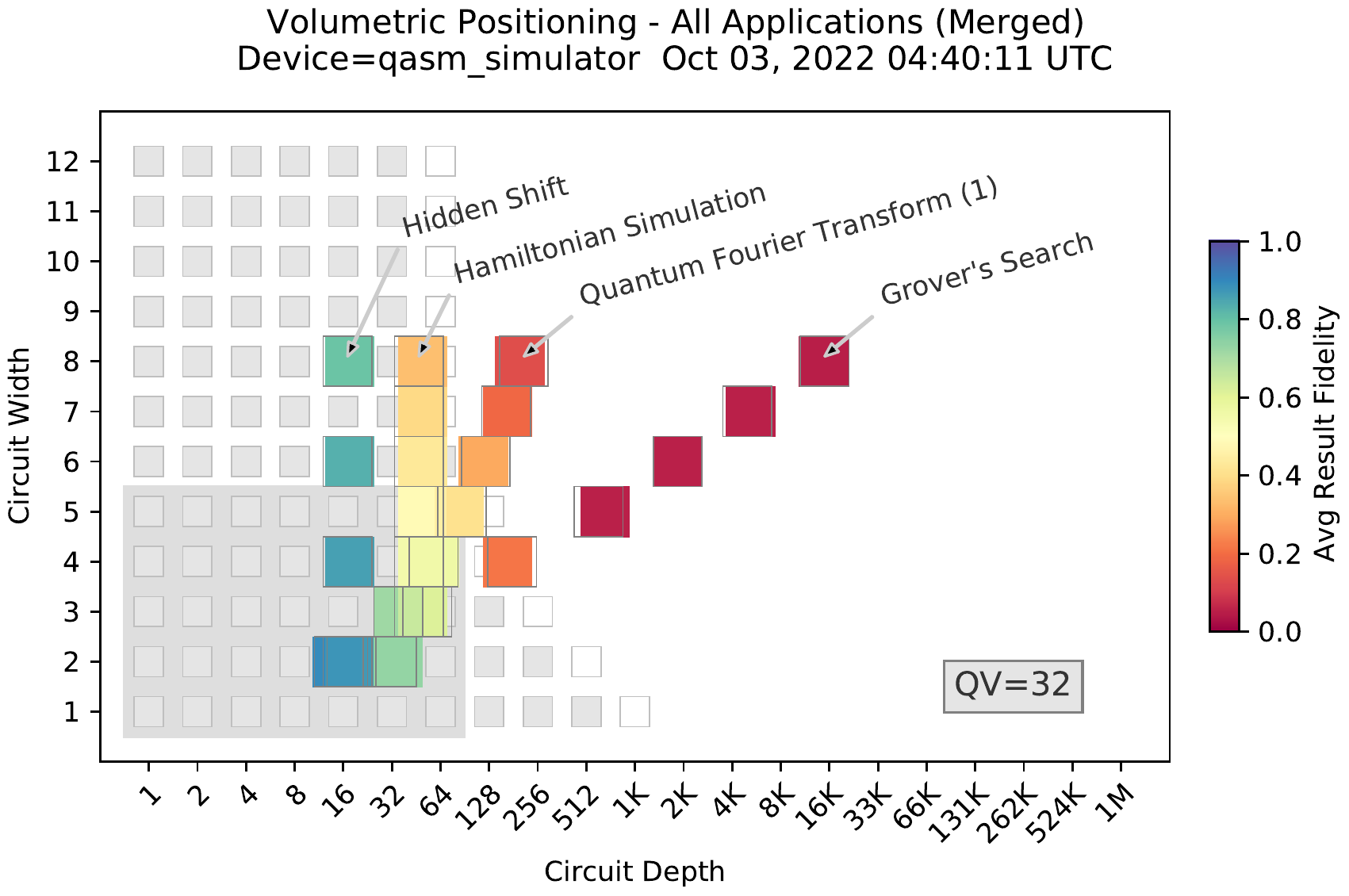}
\caption{\textbf{Application-Oriented Benchmarks.} Here, we present one way to illustrate the result fidelity obtained by executing several application-oriented benchmarks up to 8 qubits on a noisy quantum computer simulator. The plot shows the average result fidelity as a function of circuit width and the circuit depth plotted on a volumetric background to visualize the `profile' and result fidelity of the benchmark circuits.
\emph{(plot produced by QED-C benchmark suite)}
}
\label{fig:qedc_benchmark_profile_1}
\end{figure}

Component and system-level metrics offer valuable insights into overall system capability, but predicting the effectiveness of a machine with a certain level of general performance for a specific application class can be challenging~\cite{murphy2019controlling, proctor2020measuring}.
To address this, application-focused benchmarks run well-defined programs tailored to provide application-specific performance metrics.

Due to its relative maturity, benchmarking of QA tends to involve application-level tests using models with more than 100 qubits, which presents significant challenges for validation by comparison to the classical simulation of ideal quantum systems~\cite{mcgeoch2013, King2015TTT, Subires_2022, Sinha022ResearchQA}.
Such benchmarking work often compares the runtime performance of quantum hardware to that of classical methods~\cite{doi:10.1126/science.220.4598.671, Gey1991, Zhu2015,mcgeoch2013,coffrin2019evaluating}.
Using synthetic optimization problems, a problem instance with a known optimal solution is planted~\cite{PhysRevA.92.042325,2005.14344,2103.08464}. 

In contrast, early-stage gate model quantum computers require benchmarks that involve smaller problems and numbers of qubits.
Application-oriented benchmark frameworks typically create circuits that use well-known quantum gate combinations or algorithms, provide inputs and expected outputs, and execute them on quantum simulators or physical hardware~\cite{lubinski2023_10061574,qedc-app-benchmarks, QASMBench2020,supermarq2022, QPackScores2022, QUARK2022, Schoot_2022}.
Result quality metrics are computed using statistical differences between expected and actual measurements or proximity to an application-specific metric derived from the measurements. 

\vspace{0.3cm}

Of particular relevance is the first QED-C application-oriented benchmark suite \cite{lubinski2023_10061574,qedc-app-benchmarks}, upon which our work is based.
The QED-C suite offers a practical methodology to evaluate the performance of various quantum programs across a range of quantum hardware and simulator systems.
Its benchmark programs sweep over a range of problem sizes and input characteristics while systematically capturing key performance metrics, such as quality of result, execution run-time, and quantum gate resources consumed, as shown in~\autoref{fig:qedc_benchmark_profile_1}.
Supporting infrastructure and abstractions make these benchmarks accessible to a broad audience.
The framework also provides the structure to enable benchmarking of iterative algorithms, such as QAOA, or to execute an algorithm in a single operation, as in QA.

The QED-C benchmarks compute several important figures of merit, which we use throughout this manuscript. 
The quality of result for individual circuits is given by the ``Normalized Hellinger Fidelity'', a modification of the standard ``Hellinger Fidelity'' that scales the metric to account for the uniform distribution produced by a completely noisy device.
Resource consumption is quantified as the total number of gate layers, or ``Circuit Depth'', which can be ``Algorithmic'' or ``Normalized'' (transpiled to a normal basis).
Execution time is captured as ``Elapsed Quantum Execution Time'' (wall clock) or ``Quantum Execution Time'' (reported by quantum provider service) with more granularity possible in some systems.
These metrics are defined in detail in our prior work~\cite{lubinski2023_10061574}.

The framework to capture run-time metrics is fundamental to our new work in benchmarking algorithms for combinatorial optimization.
Although execution time was included in the first \mbox{QED-C} benchmarks, benchmarking iterative and hybrid algorithms can provide a more complete picture of a quantum computer's run-time performance.
The time required to execute circuits within a quantum application is an essential metric of performance \cite{johnson_faro_2021, Bertels_2020, M_ller_2017, cao_hirzel_2020},
often cited when comparing quantum and classical computations.
For example, quantum and classical computing times were central to recent demonstrations of quantum advantage~\cite{Arute2019-mk, Zhong2020-rk}.


\subsection{Quantum Benchmarks for Optimization Tasks}
\label{sec:optimization_benchmarks}

Much background material is available on benchmarking classical solutions to optimization problems~\cite{mcgeoch2013,coffrin2019evaluating,Lee_2021} and comparing quantum methods to classical solutions \cite{2018APS..MARR15007W,crooks2018,Tasseff2022QAPotential}.
This work is often oriented toward improving the performance of quantum algorithms, separate from the analysis of specific hardware platforms.
With QA and QAOA, this approach generally involves comparing different implementations or tuning strategies of the quantum algorithm rather than performing runtime comparisons against classical algorithms.

Quantum computation based on quantum annealing techniques has been used to address combinatorial optimization problems for more than a decade~\cite{mcgeoch2013, Djidjev2018EfficientQA, Tasseff2022QAPotential, Trummer2015, Tasseff2022Emerging, Albash_2018,coffrin2019evaluating,doi:10.1126/science.abo6587}.
In these studies, the primary metrics of interest are expected solution quality and \emph{operation counts} (i.e., circuit depth), which are used as approximate runtime measurements. 
Within this context, two main threads of execution time analysis have been used in comparing the performance of the quantum annealing algorithm to classical heuristics.   
The {\em time-to-solution} (TTS) metric determines the expected wall-clock time required to solve a problem to optimality~\cite{R_nnow_2014, Mandr__2018, Albash_2018, Perdomo_Ortiz_2019,doi:10.1126/science.abo6587}.
In contrast, the {\em time-to-target} (TTT) metric determines the expected wall-clock time required to solve a problem to a specified {\em target} solution quality ~\cite{mcgeoch2013, King2015TTT, Trummer2015, Tasseff2022QAPotential}.
The typical approach is to measure how TTS or TTT changes as a function of problem size for both the classical and quantum methods (i.e., a scaling advantage) so that one can forecast at what system sizes the quantum solution approach is likely to be faster than the classical one.
In this work, we use these concepts to inform our analysis of the trade-off between the quality of the solution and the time it takes to execute the quantum algorithm.

\vspace{0.3cm}

Significant challenges exist for benchmarking quantum solutions on real-world hardware, as quantum computer noise characteristics and runtime overhead introduce additional requirements.
Due to the implementation complexity of configuring optimization tasks for benchmarking quantum hardware, several software frameworks have emerged to support the evaluation of the same (or similar) optimization methods implemented on different platforms.
Benchmark frameworks such as SupermarQ \cite{supermarq2022} and QPack Scores \cite{QPackScores2022} include one or more QAOA applications in their sample benchmarks, while QUARK \cite{QUARK2022} considers specific optimization problems arising in industry.
The Q-score metric \cite{Schoot_2022} is claimed to apply to quantum processors in several categories, measuring the size of the largest graph for which the solver outperforms random guessing within a fixed time limit. 
All references present results that measure solution validity, feasibility, and run-time on several backend quantum computers, some on both gate model and quantum annealing devices.
We used much of this work to guide our development of new benchmarks in the QED-C suite based on combinatorial optimization problems.



\section{QED-C Benchmark Framework Enhancements}
\label{sec:qedc_benchmarks_iterative}

In this section, we describe the enhancements to the QED-C benchmark suite that collect and analyze application-specific quality and temporal metrics associated with hybrid quantum applications, e.g., QAOA and QA, where the trade-off between the quality of solution and utilization of resources (here execution time) is essential.
Our effort has two primary goals:
1) to integrate and enhance critical concepts from other optimization-centric benchmark efforts into the QED-C benchmark suite as a standard feature and
2) to present the results and analysis in ways recognizable by practitioners in the operations research field who are already familiar with benchmarking classical solutions to optimization problems.

This section describes specific features of the framework we developed for cross-paradigm quantum optimization heuristics benchmarking.  
The enhancements to our original benchmark framework are driven by specific features and challenges of optimization problems and heuristic performance evaluation, distinct from the simple test scenarios used in our initial benchmark suite.
The OR community has developed methodologies and tools for evaluating computational performance in this context, some of which we have adapted to the quantum scenario.
See Appendix \ref{apdx:theory} for a discussion of the theoretical foundations.

Several features distinguish our benchmarking framework from others.
Users can evaluate both execution time and solution quality in detail and explore the trade-offs (as opposed to fixing a specific TTS or TTT metric).
The platform supports benchmarking of quantum computing hardware that can run quantum annealing or gate model algorithms.
It also provides the ability to select problems and inputs of interest beyond our simple illustrations using Max-Cut inputs.
Presentation of benchmark results is aligned with standard methodologies of Operations Research and the QED-C framework. 
As quantum computers grow in size, the benchmark framework will be able to support testing on a wide variety of optimization problems.

\vspace{0.3cm}

We focus on its application to a combinatorial optimization problem to demonstrate the key enhancements, using Max-Cut as a specific example.
Unlike the simple algorithms used in the initial benchmark suite, where circuit execution fidelity is the key metric, the enhanced benchmark must derive a solution quality metric that is application-specific and accounts for the fact that solutions to optimization problems are often approximate.
Results from its execution on a classically-implemented circuit-based quantum simulator illustrate how key metrics are collected and presented.

The QED-C benchmark framework includes shared functions to manage the execution of benchmark algorithms over a range of problem definitions, collect metrics during execution, and present results consistently across backend targets.
Both the Max-Cut QAOA and QA benchmark algorithms operate on a target system $\av{backend\_id}$, sweeping over a range of problem sizes $[\av{min\_size},\av{max\_size}]$, to solve a $\av{problem}$ defined by input $\av{args}$. 
An inner loop, controlled by the $\av{max\_restarts}$ argument, provides the ability to execute the benchmark algorithm multiple times at each problem size. 

For each problem size tested, we consider a set of random 3-regular graphs using the $networkx$ package~\cite{networkx} and determine the maximum cut size for each using the $gurobi$ package~\cite{gurobi}.
These are used to generate the quantum circuits for testing and to determine solution quality after execution respectively.

A key practical difference between the QAOA and QA algorithms lies within the restart loop, where the specific solvers are applied to the input, and the quality of the solution is evaluated over increasing execution times.
A gate model device iterates through a series of quantum circuit executions, testing parameter values, to find a set that yields a low-energy state.  
In contrast, a quantum annealing system gradually attempts to reach its lowest energy state as a transverse Ising model undergoing quantum mechanical evolution.
Comparing the evolution of the state over time in these systems requires different data collection and presentation.
Below, we detail the related algorithms used to benchmark these solutions and highlight differences between them and how they impact the results, omitting some details for brevity.


\subsection{Benchmark Algorithm for QAOA}
\label{sec:benchmark_algorithm_qaoa}


\begin{algorithm}[t!]
    \scriptsize
    \caption{Benchmark Algorithm for QAOA}
    \label{alg:benchmark_execution_loop_qaoa}
    
    \begin{algorithmic}[1]

    \State $target \gets \av{backend\_id}$
    \State $initialize\_metrics()$
    \For{$size \gets\av{ min\_size}, \av{max\_size}$} 
        \State $circuit\_def \gets  define\_problem(\av{problem}, size, \av{args},\av{rounds})$
        \For{$restart\_id \gets 1, \av{max\_restarts}$} 
            \State $cost\_function \gets  define\_cost\_function(problem)$
            \State $circuit,num\_params \gets create\_circuit(circuit\_def)$
            \State $cached\_circuit \gets compile\_circuit(circuit)$
            \State $params[{\vv \beta, \vv \gamma}] \gets random(num\_params)$
            \While{$minimizer()\  not\  done$}   \Comment{$minimizing$}
                \State $circuit \gets apply\_params(cached\_circuit,params)$
                \State $\ar{counts} \gets execute(target, circuit, \av{num\_shots})$
                \State $energy,quality \gets cost\_function(counts)$
                \State $store\_iteration\_metrics(\ar{quality},\ar{timing})$
                \State $params[{\vv \beta, \vv \gamma}] \gets optimize(params[{\vv \beta, \vv \gamma},energy])$
                \State $done \gets True \ if\ lowest(energy)\ found$
                \State $done \gets True \ if\ iteration\_limit\_reached()$
            \EndWhile
            \State $compute\_and\_store\_restart\_metrics()$
        \EndFor
        \State $compute\_and\_store\_group\_metrics()$
    \EndFor

   \end{algorithmic}
\end{algorithm}

The QAOA benchmarking method is defined in Algorithm~\ref{alg:benchmark_execution_loop_qaoa}.
Nested within the first and second for loops is the QAOA algorithm, which defines a cost\_function based on the problem specifics and a gate model quantum circuit that implements the Hamiltonian associated with the problem and is parameterized by variables $\vv \beta$ and $\vv \gamma$.
The quantum circuit used with QAOA can be replicated by some number of $\av{rounds}$ (often referred to as $p$ in code). 

Starting with a random or fixed set of parameters $\av{params}$ (corresponding to $\ket{\vv \beta, \vv \gamma}$ from~\autoref{eq:qaoa_ansatz}. the quantum circuit is executed $\av{shots}$ times to obtain the measurement $\ar{counts}$ and compute a value for the cost function.
Classical optimizer code explores the parameter landscape by varying the set of parameters to obtain measurement counts representing the Hamiltonian's lowest energy state, iterating until either the lowest energy is determined or an iteration limit is reached.
A relevant $\ar{quality}$ metric is calculated and stored along with metrics that track the quantum and classical $\ar{timing}$ information. 
Although we use random starting parameters for the benchmark, users may have some information about a reasonable starting point in practice, which could result in a better or faster solution (see \autoref{sec:fixed_angle_conjecture}). 

The results of this algorithm's execution can be affected by a number of factors unique to QAOA, such as the number of shots and rounds, the type of classical optimizer employed, and, most importantly, the noise level in the target system. 
The quality of the results can also be constrained by limiting the number of iterations the classical optimizer performs. 
In the QAOA benchmark, this is a configurable option, but we set this limit to 30 by default to avoid runaway execution on costly hardware.  

The Max-Cut benchmark can be executed using two different methods, enabling the study of these factors independent of the complete QAOA algorithm.
Method (1) executes one instance of the ansatz circuit for a specific problem using configurable shots and rounds, permitting detailed analysis of these factors.
Method (2) executes the complete QAOA algorithm and provides the option to specify the classical optimizer, with COBYLA as the default.
Additionally, one of several variants of the cost function may be selected. These variants are described below in~\autoref{subsec:app_specific_result_quality}.


\subsection{Benchmark Algorithm for QA}
\label{sec:benchmark_algorithm_qa}

\begin{algorithm}[t!]
    \scriptsize
    \caption{Benchmark Algorithm for QA}
    \label{alg:benchmark_execution_loop_qa}
    
    \begin{algorithmic}[1]

    \State $target \gets \av{backend\_id}$
    \State $initialize\_metrics()$
    \For{$size \gets\av{ min\_size}, \av{max\_size}$} 
        \For{$restart\_id \gets 1, \av{max\_restarts}$} 
            \State $compute\_quality \gets  define\_compute\_quality(\av{problem})$
            \For{$a\_time \gets\av{min\_anneal\_time}, \av{max\_anneal\_time}$}
                \State $embedding \gets define\_problem(problem, size, \av{args})$
                \State $sampler \gets create\_sampler(target, embedding)$
                \State $\ar{samples} \gets sample\_ising(sampler, a\_time, \av{reads})$
                \State $quality \gets compute\_quality(samples)$
                \State $store\_iteration\_metrics(\ar{quality},\ar{timing})$
            \EndFor
            \State $compute\_and\_store\_restart\_metrics()$
        \EndFor
        \State $compute\_and\_store\_group\_metrics()$
    \EndFor

   \end{algorithmic}
\end{algorithm}

The QA benchmarking method is described in Algorithm~\ref{alg:benchmark_execution_loop_qa}.
The core of the QA benchmark algorithm is within the first and second \texttt{for} loops. It uses a special metrics collection loop unique to the QA benchmark.

From the user's perspective, the convergence to a solution using a quantum annealer is performed in a single step, entirely within the quantum system. 
A Hamiltonian describing the problem is embedded into the quantum components of the device and is evolved in time to settle to the lowest energy state, representing an optimal solution.

In our QA benchmark, the QA algorithm is executed multiple times \emph{from the start} within this particular collection loop. The $\av{anneal\_time}$ is initialized to 1$\mu$s and doubled after each execution until it reaches 256$\mu$s (the range of annealing times is a configurable option in the benchmark.)
Each time, the problem is mapped to an initial state (equal superposition with respect to the problem basis) on the hardware, and the system is set to anneal towards a solution.
Longer annealing times are associated with higher solution quality.

Using this approach, we are able to provide a measure of the time versus quality trade-off comparable to the QAOA benchmark by capturing the quality of the solution obtained at each value of the annealing time.
It is impossible to monitor the solution's quality as it evolves within a single execution of the QA algorithm.

For each of the executions performed in the benchmark, the algorithm queries ($\av{reads}$) $\ar{samples}$ the same number of times as we do shots in QAOA. 
As with benchmarking QAOA, a relevant $\ar{quality}$ metric is calculated after each execution and stored along with metrics that track the quantum and classical $\ar{timing}$ information.

The results of this algorithm's execution are affected primarily by the number of shots or samples, the annealing time used, and the noise level in the target execution system. 
The Max-Cut benchmark can be executed using two different methods, enabling the study of these factors independently of the entire QA algorithm.
Method (1) executes at a single, configurable annealing time for each problem size using configurable shots, permitting detailed analysis of these factors.
Method (2) executes the complete QA described in this section algorithm.
Note that in this case, Method (1) is the same as Method (2), with the range collapsed to a single value.


\section{Analysis of Max-Cut Benchmark Metrics}
\label{sec:analysis_benchmark_metrics}

In this section, we discuss our analysis of the data collected as the benchmarks execute. 
First, we discuss the application-specific quality metrics explicitly computed for the Max-Cut problem, produced in common for both QAOA and QA algorithms.
The benchmarking framework generates optimality gap plots and cut-size distribution plots as practical tools to visualize these additional metrics succinctly.
Furthermore, metrics other than the approximation ratio can also be used to assess result quality, which we discuss in this section.

We follow this with our visualization of the trade-off between execution time and the quality of the result that can be obtained. While displaying these data in simple line charts is typical, we generate informative plots that use color to represent the quality of execution, horizontal length to represent the execution time of individual iterations, and vertical/horizontal position to indicate problem size and the cumulative execution time. 
The section ends with our analysis of the effect of shot count and the number of rounds on the quality of result for the QAOA implementation of the benchmark.


\begin{figure}[t!]
\includegraphics[width=0.84\columnwidth]{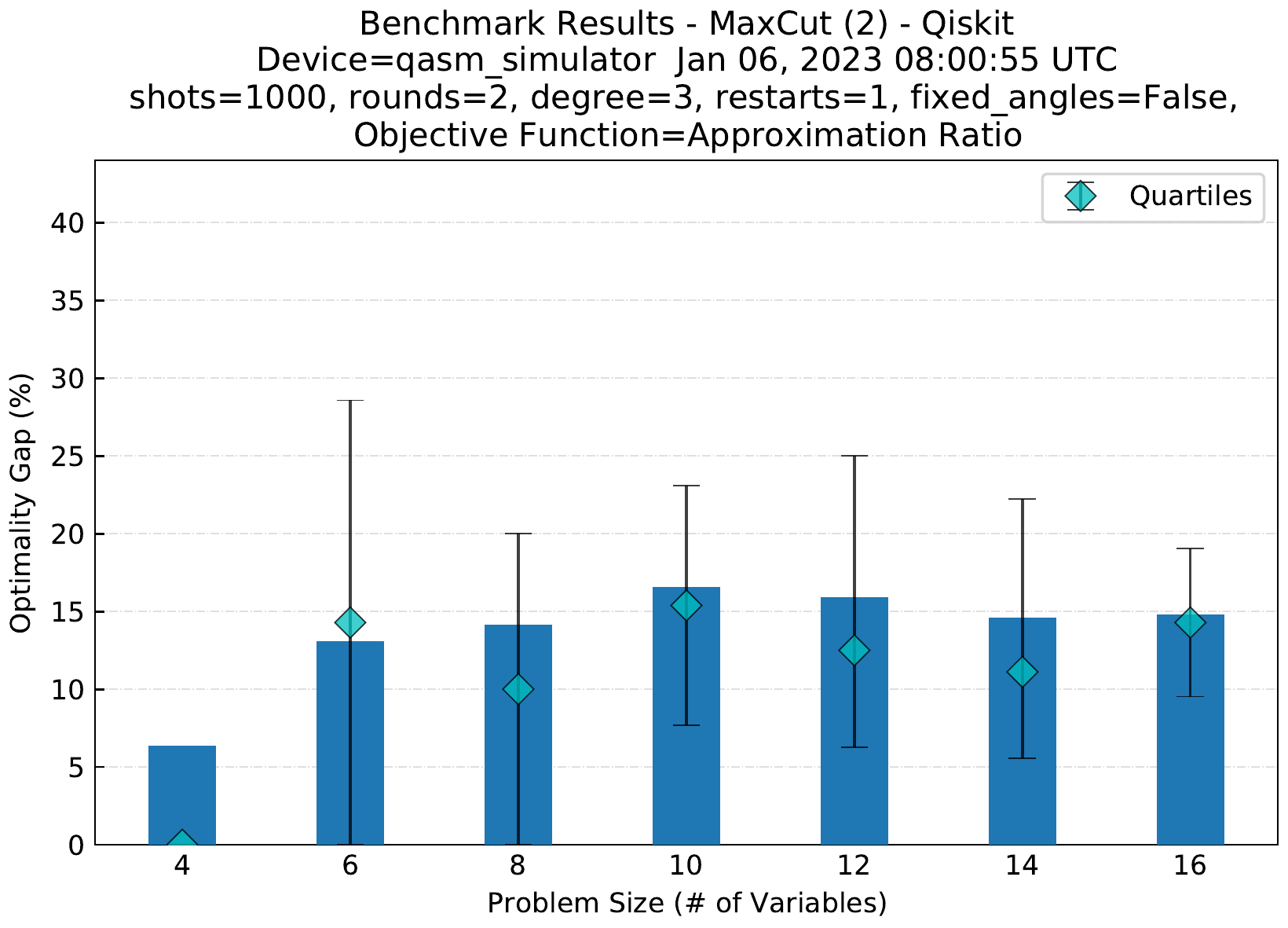}
\caption{\textbf{Showing Final Optimality Gap.} As the QAOA program executes, the minimizer finds an optimal solution to the Max-Cut problem, represented by the approximation ratio computed after the final iteration. The optimality gap for each problem size is computed from these values and shown in this bar chart, along with quartile marks showing the distribution of the final measurement results.
In some cases, the quality of the results could improve with additional execution time, but we limit the benchmark to 30 iterations to conserve computing resources.}
\label{fig:optgaps_barchart}
\end{figure}

\subsection{Application-Specific Result Quality Metrics}
\label{subsec:app_specific_result_quality}

The QAOA and QA algorithms attempt to converge on a solution to an optimization problem by finding the lowest energy state of a Hamiltonian after a sequence of parameter tests (QAOA) or annealing operations (QA) on a target system.
In both cases, the result is an energy value computed as a function of the distribution of the energy samples obtained at the end of the execution.

This energy value can be compared against the precomputed optimal solution (i.e., the lowest energy of the Hamiltonian).
The ratio between the actual and expected energy values, or the approximation ratio~\cite{Zhou_2020,farhiQuantumApproximateOptimization2014, WillschBenchmarking2020}, is the metric commonly used to quantify the quality of the results for the Max-Cut problem, as discussed in~\autoref{sec:maxcut_problem}.
It is formally defined as follows.

Let $M$ denote the number of shots so that for given values of (vectors) $\vv \beta, \vv \gamma$, we obtain $M$ cut-sizes, one corresponding to each of the bit-strings obtained from computational basis measurements.
Let these energies be denoted by $E_1,\dotsc, E_M$, arranged in non-decreasing order. Since these energies are $\leq 0$, $|E_1|, \dotsc, |E_M|$ are non-negative integers arranged in non-increasing order. Then, the energy expectation value is approximated by
\begin{align}
    F \approx \frac{\sum_{i=1}^M E_i }{M}.
    \label{eq:energy_def}
\end{align}

Normalizing the result of this computation to the range $[0,1]$ is convenient. Hence, we define the approximation ratio in
\begin{align}
    \text{Approximation Ratio } r &= \frac{F_{\vv \beta, \vv \gamma}}{|\emin|},
    \label{eq:all_metrics}
\end{align}
where $\emin<0$ is the actual ground state energy of the problem Hamiltonian.

Optimization performance studies typically use the complement of this, the optimality gap.
In \autoref{fig:optgaps_barchart}, we show the final optimality gap computed at the end of the execution of the benchmark algorithm for each of the problem sizes tested.
In this plot, we include quartile bars, which provide information on the width of the distribution in addition to the mean. 
Although these results were produced using the QAOA algorithm on a classically implemented quantum simulator, the results of our QA algorithm are also plotted in this fashion.

The approximation ratio is a valuable measure of a quantum computing system's ability to solve the Max-Cut problem.
The higher the mean of the cut sizes found in the distribution, the more likely the algorithm will produce a cut size that is the maximum.
However, we note that the final output of these quantum algorithms is not the approximation ratio but the best-measured cut (i.e., the cut corresponding to the largest cut size) obtained across all iterations of the algorithm.

The best-measured cut is often a poor measure of the quality of the result because it is numerically unstable, particularly with smaller numbers of samples.
However, it has inspired a few other objective functions, such as the Conditional Value at Risk or CVaR~\cite{barkoutsosImprovingVariationalQuantum2020}, and the Gibbs objective function~\cite{LiQuantum2020}.
Both metrics focus on the tail end of the distribution rather than treating all measurements equally.

\begin{figure}[t!]
\centering
\includegraphics[width=0.90\columnwidth]{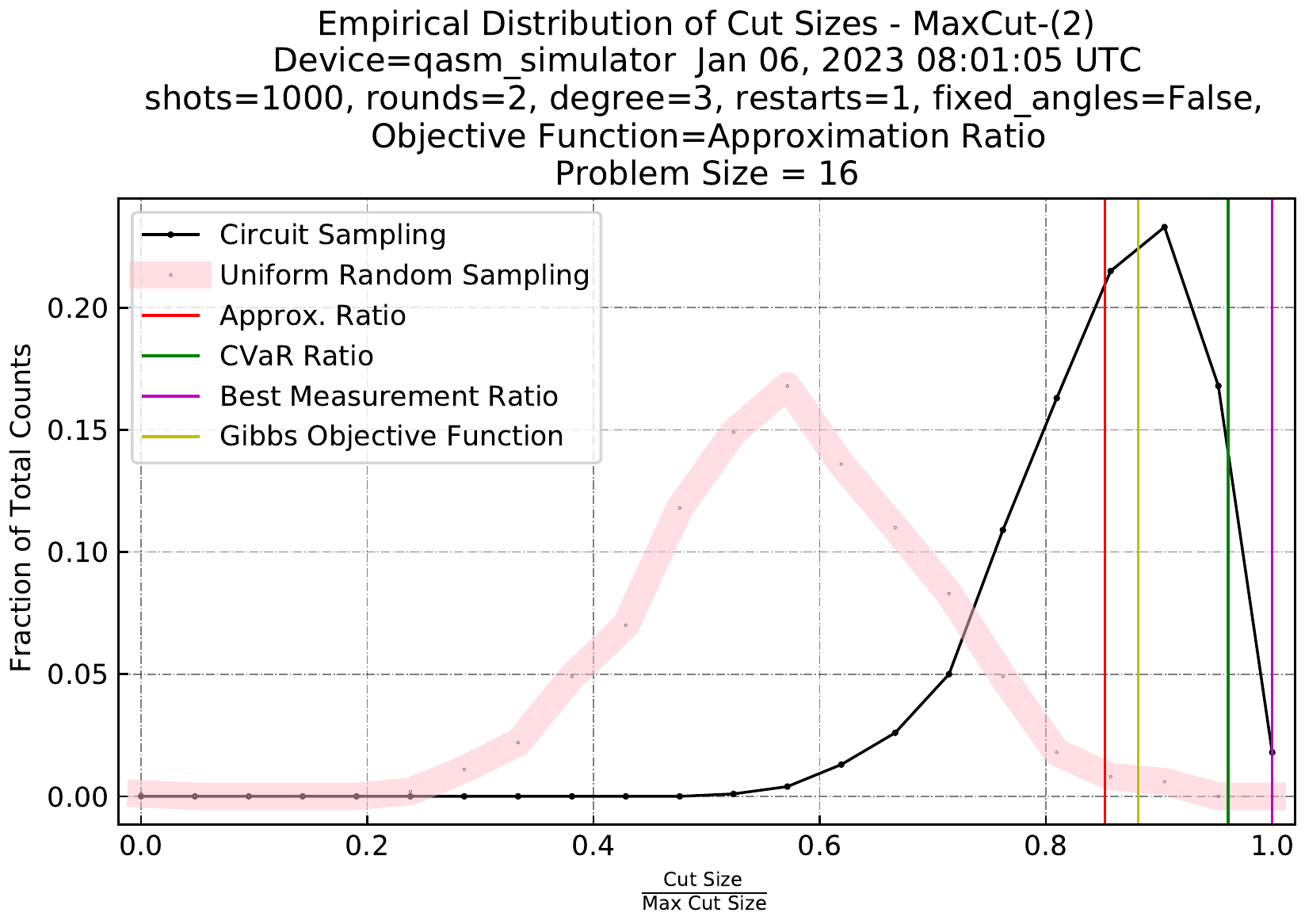}
\caption{{\bf Cut Size Distribution:} The quality of the final output of QAOA can be understood by inspecting the distribution of the cut size values obtained at the final optimizer iteration. A distribution peaked closer to the right indicates higher result quality. Also plotted here are the various metrics (vertical lines) and the distribution corresponding to a uniform random sampling of bit-strings (pink line).}
\label{fig:cutsize_width=20}
\end{figure}

To illustrate how these different quality metrics relate to each other, in~\autoref{fig:cutsize_width=20}, we illustrate the distribution of cut sizes produced from the execution of our benchmark on a noiseless quantum simulator at a problem size of 16 with 1000 samples taken (shots).
The distribution obtained from our benchmark is shown with a black line.
A wide pink line shows a simulated distribution that would be obtained by executing the algorithm on a computing system that returns uniformly random results. 
A distribution that peaks closer to the right indicates a higher result quality.
In this case, the best-measured result is shown at 1.0, indicating that the algorithm returned the expected optimal Max-Cut.
The CVar and Gibbs ratios fall between the approximation and best-measured ratios.

Formally, CVaR~\cite{barkoutsosImprovingVariationalQuantum2020}, for a chosen value of parameter $\alpha\in (0,1]$, is defined as
\begin{align}
    \textrm{CVaR}_\alpha (\vv \beta, \vv \gamma) &= \frac{1}{\ceil{\alpha N}} \sum_{i=1}^{\ceil{\alpha N}} E_i, \label{eq:CVaR_def}
\end{align}
where $\ceil .$ denotes the ceiling function. CVaR$_\alpha$ denotes the mean value of (the negative of) cut-sizes over the lower $\alpha$-tail of the measured energy distribution. The limit $\alpha\rightarrow 0$ corresponds to the ground state energy value (i.e., $E_1$), while $\alpha=1$ corresponds to the energy expectation value $F_{\vv \beta, \vv \gamma}$. The value of the metric depends on the choice of $\alpha$.
While the value of this parameter can be configured, we default to $\alpha=0.1$ in all plots and analyses.

Another choice is the Gibbs objective function~\cite{LiQuantum2020}, which is defined as 
\begin{align}
f_\eta(\vv \beta, \vv \gamma) = \ln \bra{\vv \beta, \vv \gamma} e^{-\eta H} \ket{\vv \beta, \vv \gamma},
\end{align}
with $\eta>0$, and where $H$ denotes a Hamiltonian whose ground state is sought. The parameter $\eta$ determines the relative weights of the low energy states of $H$ in the expression. The parameter $\eta$ tunes $f_{\eta}$ between two extreme values (similar to $\alpha$ in CVaR): $f_{\eta=0}(\vv \beta, \vv \gamma) = F_{\vv \beta, \vv \gamma}$, while $f_{\eta \rightarrow \infty} = \emin$, i.e., the lowest measurable energy of $H$ in the state $\ket{\vv \beta, \vv \gamma}$.

We normalize each of these objective functions so that they lie in the range $[0,1]$ and thus define the following quantities:
\begin{align}
    \begin{split}
    \text{CVaR}_{\alpha}\text{ Ratio} &= \frac{\text{CVaR}_{\alpha}}{|\emin|},\\
    \text{Gibbs Ratio} &= \frac{f_\eta}{\eta |\emin|}, \\
    \text{Best Measurement Ratio} &= \frac{E_1}{|\emin|}.
    \end{split}
    \label{eq:all_metrics_2}
\end{align}

In our benchmarking framework, the objective function may be set to any of these.
The approximation ratio is commonly used in studies of quantum computing solutions to optimization problems, and the other ratios appear less often in the literature.
These are measures of the quality of the solution where a value of 1.0 is optimal. 

\vspace{0.3cm}

\begin{figure}[t!]
    \centering
    \includegraphics[width=0.90\columnwidth]{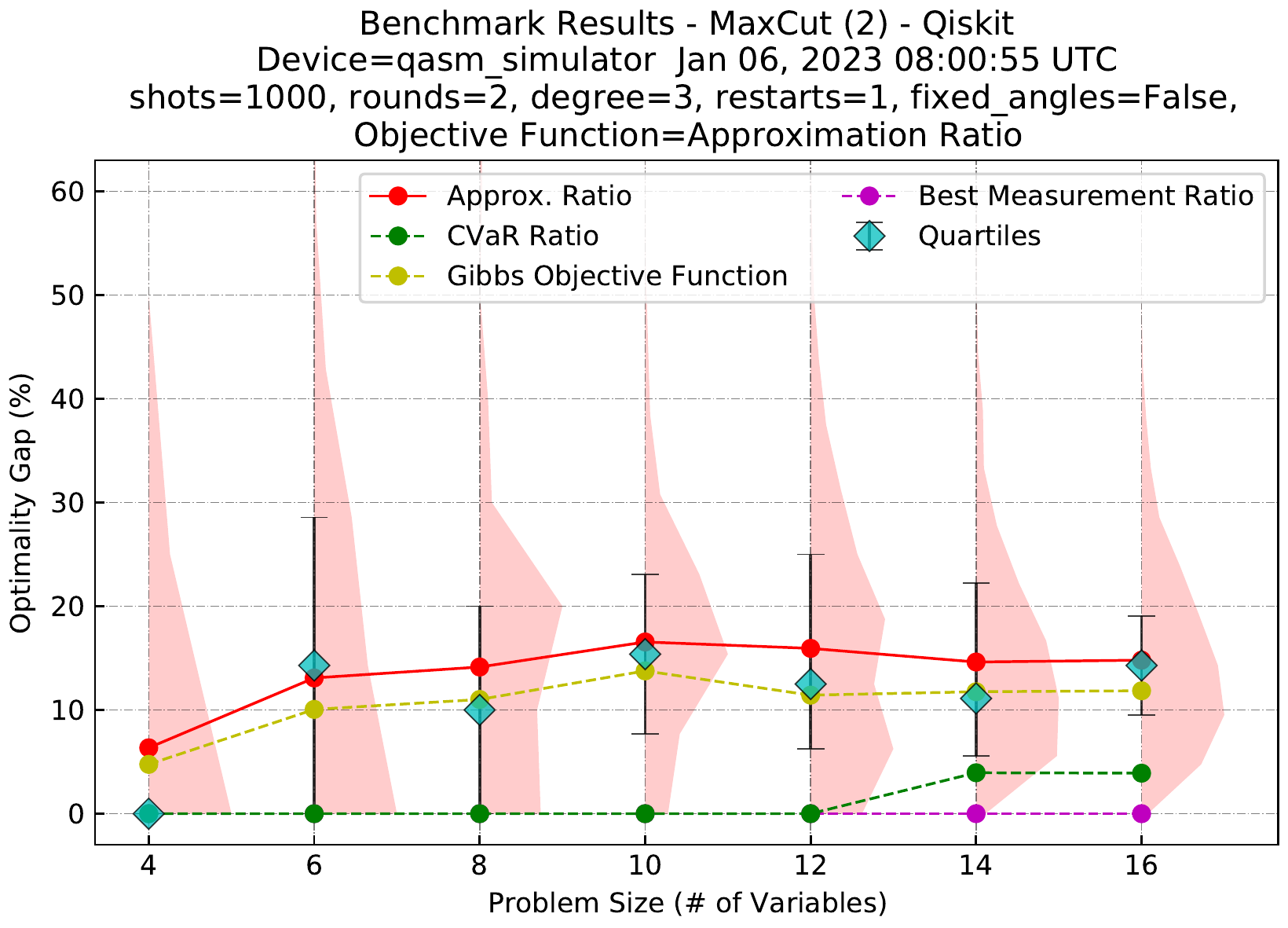}
    \caption{\textbf{Detailed optimality gaps plots.} A variety of metrics can be used to assess the quality of the final distribution of outputs of QAOA. For each implemented problem size or circuit width (along the X-axis), we plot the optimality gap (along the Y-axis). The obtained distributions of optimality gaps are shown as pink half-violin plots. The optimality gap values regarding the CVaR ratio, approximation ratio, Gibbs Ratio, and Best Cut ratio are shown as line plots.}
    \label{fig:optgaps_detailed_AR}
\end{figure}


Our framework also generates `detailed optimality gap plots' (e.g., \autoref{fig:optgaps_detailed_AR}). For each problem size, the empirically obtained distribution of cut sizes is shown using a half-violin plot. (The plotted distribution is that of the quantity $1 - \frac{\text{Cut Size}}{\text{Optimal Cut Size}}$, so it is normalized to be between [0,1]). 
The four metrics in~\autoref{eq:all_metrics}-\ref{eq:all_metrics_2} are shown in terms of their optimality gap, i.e., $(1-metric\_value)*100$. This provides a detailed snapshot of the quality of the result as a function of problem size in terms of various quality metrics.




\subsection{Result Quality and Time of Execution}
\label{subsec:area_plots}

\begin{figure*}[ht!]
\centering
    \hspace*{\fill}
    \begin{subfigure}{0.44\textwidth}
        \includegraphics[width=\textwidth]{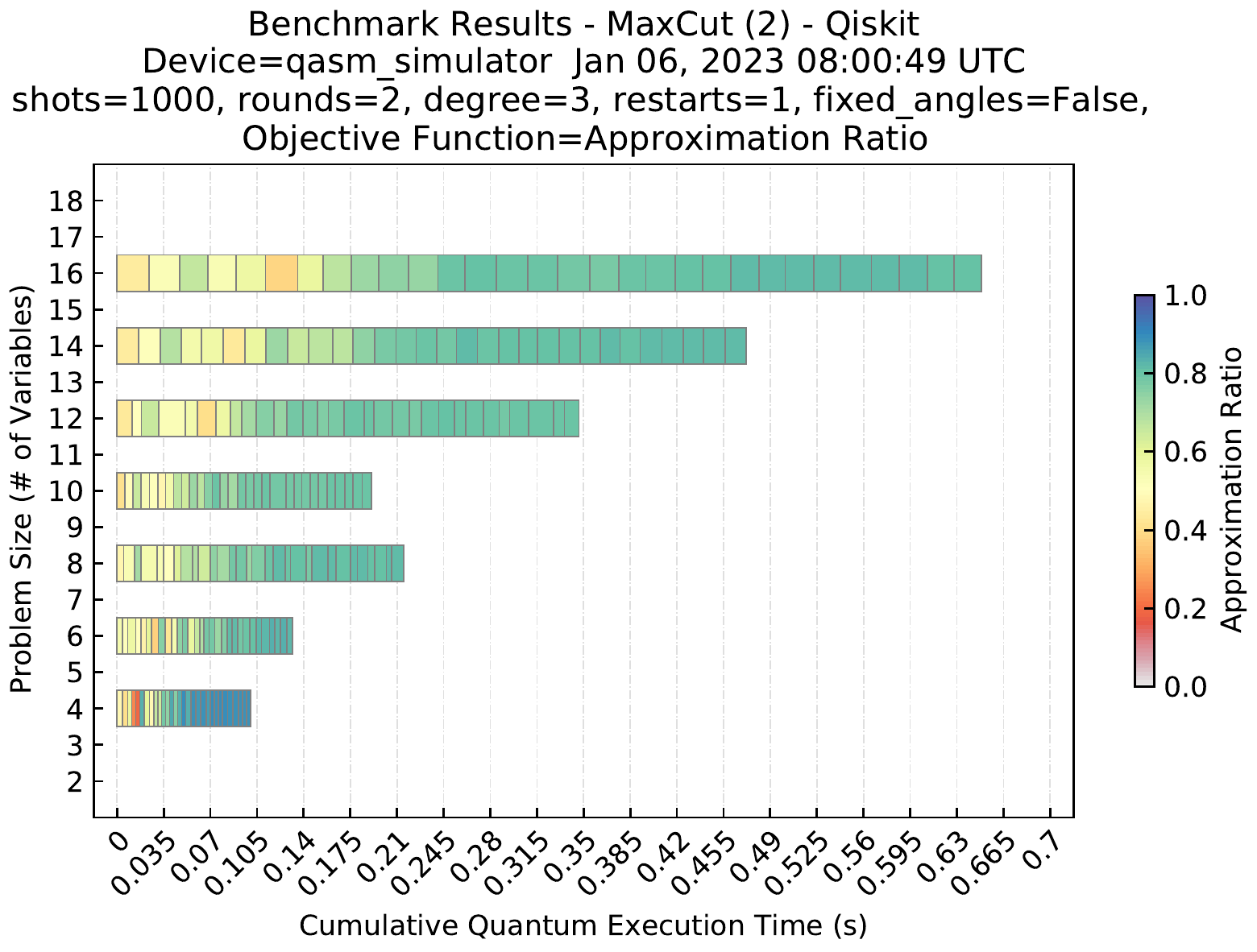}
        \caption{}
        \label{fig:area_plot_quantum}
    \end{subfigure}
    \hspace*{\fill}
    \begin{subfigure}{0.44\textwidth}
        \includegraphics[width=\textwidth]{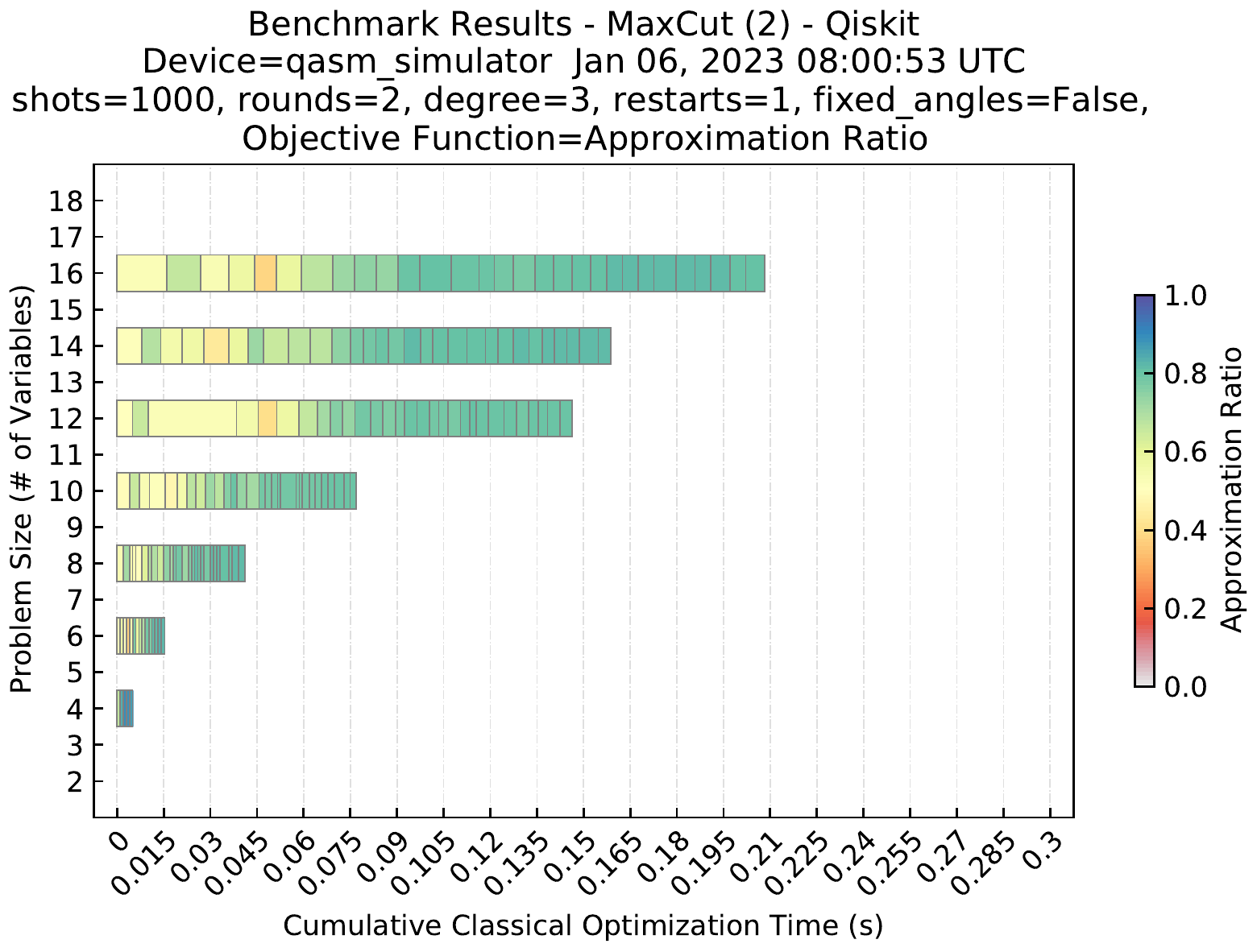}
        \caption{}
        \label{fig:area_plot_classical}
    \end{subfigure} 
    \hspace*{\fill}
\caption{\textbf{Iterative Execution of QAOA Max-Cut Algorithm.} In QAOA, a classical minimizer function iteratively executes an ansatz, varies its parameter values, and computes a cost function to converge to an optimal solution.
In (a), each horizontal row represents successive iterations at each problem size (number of qubits), where the position on the X-axis represents the cumulative quantum execution time, and the color tracks the approximation ratio of each iteration of the optimizer. In (b), the X-axis represents the cumulative classical execution time (optimizer). Both of these times contribute to the total elapsed time that a user experiences.
}
\label{fig:iterative_execution}
\end{figure*}

In this section, we introduce a new method for visualizing the relationship between solution quality and execution run-time in the results from our Max-Cut benchmark.
The methodology is inspired by the typical visualizations used in Operations Research (e.g., the performance profile in~\autoref{fig:opt_intro_examples_1}). Still, it is enhanced in ways that yield valuable insights about execution time unique to hybrid quantum computing algorithms.
Some aspects of this approach are especially relevant in the early stages of quantum computing maturity. They provide critical information about bottlenecks and other drag factors that impact system throughput more than realized.

Our approach can be applied to both QAOA and QA, although the visuals vary slightly in ways that mirror algorithm differences.
In~\autoref{fig:iterative_execution}, we illustrate the time versus quality trade-off for the QAOA algorithm using a novel performance profile referred to as an `area plot'.
Similar to the volumetric plot of~\autoref{fig:qedc_benchmark_profile_1}, it shows the circuit width (problem size) on the Y-axis and uses the color of rectangles to illustrate a metric score.
In the area plot, the horizontal width of an individual rectangle represents the execution time for a single ansatz, and its location along the X-axis indicates the cumulative execution time, including prior iterations.

A key difference between the QAOA and QA benchmarks is in evaluating the execution times to be plotted on the X-axis.
With QAOA, the algorithm inherently executes in a series of iterations, and the execution time accumulates with each, represented by stacked rectangles.
With QA, however, the complete algorithm is executed in a single step. To evaluate how well the algorithm performs at different times, the algorithm is executed from the start each time with different annealing times. \autoref{fig:qa_area_plot_intro} shows how we visualize the re-initialization with rectangles that overlap instead of being stacked.
More detail about QA execution can be found in~\autoref{sec:annealing_hardware_execution}.

\vspace{0.3cm}

In the area plots shown here, we use the approximation ratio as the default figure of merit to gauge the quality of the result. 
The approximation ratio ranges from 0 to 1.0, but for QAOA, it usually starts above 0.5 and oscillates as the optimizer converges to a solution.
Due to the annealing computer's nature, the QA's starting point is often above 0.9. We use a different color scale for QA to emphasize the fundamental difference between the algorithms.
However, when running the benchmarks, any objective functions in the previous section can be selected as the figure of merit shown in the plots.

The benchmarking framework collects multiple measures of execution time.
The first plot of~\autoref{fig:iterative_execution} shows the cumulative quantum execution time or the time spent executing the quantum processor.
The second plot shows the cumulative classical execution time, primarily consisting of the time taken by the optimizer in QAOA or the setup time in QA.
In~\autoref{fig:qaoa_area_plot_intro}, we show the cumulative elapsed quantum execution time, which is the total wall clock time that includes both of these plus other setup times such as compilation or time to load the program into the quantum processor.
We include a detailed analysis of these and other essential times related to QAOA and QA algorithms in Appendix~\ref{apdx:exec_time_analysis}

The quality vs time visualization we use here, the area plot, significantly enhances the information presented to users about executing a hybrid quantum algorithm such as QAOA or QA.
For example, this QAOA and QA evolution analysis can provide information about the \emph{incremental time units consumed by execution}.  With some of the newer hybrid systems and the use of error mitigation, it is extremely valuable to inform the user of the bottlenecks or anomalies in the execution.

For example, in the plots of~\autoref{fig:iterative_execution}, the width of several of the rectangles representing the time of each iteration is not uniform. 
With quantum computing in its early stages of maturity, the execution pipeline often contains many steps that involve non-deterministic classical computation, some of which are unique to quantum computing, such as error mitigation.
These plots effectively convey a measure of the level of unpredictability in the execution times that may contribute to throughput degradation.

Other types of information unique to quantum are also transmitted in these plots. For example, QAOA can require classical pre-processing, specifically compilation, and transpilation to a target topology and gate set from an intermediate representation. In contrast, QA requires embedding the problem graph onto the device topology before execution. With gate model computers, intermittent calibration processes can sometimes interrupt program execution and appear as rectangles with larger widths.
In addition, there is often a start-up cost associated with executing any circuit component of these algorithms.

While we only show a few examples here, the area plots allow users to view all of these things at a glance and can assist them in quickly interpreting how execution time impacts the quality of results and overall throughput of the quantum algorithm.


\subsection{Factors Affecting QAOA Ansatz Fidelity}
\label{subsec:method1}

Several factors can impact the results obtained from executing the QAOA algorithm.
Here, we use Method (1) of the QAOA benchmark to analyze how the number of shots and rounds can affect a quantum computing system's ability to execute the ansatz circuit used in the algorithm.

Each iteration of the QAOA algorithm involves repeatedly measuring a parameterized circuit, executed with parameter values $\ket{\vv \beta, \vv \gamma}$ determined by a classical optimizer routine. The algorithm's success relies on the ability of the quantum subroutine to compute an accurate value of the objective function. If the measurement probabilities obtained by the quantum subroutine do not match sufficiently well with the probabilities from the ideal distribution $P_{\textrm{ideal}}(s)= |\langle s \ | \ \vv \beta, \vv \gamma\rangle|^2$, the effectiveness of the classical optimizer can be negatively affected. 

\begin{figure}[t!]
\includegraphics[width=0.88\columnwidth]{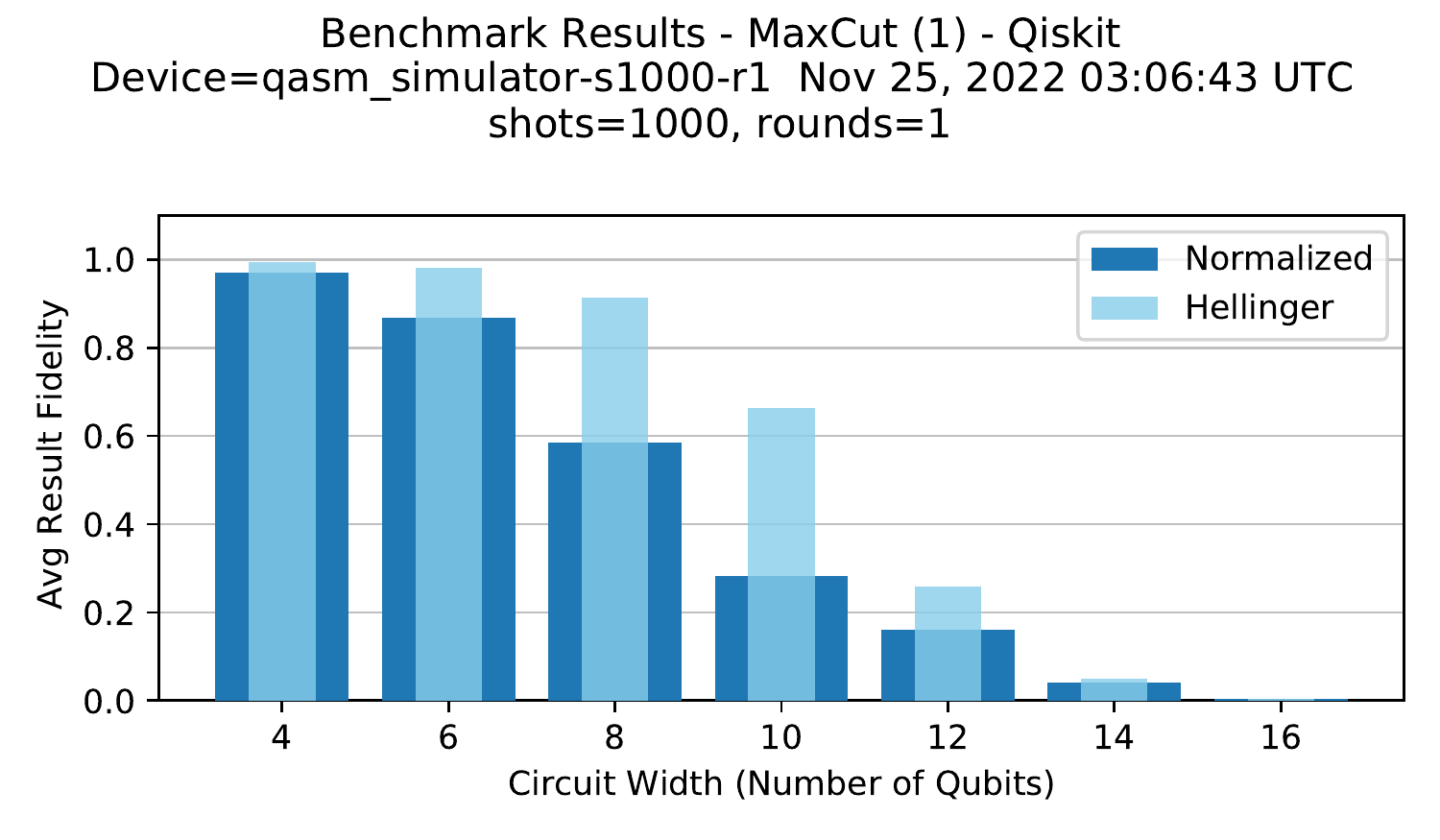}
\includegraphics[width=0.88\columnwidth]{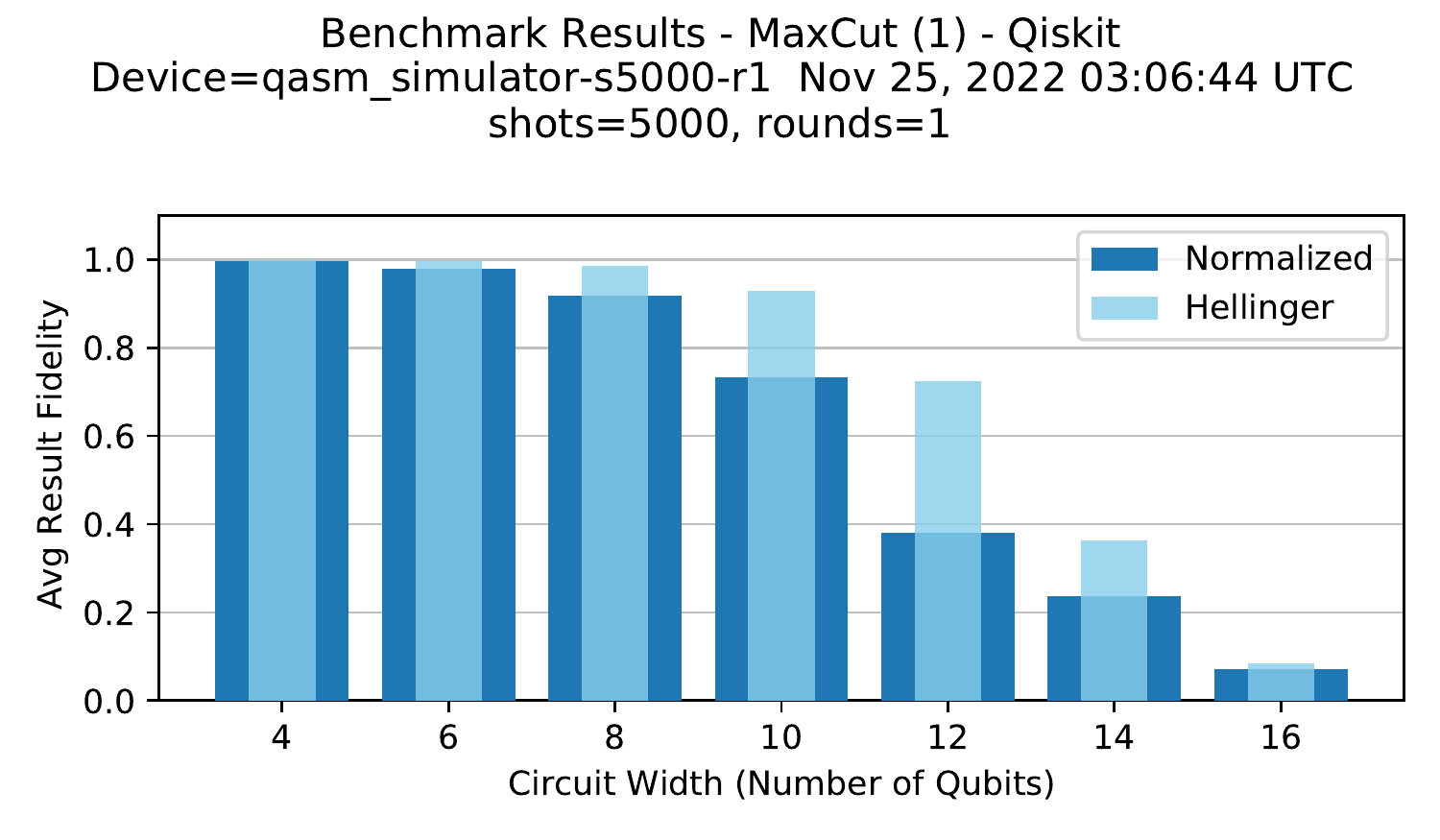}
\caption{\textbf{Impact of Shots on Fidelity of Ansatz Execution.} Here, we illustrate the difference in fidelity when executing the same Max-Cut ansatz circuit at circuit widths ranging from 4 to 16 qubits, with 1000 shots and again with 5000 shots, on an ideal quantum simulator. For each problem size, we use a single graph, which defines an \emph{instance} of the Max-Cut problem. The resulting fidelity is greater when using a larger number of shots.
}
\label{fig:method1_fidelities}
\end{figure}

\begin{figure}[t!]
\includegraphics[width=0.80\columnwidth]{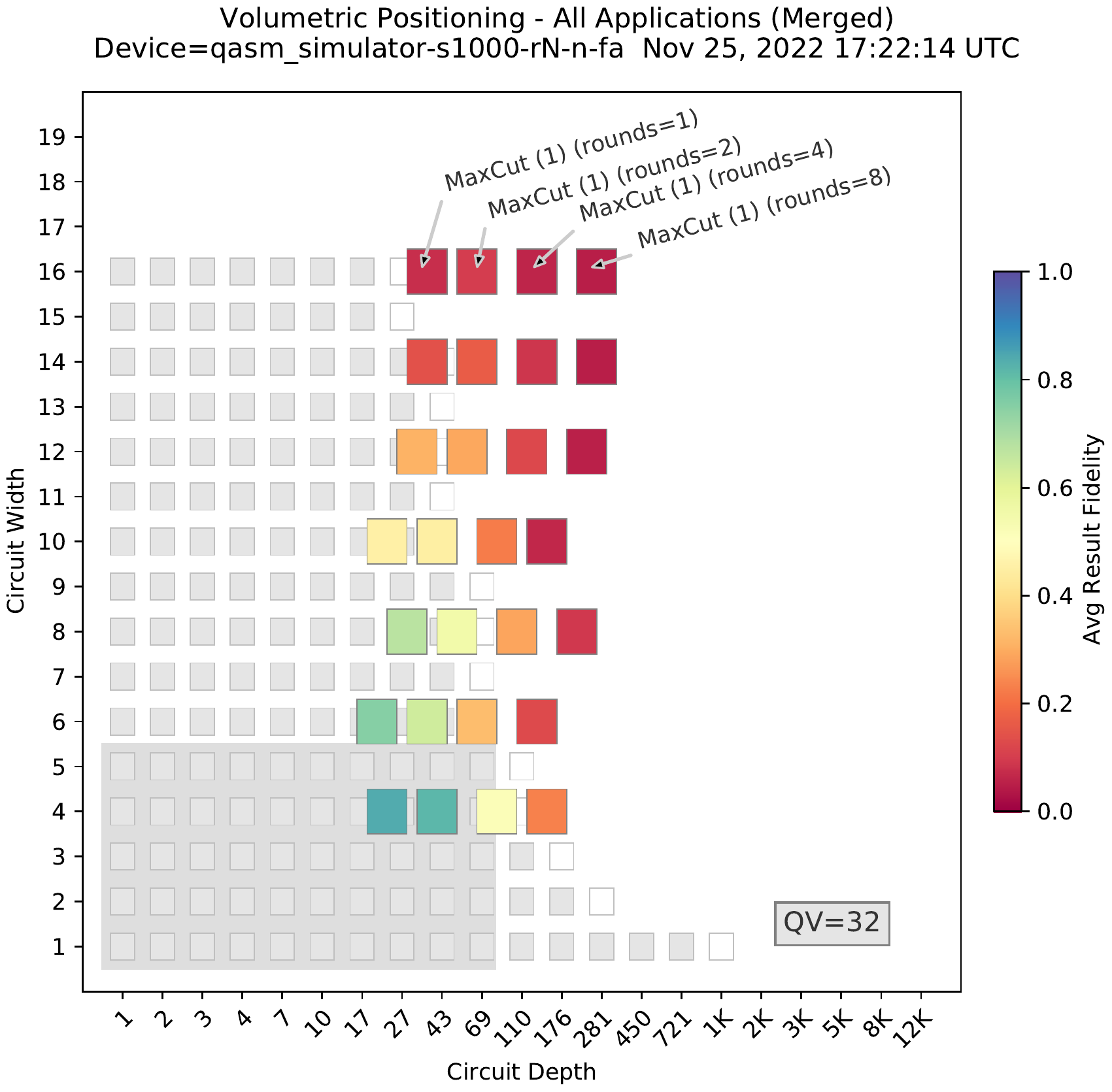}
\caption{\textbf{Volumetric Presentation of Fidelity and Impact of Rounds.} The fidelity metrics obtained for the execution of any quantum circuit are influenced by both the width of the circuit and its depth or its total number of quantum gate layers. Here, the Max-Cut ansatz circuit with varying rounds is executed on a noisy quantum simulator with a quantum volume of 32 (one- and two-qubit gate error rates 0.003 and 0.03, respectively). The (normalized) result fidelity at a specific width and depth is represented by the color shown in the rectangle at that location and degrades with increasing rounds (depth) or problem size (width).}
\label{fig:volumetric_plot_method1}
\end{figure}

Even in a noiseless simulator, perfect fidelity can be achieved only within the limit of an infinite number of shots. On quantum hardware, noise and decoherence can exacerbate the drop in fidelity, as can limited connectivity between qubits.
To quantify circuit fidelity, we use both the Hellinger fidelity and the normalized Hellinger fidelity as defined in our initial work on application-oriented benchmarks~\cite{lubinski2023_10061574}. The normalized fidelity is most useful in our context, recalling that a circuit fidelity of $0$ corresponds to a uniformly random probability distribution. In contrast, a fidelity of $1$ corresponds to the ideal distribution.

In~\autoref{fig:method1_fidelities}, the number of measurements per iteration (which we call the number of `shots') is shown to affect the circuit fidelity significantly. For example, on the noiseless simulator used here, the normalized circuit fidelity falls below $0.6$ at eight qubits with 1000 shots, while it does so at twelve qubits with 5000 shots. 
As the circuit's width increases, the number of shots required to distinguish between the ideal and random distributions increases.
With the variant of QAOA used here, larger problems require a larger number of shots to effectively capture the cut sizes in the resulting larger distributions.


\autoref{fig:volumetric_plot_method1} illustrates how circuit fidelity is impacted as one of the arguments for the QAOA ansatz definition, the number of $\av{rounds}$ (referred to as $p$ in code), is increased from 1 to 8.
Execution fidelity is expected to degrade not only as circuits become wider (i.e., comprise more qubits) but also deeper (i.e., have a larger number of gate layers). More rounds result in deeper circuits.
The volumetric plot uses a color scale to represent the fidelity at each circuit width and depth tested.
In this case, the circuit was executed with 1000 shots on a quantum simulator with noise characteristics that mimic a typical quantum computer (one- and two-qubit gate error rates of 0.003 and 0.03, respectively) and with a quantum volume equal to 32 (the region shown in the dark rectangle).
As the `rounds' parameter grows, the circuit becomes correspondingly deeper, and the result fidelity degrades as a function of depth.
A consequence is that the theoretical benefit of using a larger number of rounds is countered by the lower fidelity that results from executing a deeper circuit with a larger gate count.

\begin{figure*}[t!]
\centering
    \begin{subfigure}{0.32\textwidth}
        \includegraphics[width=\textwidth]{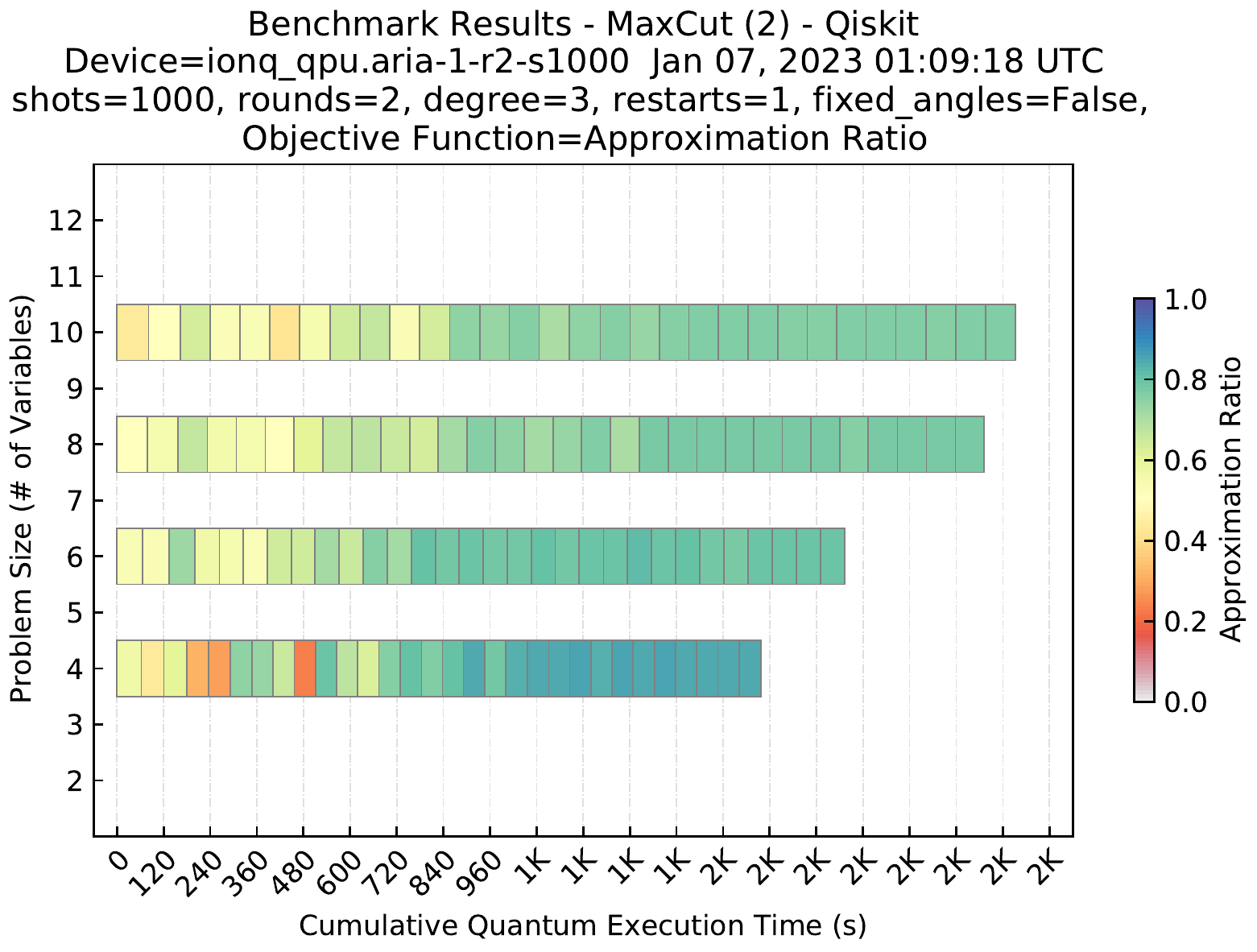}
        \caption{}
        \label{fig:ionq_qpu_aria_exec_area}
    \end{subfigure} \hfill
    \begin{subfigure}{0.32\textwidth}
        \includegraphics[width=\textwidth]{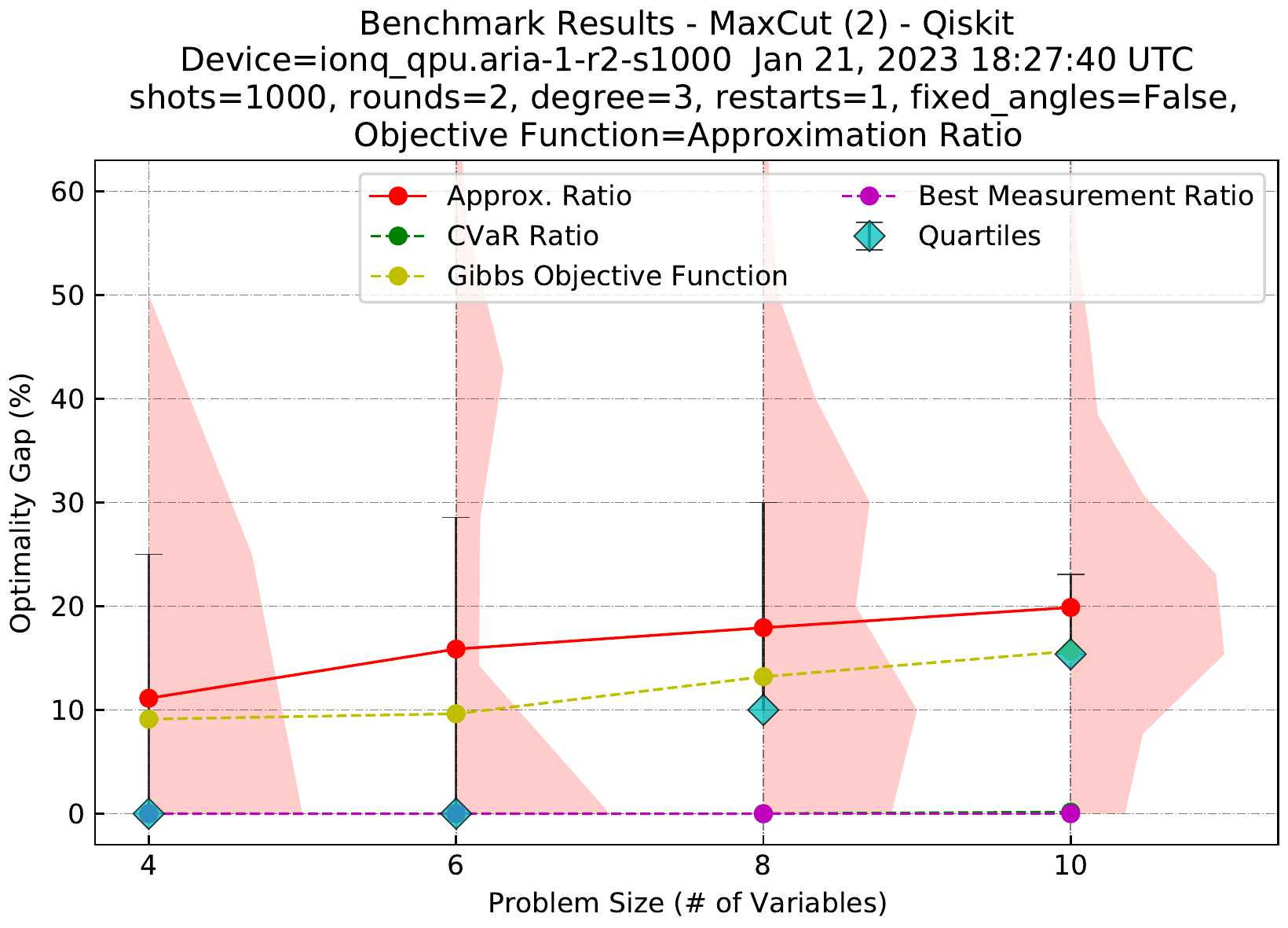}
        \caption{}
        \label{fig:ionq_qpu_aria_optgaps_detail}
    \end{subfigure} \hfill
    \begin{subfigure}{0.32\textwidth}
        \includegraphics[width=\textwidth]{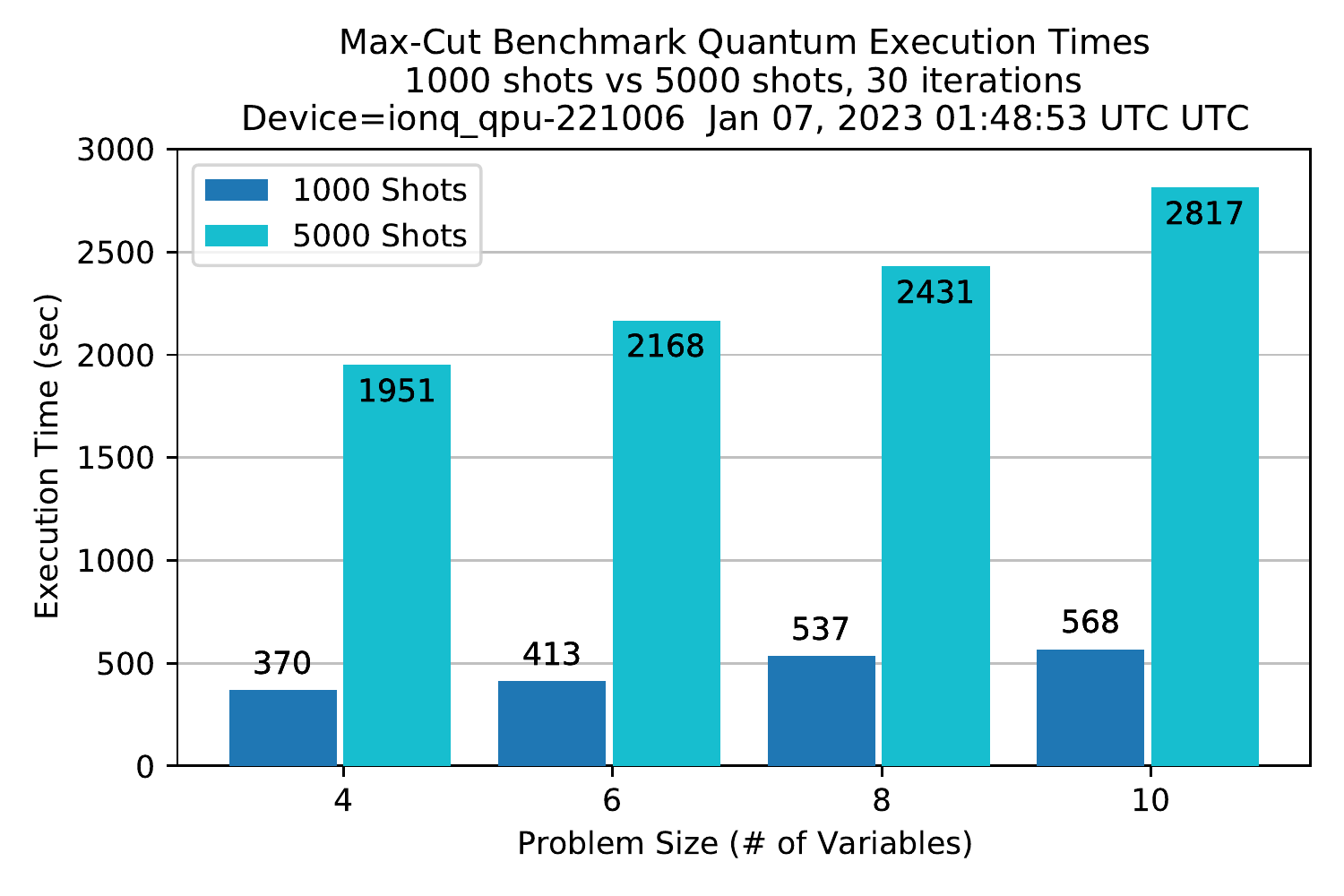}
        \caption{}
        \label{fig:ionq_qpu_shot_times}
    \end{subfigure}
\caption{{\bf Execution on Ion Trap System.} These plots present results from executing the Max-Cut benchmark on the IonQ Aria QPU at different problem sizes (30 iterations, each with 1000 shots).
The area plot (a) shows the approximation ratio improving for each problem size as the cumulative quantum execution time increases (to around 2400s at ten qubits).
The violin plot (b) shows the final optimality gap for each computed ratio and illustrates how the approximation ratio declines with larger problem sizes (to around 20\% at ten qubits).
Plot (c) presents data from a different system, IonQ Harmony, comparing the total quantum execution times, using 1000 shots (568s at ten qubits) and again using 5000 shots (2817s at ten qubits). This indicates that the cost of executing the Max-Cut algorithm on these systems is nearly proportional to the number of shots used.
(\emph{Data collected via cloud service}.)
}
\label{fig:ionq_qpu_aria_runs}
\end{figure*}


\section{Execution on Quantum Hardware}
\label{sec:execution_on_hardware}

This section presents results from executing the Max-Cut benchmark on several representative quantum hardware systems based on underlying quantum technologies. 
Our objective is to demonstrate the robust capability of the benchmark framework to accurately capture key performance metrics that highlight fundamental distinctions between technologies.

This presentation can serve as a valuable resource for providers of these systems, providing insight into incremental performance improvements across successive generations of hardware.
It can also equip users with tools to form a comprehensive understanding of the trade-offs inherent in utilizing these emerging technologies.
Particular attention is placed on the analysis of solution quality versus execution run-time.

However, we emphasize that the results in this section should not be taken as representative of the comparative performance of these quantum platforms in general. They are designed for illustrative purposes and are not intended to be a formal comparison between quantum systems.
Furthermore, we use the respective manufacturers' default software and parameter settings. Together with the quantum systems, these software tool sets are developing rapidly; therefore, our conclusions about quantum system performance represent a snapshot of progress over time. We hope that users will use the framework to create more thorough benchmark tests and utilize them to draw conclusions.

\vspace{0.3cm}

Several critical factors affect benchmark algorithm outcomes on current quantum computing hardware.
Errors from gate infidelity and decoherence can lead to significant differences between the obtained measurement distribution and an ideal system, especially with larger circuits. These errors accumulate in iterative algorithms like QAOA, reducing the quality of the results. Noise in Quantum Annealing (QA), such as thermal energy and control line fluctuations, can disrupt qubit states. Inefficient mappings or embeddings onto specific hardware topologies worsen these effects.

Apart from purely quantum computation, the quality of solutions returned by QAOA depends significantly on the quality of classical computations, such as compilation and optimization of beta and gamma for each round.  Similarly, the quality of solutions returned by QA depends on the classical operations of minor embedding and post-processing and the user parameters that control the quantum computation.
In this sense, our benchmark framework should be viewed as a tool to evaluate the performance of the quantum {\em system} performance in combination with algorithmic choices and parameter settings rather than the performance of a standalone circuit. 

When quantum optimization applications are run on hardware, the quality of the result will degrade compared to a simulator, as the programs will be negatively impacted by noise.
To illustrate the practical limits to execution on hardware, the benchmarks are executed on three different classes of quantum computers: ion trap, superconducting transmon, and quantum annealing system.

\begin{figure*}[t!]
    \begin{subfigure}{0.32\textwidth}
        \includegraphics[width=\textwidth]{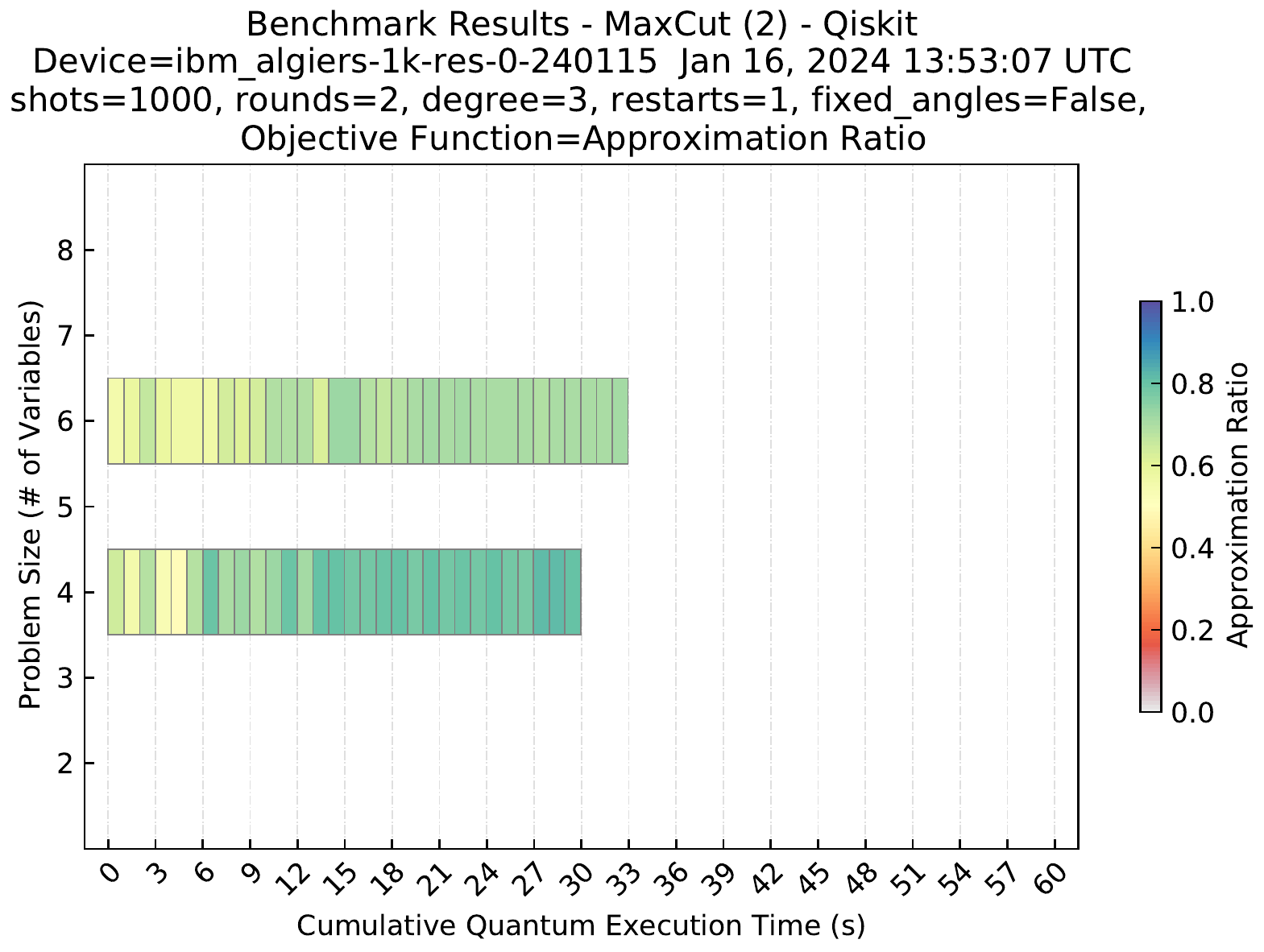}
        \caption{}
        \label{fig:ibm_exec_area}
    \end{subfigure} \hfill
    \begin{subfigure}{0.32\textwidth}
        \includegraphics[width=\textwidth]{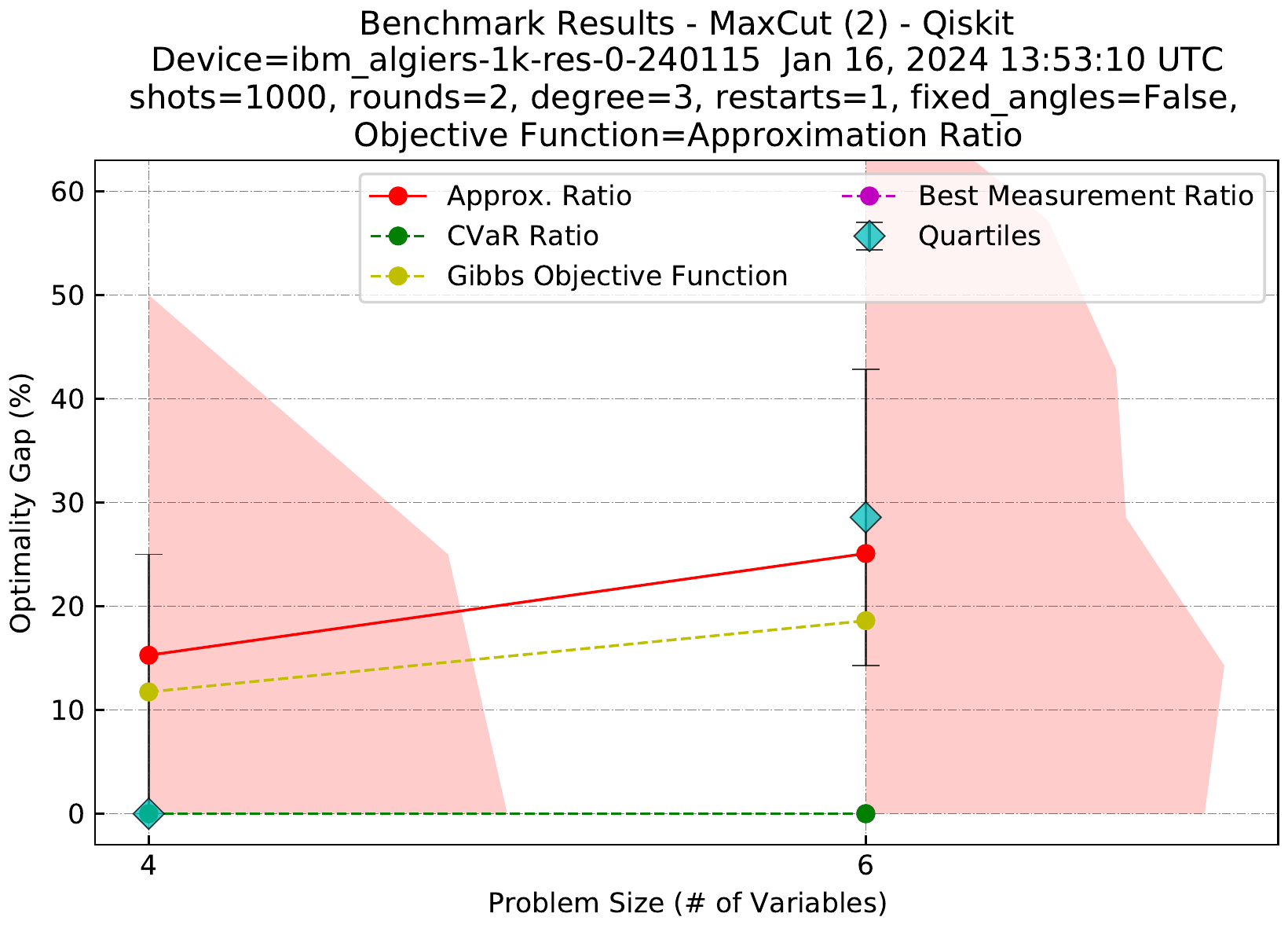}
        \caption{}
        \label{fig:ibm_optgaps_detail}
    \end{subfigure} \hfill
    \begin{subfigure}{0.32\textwidth}
        \includegraphics[width=\textwidth]{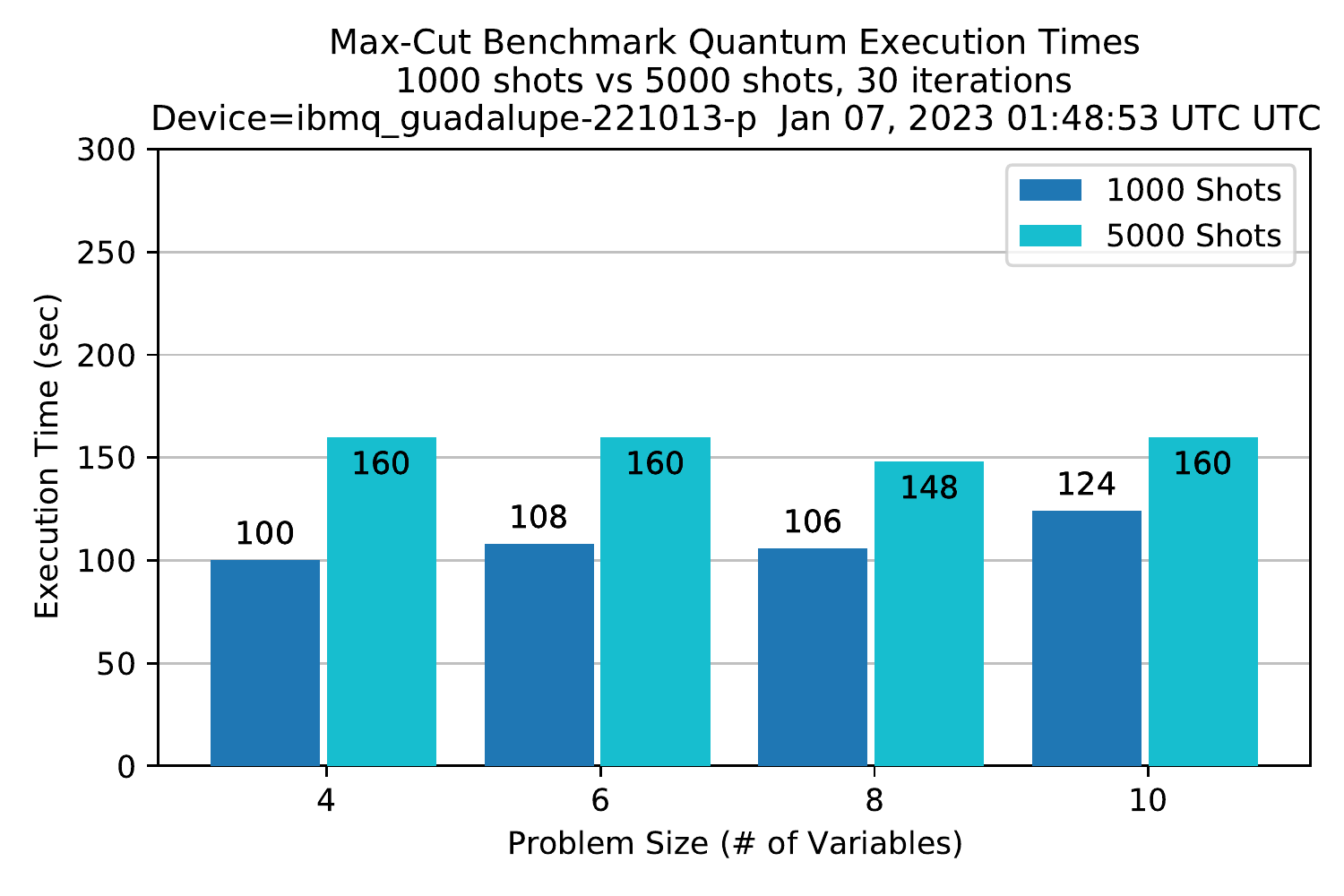}
        \caption{}
        \label{fig:ibm_qpu_shot_times}
    \end{subfigure}
\caption{{\bf Execution on Superconducting Transmon System.} 
These plots present results from executing the Max-Cut benchmark on the IBM Quantum \emph{ibm\_algiers} system at different problem sizes (30 iterations, each with 1000 shots).
Execution was performed using the Sampler primitive without error mitigation (resilience level 0).
The area plot (a) shows the approximation ratio improving for each problem size as the cumulative quantum execution time increases (to~\textasciitilde~33s at six qubits).
The violin plot (b) shows the final optimality gap for each computed ratio and illustrates how the approximation ratio declines with larger problem size (to~\textasciitilde~25\% at six qubits.)
Plot (c) presents data from a different system, IBM Quantum \emph{ibm\_guadalupe}, comparing the total quantum execution times, using 1000 shots (124s at ten qubits) and again using 5000 shots (160s at ten qubits). This indicates that the cost of executing the Max-Cut algorithm on this system is only marginally increased using more shots.
(\emph{Data collected via cloud service}.)
}
\label{fig:algiers_run_1}
\end{figure*}


\subsection{Ion Trap Systems}
\label{sec:hardware_results}

We first show results from executing the Max-Cut benchmark on two remotely accessed gate model quantum computing systems that use ion trap technology.
Quantum computers based on ion traps offer all-to-all connectivity and high fidelity, but this advantage is offset by longer execution times than with other technologies.

\autoref{fig:ionq_qpu_aria_runs} presents results obtained using the IonQ Aria QPU, a second-generation ion trap computer, and its first-generation predecessor, IonQ Harmony. 
At each problem size, from 4 qubits to 10 qubits, we executed the benchmark on both systems with a maximum of 30 iterations using 1000 shots each.

The first plot (a) uses the area plot of~\autoref{subsec:area_plots} to illustrate, for each problem size, how the approximation ratio improves as the cumulative quantum execution time increases with each iteration (to~\textasciitilde~2400s at ten qubits on Aria).
At larger problem sizes, the ansatz circuit becomes deeper, which increases the total execution time and lowers the quality.
The degradation in the final result quality is visible in the violin plot (b), which shows the final optimality gap for each computed ratio increasing with problem size (to~\textasciitilde~20\% at ten qubits.)
However, note that the best measurement ratio gap is 0\%. 

Plot (c) of this figure presents data generated using IonQ Harmony, on which we executed the Max-Cut benchmark twice, once using 1000 shots and again using 5000 shots.
For all problem sizes, the increase in the cumulative quantum execution time is roughly proportional to the rise in the number of shots.
For example, at ten qubits, the time increases from 568 seconds to 2817 seconds, a factor of $4.959$.
The height difference between the bars at a specific problem size represents the difference in time spent executing an additional 4000 shots over 30 iterations since both runs share the same initialization time.
From this, we can compute the execution time per shot. For example, at six qubits, this is $(2168 - 413) / (4000 * 30)$ or 14.6  ms/shot for IonQ Harmony.
These results indicate that the total cost to a user to execute the Max-Cut or similar algorithms on this class of quantum computers includes not only the financial cost, which depends on the shot count but also a reduction in total throughput that is a consequence of the longer execution times.

We observe that execution times on IonQ Aria are longer than on IonQ Harmony.
For example, at ten qubits, we see approximately 2400 seconds and 568 seconds approximately, respectively.
The Aria device applies an error mitigation scheme to measurement data to improve results. Still, we did not investigate the reasons behind the increased execution time and improvement in quality on Aria.

\begin{figure*}[t!]
\centering
    \hspace*{\fill}
    \begin{subfigure}{0.38\textwidth}
        \includegraphics[width=\textwidth]{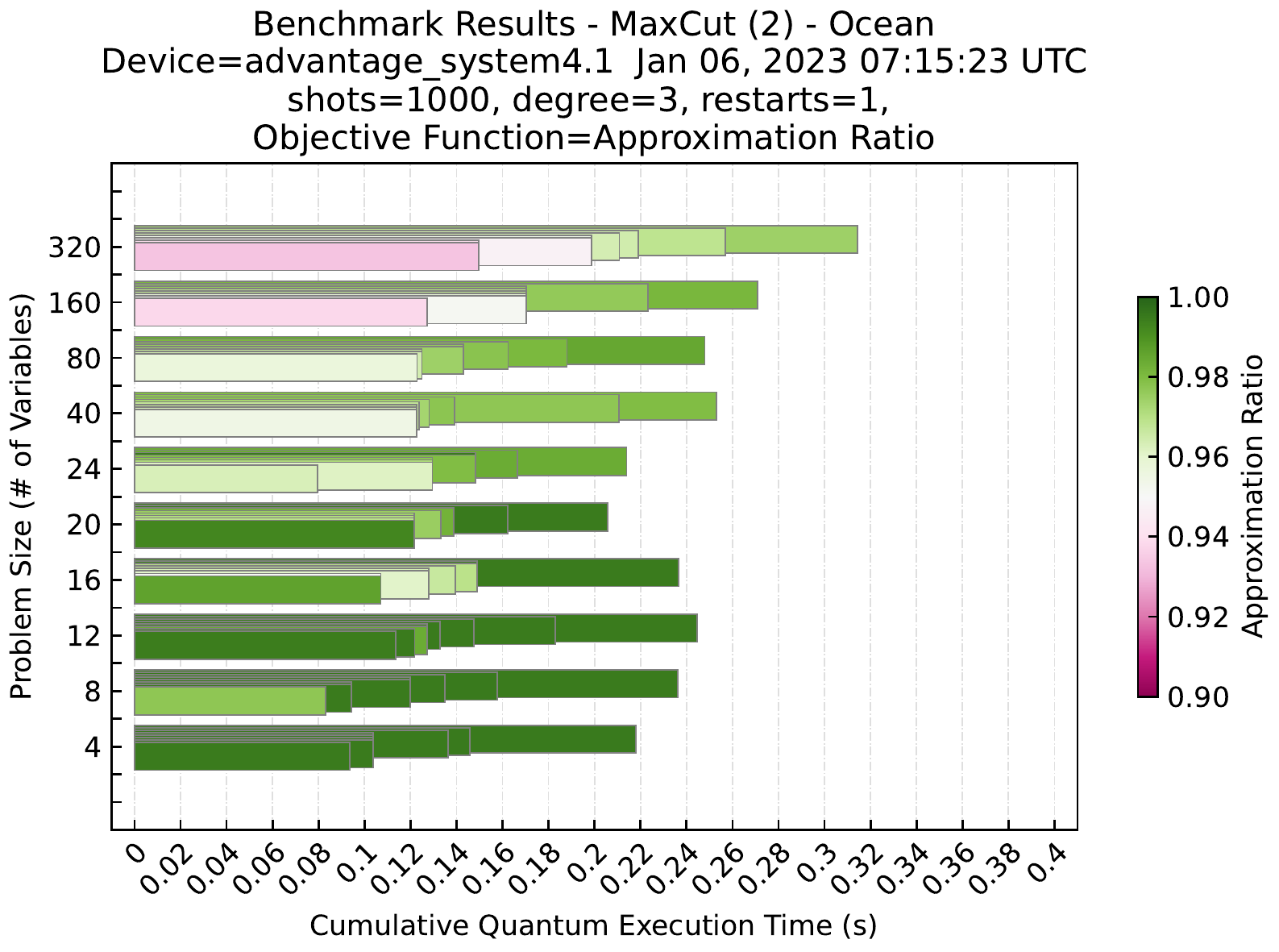}
        \caption{}
        \label{fig:dwave_exec_area}
    \end{subfigure}
    \hspace*{\fill}
    \begin{subfigure}{0.38\textwidth}
        \includegraphics[width=\textwidth]{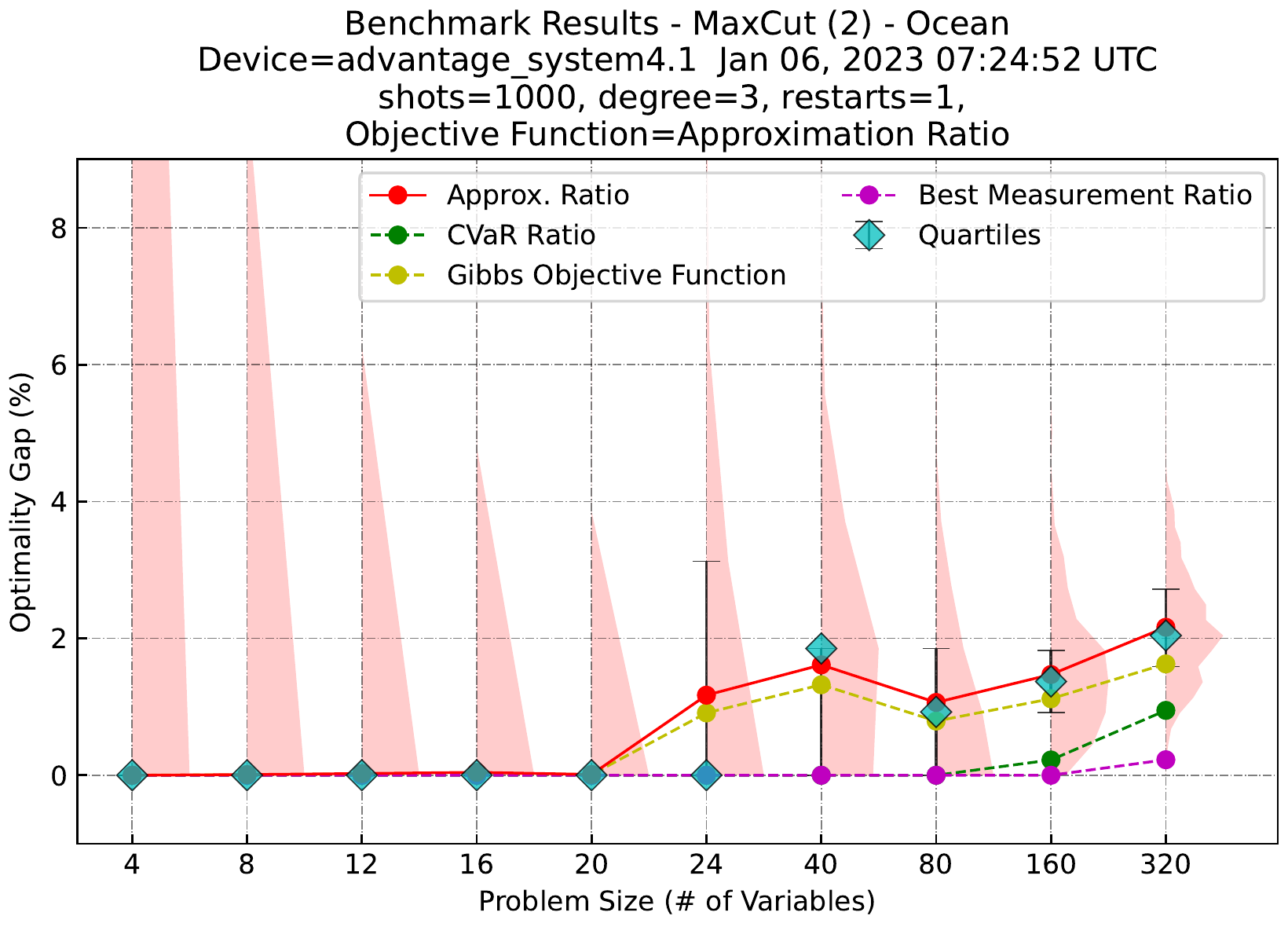}
        \caption{}
        \label{fig:dwave_optgaps_detail}
    \end{subfigure} 
    \hspace*{\fill}
\caption{{\bf Execution on Quantum Annealing System.} 
These plots present results from executing the Max-Cut benchmark on the D-Wave \emph{advantage\_system4.1} quantum annealing system at different problem sizes (each with 1000 shots).
The problem is embedded once and executed over a range of annealing times from 1 ms to 256 ms to evaluate the impact of quantum annealing time on the resulting quality.
The area plot (a) shows the approximation ratio improving for each problem size as the cumulative quantum execution time increases (to~\textasciitilde~0.31s at 320 variables).
The violin plot (b) shows the final optimality gap for each computed ratio and illustrates how the approximation ratio declines with larger problem sizes (to~\textasciitilde2\% at 320 variables.)
(\emph{Data collected via cloud service}.)
}
\label{fig:dwave_qpu_results_1}
\end{figure*}


\subsection{Superconducting Transmon Systems}
\label{sec:superconducting_transmon_systems}

Here, we present results from executing the Max-Cut benchmark on two-gate model quantum computing systems that use superconducting transmon technology and can be remotely accessed.
These results highlight and quantify the characteristics of quantum program execution inherent in hardware implementations built on this technology.
Quantum computers based on superconducting transmons execute more quickly than other technologies.
However, this advantage is offset by reductions in fidelity that can result from the introduction of swap operations to compensate for limited connectivity between qubits.

Execution on both of the systems was performed using the Qiskit Sampler primitive run through the IBM Cloud Qiskit Runtime service~\cite{ibmqcloud2023}. 
Error mitigation was turned off by setting the ``resilience\_level'' execution argument to 0. 
While the Sampler offers automatic error mitigation, we elected not to enable it in our hardware demonstrations.
Users are encouraged to gauge for themselves the impact of selecting this option on both the quality of the result and the total execution time.

The plots in ~\autoref{fig:algiers_run_1} present results from executing the Max-Cut benchmark on the IBM Quantum \emph{ibm\_algiers} system at different problem sizes (30 iterations, each with 1000 shots).
The area plot (a) shows the approximation ratio improving for each problem size as the cumulative quantum execution time increases with each iteration (to~\textasciitilde33s at six qubits).
With deeper ansatz circuits at larger problem sizes, the total execution time grows longer, and the result quality declines.
However, note this system's overall run-time performance.  Each iteration of 1000 shots requires ~\textasciitilde1s at four qubits and ~\textasciitilde1.1s at six qubits. 
This results in completing the QAOA execution, limited to 30 iterations, in 30 and 33 seconds, respectively.

The violin plot (b) shows the final optimality gap for each computed ratio increasing with larger problem size (to~\textasciitilde~25\% at six qubits.) as the approximation ratio declines. 
However, the best measurement ratio gap is 0\%, indicating that the algorithm could identify the maximum cut at both problem sizes.

Plot (c) presents data from the execution of the benchmark on a different system, IBM Quantum \emph{ibm\_guadalupe}.
This is an earlier generation system with execution times that are longer than on \emph{ibm\_algiers}.
However, we use these results to illustrate an important aspect of quantum program execution on superconducting transmon computers.
Here, we compare the total quantum execution times obtained when running the benchmark using 1000 shots (124s at ten qubits) and again using 5000 shots (160s at ten qubits).
Since both runs perform similar initialization processing, the height difference between the bars at each problem size represents the difference in time required to execute an additional 4000 shots over 30 iterations.
This lets us determine the time to execute a shot on this quantum computing device.
For example, at six qubits, the execution time per shot can be computed as $(160 - 108) / (4000 \times 30)$ or 0.43 ms/shot.

These results suggest that using more shots to execute quantum programs only marginally increases the cost of executing the Max-Cut optimization or similar algorithms on this system.
Users may take advantage of significantly higher throughput on this class of devices to execute the circuit repeatedly to search for an optimal result while still achieving a shorter total execution than with alternative technologies.
Additionally, the ability to execute more shots could be explored to achieve higher-quality results.


\subsection{Quantum Annealing System}
\label{sec:annealing_hardware_execution}

An important focus of our work was to structure the QED-C benchmark suite to enable the execution of some benchmarks on back-end systems implemented using quantum technologies other than gate model quantum circuits.
In this section, we describe the execution of the Max-Cut benchmark on a D-Wave Advantage system, accessed through LEAP Solver ``advantage\_system4.1'', as a way to illustrate the framework's support for test orchestration with varying parameters, capture of relevant performance metrics, and presentation of results consistently across quantum technologies.

The plots in \autoref{fig:dwave_qpu_results_1} present results from executing the Max-Cut benchmark on the D-Wave \emph{advantage\_system4.1} quantum annealing system at problem sizes ranging from 4 to 320 variables (each executed with 1000 shots).
As in the corresponding gate-model displays,
the area plot (a) shows the approximation ratio improving for each problem size as the cumulative quantum execution time increases (to~\textasciitilde~0.31s at 320 variables).
Similarly, the violin plot (b) shows the final optimality gap for each computed ratio, and we see the approximation ratio declining with larger problem sizes (to~\textasciitilde2\% at 320 variables.)

However, the QA benchmark implementation (described in~\autoref{sec:benchmark_algorithm_qa}) differs from the QAOA benchmark, affecting how the results are presented. 
At each problem size, the QA algorithm is re-executed from the beginning, doubling the annealing time in steps from 1 to 128 $\mu$s.
In the annealing version of the plot (a), for each problem size, the rectangles are drawn overlaid to highlight this difference from the QAOA benchmark algorithm.
Each rectangle represents a complete execution but with a larger anneal time.

In this way, we illustrate the quality vs. time trade-off for quantum annealing in the same way we do for the QAOA algorithm but unambiguously convey the difference between the algorithms.
The intent is to permit a user to quickly see the level of quality that can be expected for a specific annealing time, which is essential to evaluate the overall cost of the quantum annealing solution.

The total quantum execution time for the annealing algorithm increases only slightly as the problem grows, from ~\textasciitilde0.22s at 4 variables. to ~\textasciitilde0.32s at 320 variables. 
This is because the quantum execution reported in the plot includes the non-quantum operations of chip programming and readout in addition to the actual annealing time, which is small relative to these.
This suggests that while increasing the annealing time of the computation may increase the financial cost, it does not impact the throughput that can be achieved when using quantum annealing.

At the problem sizes tested in this benchmark, the approximation ratios obtained with the QA benchmark are above 0.90 in all cases. 
Therefore, the QA area plots use a different color scale to represent the approximation ratio, making the evolution of the quality visible in this different range of values.
Note also that the QA algorithm benchmark identifies the best cut size for problems with up to 160 variables.


\subsection{Discussion of Hardware Results}
\label{sec:hardware_summary}

The Max-Cut extension to the QED-C benchmark suite enables user control over problem definition and size, shots, rounds, restarts, initial angles, and the choice of optimizer and its parameters.
The number of possible combinations of these settings is large, and we explored a limited subset.
In the hardware tests above, we execute benchmark problems of different sizes on three distinct classes of quantum computers and explore how the quality of the solution varies as execution time increases. In two of the tests, we also varied the number of shots (1000 and 5000) to gauge the impact of this parameter on system throughput.

In other tests, using simulations implemented with noise characteristics of these target systems, we found that setting the rounds parameter to 2 provides a good balance between the QAOA algorithm's effectiveness and the degradation from noise in longer circuits. Other tests indicated that at the small scale to the tests run using QAOA on gate model hardware, a setting of 1000 shots and two rounds offers the best default configuration for executing on quantum hardware to minimize utilization of hardware resources during benchmarking.

For interested readers, we reference Appendix~\autoref{apdx:result_quality_assessment}, in which we discuss more extensive demonstrations executed to determine how the many different parameters affect result quality.
The results suggest that multiple restarts could improve benchmark results. 
Importantly, using fixed initial angles combined with multiple restarts could be an effective way to provide a standard optimization benchmark that does not require complete QAOA execution to evaluate the effectiveness of a target system, reducing the resources necessary for benchmarking.
Also presented in that section is a parameter-tuning strategy developed with the help of these benchmarks to identify the best combination of several of the parameters for QAOA execution.

\vspace{0.3cm}

In the results presented above, we depict cumulative quantum execution time using area plots to illustrate how the quality of the result improves as execution progresses. This time, reported by hardware providers, reflects the quantum processor (or simulator) usage.
It holds significant importance as it directly impacts the financial cost of utilizing quantum computers, which varies substantially across systems.
Our analysis revealed significant variations in execution throughput across different classes of quantum computing technology. When evaluating the utility of a quantum computer, it is essential to consider both the financial cost and the time required to complete tasks.

Another critical factor to consider in evaluating the total cost of a quantum computing solution is other time costs beyond execution time in the quantum processing.
Preparing the quantum program for execution involves resources overhead. This can include compilation time, transpilation to the target topology and gate set, loading the compiled program, and data transfer into and out of the quantum system. 
These overhead components directly impact the optimization application's total throughput, as every execution will include some or all of them.

These throughput factors can vary widely between vendors, particularly within the execution environment, and uniquely between users. For example, pre-compilation of the program or co-location of the classical and quantum processors can dramatically impact total throughput. 
There are also numerous vendor-specific hardware settings that we did not test that could potentially improve results.

Furthermore, the QAOA and QA algorithms, tailored to match their corresponding gate model and annealing-based hardware paradigms, possess different properties and structures, significantly impacting computation time and solution quality. This complexity can obscure the effects of hardware and system performance. For QAOA, the choice of optimizer and its parameters can dramatically impact benchmark results. 

\vspace{0.3cm}

As stated earlier, the performance results in our study are intended for illustration purposes only and should not be taken as representative of relative performance in general.  Too many variables contribute to both the quality of the solution and the total run time for these results to be viewed as a definitive characterization of these systems' performance.
For these reasons, we do not provide an exhaustive analysis of these factors.
A full-scale study of all the factors contributing to quantum performance to tease out the hardware contribution is beyond the scope of this paper.
Instead, we propose that users include the total cost of execution, including these additional overheads, in any of their studies using the benchmark suite presented here.

\vspace{0.3cm}

It is worth noting that inherent variations can significantly influence the quality of results in the quantum algorithms employed in these benchmarks. While our study did not delve into such variations, our benchmark framework offers a platform to explore recent advancements in optimization algorithms.
For instance, the framework could be configured to incorporate innovations such as pre-computing diagonal matrices~\cite{Lykov_2023}, investigating the impact of multiple rounds~\cite{Zhu_2023, majumdar2021optimizing}, adopting a multi-angle ansatz~\cite{herrman2021multiangle, herrman2022relating, Shi_2022, chalupnik2022augmenting}, utilizing an expressive ansatz~\cite{vijendran2023expressive}, or employing a large-scale solver with few qubits~\cite{sciorilli2024largescale}.
As quantum computing hardware progresses to incorporate fault tolerance and error correction features, it becomes imperative to advance algorithms in tandem. Utilizing more focused versions of the tests illustrated here will be crucial in assessing the concurrent improvement in performance.


\section{Summary and Conclusions}
\label{sec:summary_and_conclusions}

While the current generation of quantum computers may be limited in computational power, these systems are expected to rapidly evolve and become capable of performing increasingly complicated tasks.
It is thus critical to this advancement effort to establish accurate and validated methods for measuring progress that are readily available to the developers of these systems and the users who utilize the resource in solving real-world problems.

To this end, we built on the existing open-source QED-C Application-Oriented Benchmark suite, enhancing it to support benchmarking of hybrid quantum-classical solutions to combinatorial optimization problems, often cited as a use case for quantum computing.
Multiple factors affect the ability of a quantum computer to produce solutions to combinatorial optimization problems effectively.
The algorithms used to find these solutions provide only an approximate answer, and the quality of the results is typically a function of the time available for processing, often under tight constraints.
Our benchmarks are designed to provide a mechanism to evaluate these and to provide valuable and critical insights into options for improving its performance and overall throughput.

We demonstrate the features of this framework and highlight its benefits using the Quantum Approximate Optimization Algorithm algorithm for execution on gate model systems and demonstrate its adaptability to other types of solvers by using a Quantum Annealing algorithm executed on an analog quantum computer system.
We demonstrate the capabilities of our framework using the widely studied Max-Cut problem, which offers a simple early-stage target for evaluating quantum computing solutions but can scale to larger application challenges in the future.
In future work, we plan to extend this demonstration to include other technologies, such as cold atoms.

\vspace{0.3cm}

A primary goal of this work was to structure our analysis and presentation methodology to use methods familiar to quantum computing specialists but inspired by how Operations Research views the quality of results from a solver in addressing optimization problems.
The methods are enhanced to account for specific characteristics of quantum solutions, such as statistical sampling or the iterative nature of the QAQA algorithm.
They can inform the user of bottlenecks or anomalies in execution, which is extremely valuable.

These enhanced analysis and visualization techniques can provide useful information about a quantum computing solution's throughput and a detailed understanding of its total cost of ownership.
Results from these benchmarks can inform the user about the factors that can be adjusted to improve performance on these systems.
While these techniques have been used in Operations Research for a long time, applying them effectively to quantum computing is still in the early stages.

This paper does not include a full-scale comparison between quantum computing systems of different types, nor does it address benchmarking of classical solutions to optimization problems, as numerous in-depth studies exist in this area.
The performance results in this paper are intended to illustrate the features and benefits of our benchmarking framework. 
We include benchmarking the QA algorithm on annealing hardware as a proxy for other solvers using different quantum technologies.
Our work has identified many parameters that impact how a quantum computer solves an optimization problem well. However, we did not perform an exhaustive study of these or all options that might produce optimal results.
We do not attempt to provide coverage or analysis over all possible settings of algorithm parameters or vendor-specific execution options.

While we generally show that one class of device may provide higher-quality results while another may execute more quickly, the emphasis in this paper is on how the benchmark framework is structured and how it can be used to explore quantum algorithm execution and performance.
Users can execute these benchmark programs on devices to which they have access and evaluate for themselves the total cost of ownership of this technology. This is crucial to understanding how and when quantum computers may be able to offer measurable value.

\vspace{0.3cm}

We consider this work to be forward-looking with respect to the available technology.
The complexity of discrete optimization problems motivates the development of methods, such as QAOA, QA, and others, that may be challenging in the worst-case complexity analysis yet provide value in practice.
Performance-based metrics and benchmarking tools can quantify the progress of these proposed methods and provide a way of comparing alternative solutions whose capabilities are beyond those currently available.
As quantum computers evolve, the benchmark methods we have defined here will be critical to gauge performance improvement.


\section*{Code Availability}
\label{sec:data_and_code_availability}

The code for the benchmark suite introduced in this work is available at
\href{https://github.com/SRI-International/QC-App-Oriented-Benchmarks}{https://github.com/SRI-International/QC-App-Oriented-Benchmarks}.
Detailed instructions are provided in the repository. 

\section*{Acknowledgement}

The authors acknowledge the use of IBM Quantum services for this work. The views expressed are those of the authors and do not reflect the official policy or position of IBM or the IBM Quantum team.
IBM Quantum, https://quantum-computing.ibm.com/, 2021.

We acknowledge IonQ for the contribution of access to hardware. The views expressed are those of the authors and do not reflect the official policy or position of IonQ.

We acknowledge D-Wave Systems for contributing access to both hardware and software tools. The views expressed are those of the authors and do not reflect the official policy or position of D-Wave Systems.

Contributions to this work from Los Alamos National Laboratory were conducted under the auspices of the National Nuclear Security Administration of the U.S. Department of Energy under Contract No. 89233218CNA000001.
This research used resources provided by the Los Alamos National Laboratory Institutional Computing Program and was supported by the Laboratory Directed Research and Development program under the project 20210114ER.

D.B. acknowledges NASA Academic Mission Services (contract NNA16BD14C – funded under SAA2-403506).  
P.S. acknowledges support from the NASA/USRA Feynman Quantum Academy Internship program.
Both D.B. and P.S. are supported by NSF Expeditions in Computing program CCF \#1918549.
This work used computational and storage services associated with the Hoffman2 Shared Cluster provided by the UCLA Institute of Digital Research and Education’s Research Technology Group.

D-Wave\textregistered, Ocean\texttrademark, and Advantage\texttrademark{} are trademarks of D-Wave Systems, Inc.  IBM\textregistered{}, Qiskit\textregistered, IBM Q\textregistered{} and IBM Quantum System Two\texttrademark{} are trademarks of International Business Machines Corporation.  IonQ\texttrademark{}, IonQ Harmony\texttrademark{}, and IonQ Aria\texttrademark{} are trademarks of IonQ, Inc.

We acknowledge Jerry Gamble of Verizon Corporation for his contribution to code development and editorial efforts on this manuscript.
We acknowledge Jason Necaise in the Department of Physics and Astronomy, Dartmouth College (previously with D-Wave Systems), for his contribution to code development.

We thank Mark Johnson (D-Wave), Andrew Wack (IBM), David McKay (IBM), Paul Nation (IBM), Luning Zhao (IonQ), Ananth Kaushik (IonQ), Farshud Sorouifar (Ohio State University), Amos Anderson (Quantum Circuits), Steve Reinhardt (Quantum Machines), Davide Venturelli (USRA/NASA), Filip Maciejewski (USRA/NASA), and others for providing comments on this manuscript.


\clearpage

\bibliographystyle{unsrtnat}  
\bibliography{references}



\clearpage
\appendix


\section{Methods for Combinatorial Optimization}
\label{apdx:theory}

We present a general introduction to the theoretical foundations of combinatorial optimization problems and their implications for developing the hardware demonstrations to study solver performance (a solver is an algorithm or heuristic implemented in software or hardware). 
The issues discussed here informed our decisions about the choice of inputs and performance metrics in designing the QED-C benchmarking framework. 

\subsection{Combinatorial Optimization Theory}

For concreteness, we consider the class of combinatorial optimization problems defined on $n$ integer-valued variables $x = \{ x_1, \ldots x_n\}$, containing $m$ constraint functions $c(x)$, and one objective function $f$ that is a polynomial in $x$, as follows.   
\begin{align}
    & \min: f(x)  \nonumber \\
    & \mbox{s.t.: } c_i(x) \leq 0 \;\; \forall i \in \{1,\ldots,m \} \nonumber  \\
    & x_i \in \mathbb{Z} \;\; \forall i \in \{1,\ldots,n\} 
\end{align}
Given a problem thus described, the algorithmic goal is to find an assignment of 
integer values to $x$ that obeys all the constraints and minimizes the value of $f(x)$.  For example, this simple problem, 
\begin{eqnarray}
 f(x) & =&   x_1 + 2 x_2  \nonumber \\
 c_1(x) &:&  -x_1 + 1   \leq 0 \nonumber \\
 c_2(x) &:&  -x_2 + 1   \leq 0,  \nonumber 
\end{eqnarray}
asks to find two positive integers that minimize $f(x)$;  
an optimal solution $x_1 = 1, x_2 = 1$ has the objective value $f(x) = 3$. 

This notation is general enough to cover an enormous variety of optimization problems of interest to all industry sectors. To name just a few:  
\begin{itemize}
    \item The {\bf Job Shop Scheduling} problem and its variations are ubiquitous in industry scheduling problems associated with the efficient assignment of multiple resources to multiple tasks. 
    \item The {\bf Portfolio Optimization} problem is of interest to finance.  For example, given a list of items to purchase,  select a subset of items to maximize profit and minimize risk. 
    \item The {\bf Airport Gate Scheduling} problem in the transportation industry is as follows: Given a list of airport arrival times and passenger connections, assign gates to airplanes to minimize the total distance passengers must walk to the connecting gates.  
    \item {\bf Machine Learning (ML)} is a core problem of Artificial Intelligence. Most ML techniques require access to good-quality optimization heuristics as part of an inner-loop computation that may be performed hundreds or thousands of times. The heuristic finds input/output pairs that constitute diverse samples of the near-optimal solution space of a given optimization problem.      
\end{itemize}

\vspace{0.3cm}
The complexity class \(\text{NP-OPT}\) contains optimization problems (including (A1)) that are defined in terms of an objective function with a numerical result, as opposed to decision problems with binary outcomes (e.g., Yes/No or True/False), which inhabit the more famous class \(\text{NP}\). 

Every problem \(\text{P}\) in \(\text{NP}\) can be reformulated (also called translated) as a problem \(\text{P-OPT}\) in \(\text{NP-OPT}\). For example, in the binary Satisfiability (\(\text{SAT}\)) problem, \emph{Given the Boolean expression B, does there exist an assignment of variables such that B evaluates to True?} can be translated to an equivalent problem in \(\text{NP-OPT}\): \emph{ Given B\(\text{-OPT}\), find an assignment of variables that maximizes the number of satisfied clauses.}
 The transformation guarantees that an optimal solution to \(\text{SAT-OPT}\) is a yes answer to \(\text{SAT}\). In this case, the maximum number of satisfied clauses equals the total number of clauses in $B-OPT$, then the answer to the binary problem is {\em yes}. 
 
The translated \(\text{SAT-OPT}\) problem is called \(\text{NP-HARD}\) because a polynomial-time algorithm for \(\text{SAT-OPT}\) could be used to solve \(\text{SAT}\), and by extension, every problem in \(\text{NP}\) could be solved in polynomial time. Problems that are both \(\text{NP-HARD}\) and in \(\text{NP}\) (i.e., binary decision problems) are called \(\text{NP-COMPLETE}\). The famously open question {\em Does \( \text{P} = \text{NP}\)?} captures the current unhappy state of knowledge about these problems: no polynomial-time algorithm is known to exist, and no one can prove that they cannot exist. 

\subparagraph{Solving Problems by Translation}

Among many approaches to solving problems in \(\text{NP}\) and \(\text{NP-OPT}\), solution-by-transformation has been studied for a handful of problems and algorithms. 

This approach is attractive to practitioners when a single solver for the target problem $T$ can be applied to a wide variety of problems that arise in practice: that is when the overhead cost of translating individual inputs to match the formulation for $T$ is less than the overhead cost of implementing a problem-specific solver for each new problem that arises.

The most widely studied versions of this approach involve
a subset of problems formulated as (A1), known as integer linear problems, which can be solved in polynomial time when the objective function $f(x)$ and constraints $c_i(x)$ are all linear. Another common approach is translating problems to \(\text{SAT}\) or \(\text{SAT-OPT}\), for which efficient heuristics are sometimes known.  

The quadratic unconstrained binary optimization (QUBO) problem has also been considered as a general-purpose target formulation, especially for problems defined on graphs, before quantum computing came onto the scene \cite{boros2007local, kochenberger2014unconstrained}. The emergence of quantum annealing processors that solve QUBOs natively in hardware has sparked recent interest in QUBO and its variation, the Ising Model (IM) problem, often used in physics applications.  The two problems are identical, except for the change in the domain from binary variables $x \in \{0,1\}$ (QUBO) to spin variables $s \in \{-1,+1\}$ (IM).  

The theory of \(\text{NP-COMPLETE}\)ness tells us that, in principle, any input for a problem formulated as (A1) can be transformed in polynomial time into a formulation that can be solved directly using a quantum computer.
See \cite{Glover2018QUBO, Lucas_2014} for tutorials on formulating general optimization problems expressed by (A1) as QUBOs and IMs. 
However, due to their small size, the problems currently being tested on quantum platforms are significantly restricted. 


\section{Quantum Heuristics for Optimization Problems} 
\label{apdx:heuristic_solutions}

The benchmarking framework measures performance characteristics of the two leading quantum heuristics for solving combinatorial optimization problems: the Quantum Approximate Optimization Algorithm (QAOA), which uses a gate-model quantum computer, and Quantum Annealing (QA), which uses an analog quantum computer. 
This paper presents a benchmark of these algorithms in the context of their application to solving the Max-Cut problem. 

The input for a Max-Cut problem is an undirected graph consisting of nodes or vertices ($V$) and edges ($E$).
In general, each edge of the graph can be accompanied by a `weight', but we only consider unweighted 3-regular graphs in this paper.
A cut partitions the graph's nodes into two sets. Its size is defined as the number of graph edges connecting nodes belonging to different sets. The Max-Cut problem is identifying a cut with the largest size out of all possible cuts. 

Max-Cut has emerged as a popular benchmark for quantum optimization \cite{Garey1976,Beaulieu2021MaxCut,Amaro_2022,Zhu_2020} for two reasons:
(1) it is among the most challenging combinatorial optimization tasks, even to obtain an approximate solution, i.e., APX-Hard~\cite{PAPADIMITRIOU1991425, Hastadoptimal2001},
(2) as an unconstrained discrete optimization task, it has a natural encoding as a Quadratic Unconstrained Binary Optimization (QUBO) \cite{Glover2018QUBO, Zhou_2020} or an Ising model~\cite{Lucas_2014, PhysRevE.58.5355}, ideally fitting current quantum optimization algorithms (QAOA, QA).
Although Max-Cut provides a reasonable first step for benchmarking current methods, testing more complex optimization tasks, including problems with constraints,  will be important in future work to demonstrate that quantum-accelerated optimization can impact a broad range of optimization applications.


\subsection{Quantum Approximate Optimization Algorithm}
\label{subsec:QAOA}

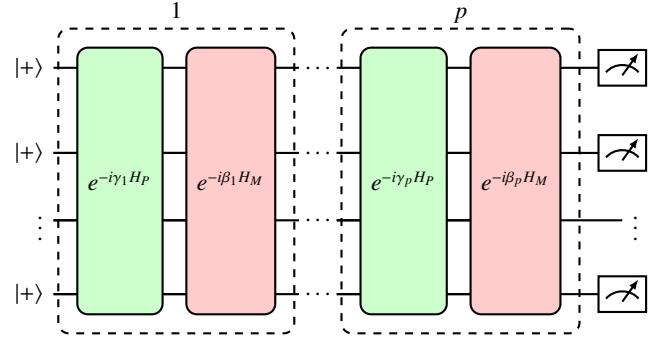
\begin{figure}[t!]
    \centering
    \begin{quantikz}[column sep=9pt,row sep={15pt}]
         \lstick{$\ket{+}$} & \gate[4,style={rounded corners,fill=green!20}]{e^{-i \gamma_1 H_P}}\gategroup[wires=4,steps=2,style={dashed, rounded corners}]{$1$} & \gate[4,style={rounded corners,fill=red!20}]{e^{-i \beta_1 H_M}} & \qw \ \ldots \ & \gate[4,style={rounded corners,fill=green!20}]{e^{-i \gamma_p H_P}}\gategroup[wires=4,steps=2,style={dashed, rounded corners}]{$p$} & \gate[4,style={rounded corners,fill=red!20}]{e^{-i \beta_p H_M}} & [5pt] \meter{}   \\
        \lstick{$\ket{+}$} &  \qw & \qw  & \qw  \ \ldots \ & \qw  & \qw & \meter{} \\ [0.1cm]
        \lstick{$\vdots$} &  \qw &\qw  & \qw  \ \ldots \ & \qw  & \qw  & \qw \rstick{$\vdots$} \\ [0.1cm]
         \lstick{$\ket{+}$} &  \qw & \qw & \qw  \ \ldots \ & \qw  & \qw & \meter{} 
    \end{quantikz}
    \caption{\textbf{The QAOA circuit} consists of $p$ repeating parameterized blocks. First, each qubit is acted upon with the Hadamard gate to obtain the $\ket{+}$ state. Each block further consists of a rotation involving the problem Hamiltonian $H_P$, followed by a rotation involving the mixer Hamiltonian $H_M$. Finally, all qubits are measured on the computational basis  $\{ \ket{0}, \ket{1} \}$.}
    \label{fig:qaoa_circuit}
\end{figure}


\begin{figure*}[t!]
    \includegraphics[width=0.13\textwidth]{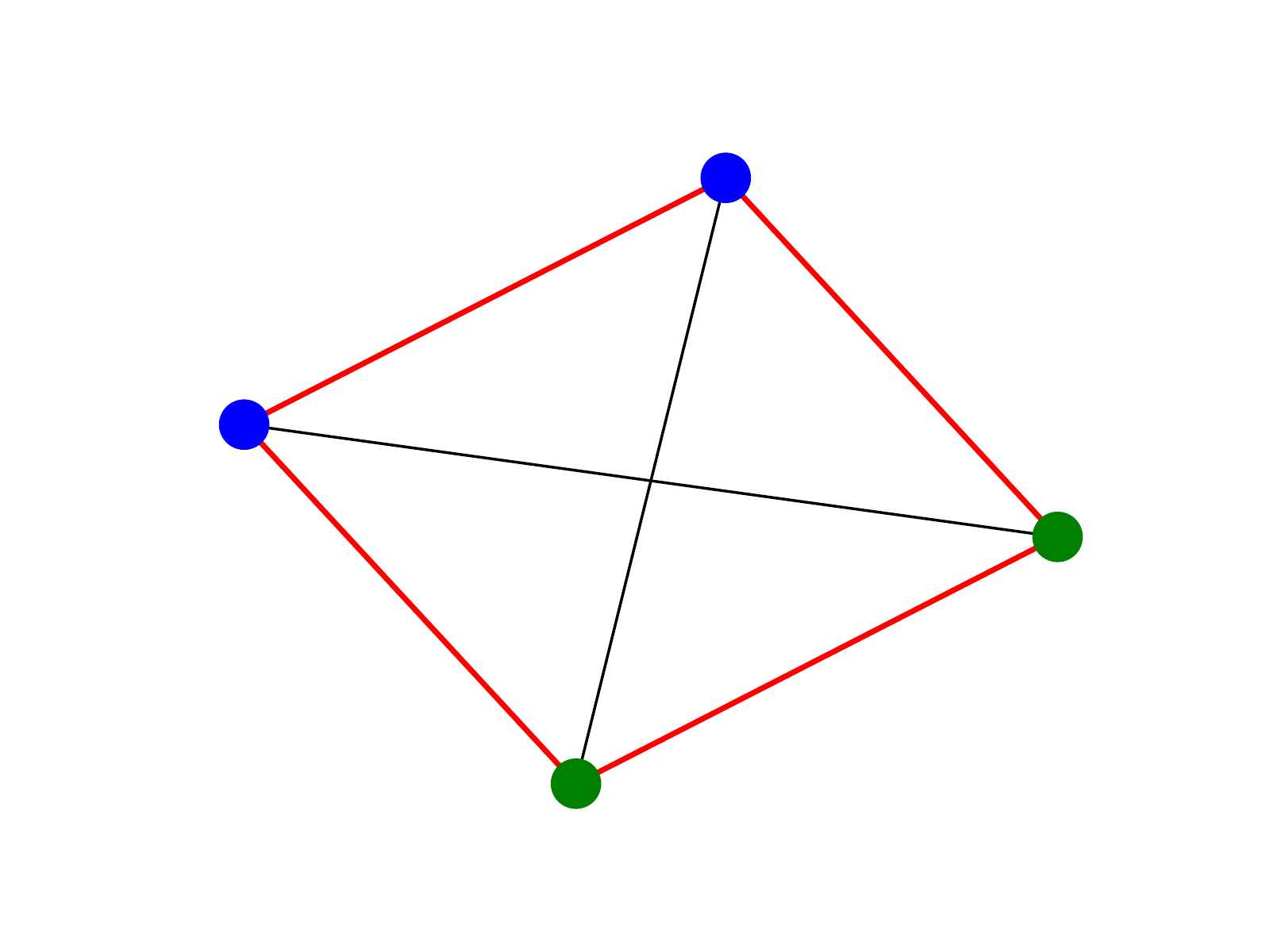}
    \includegraphics[width=0.13\textwidth]{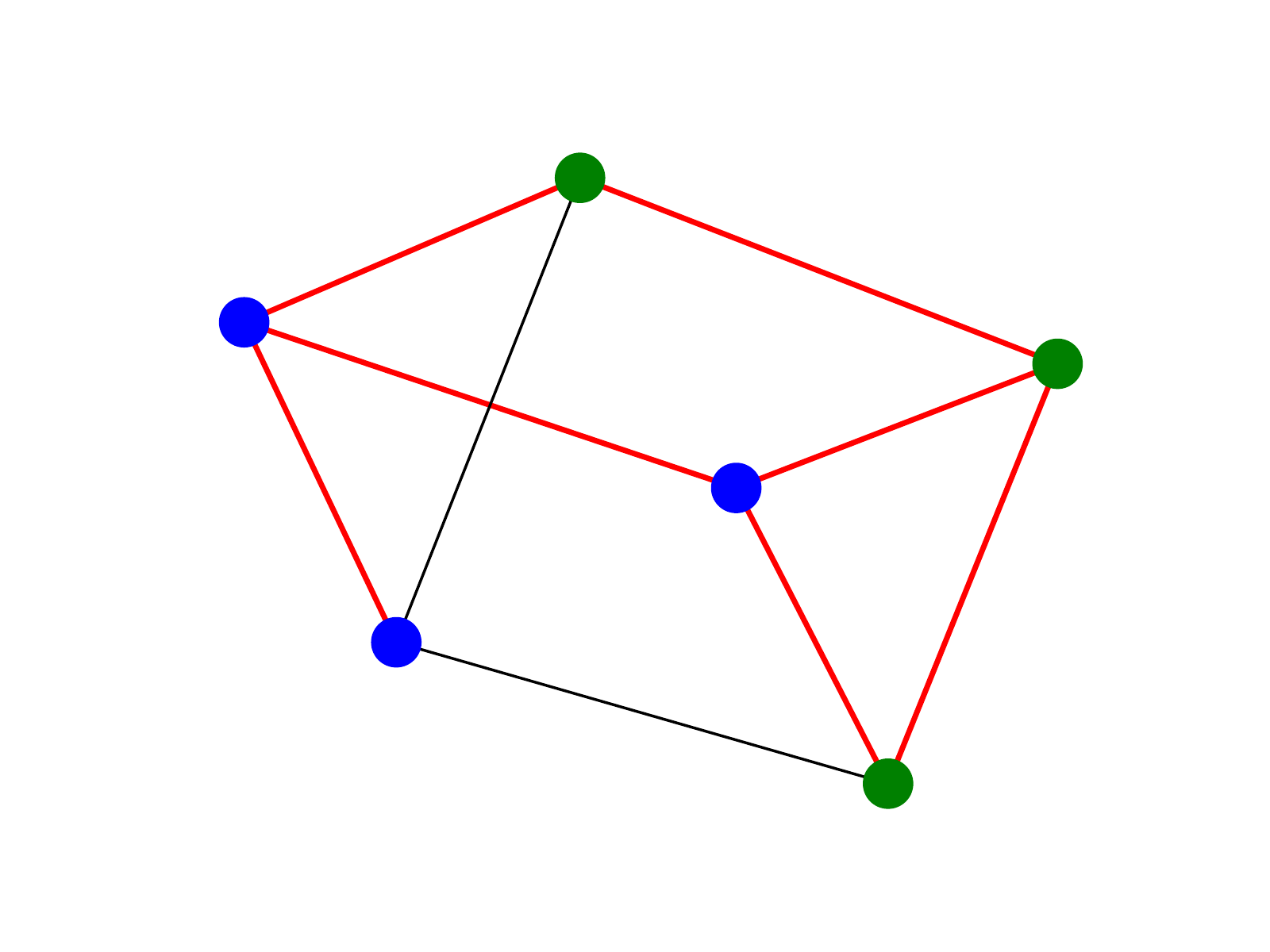}
    \includegraphics[width=0.13\textwidth]{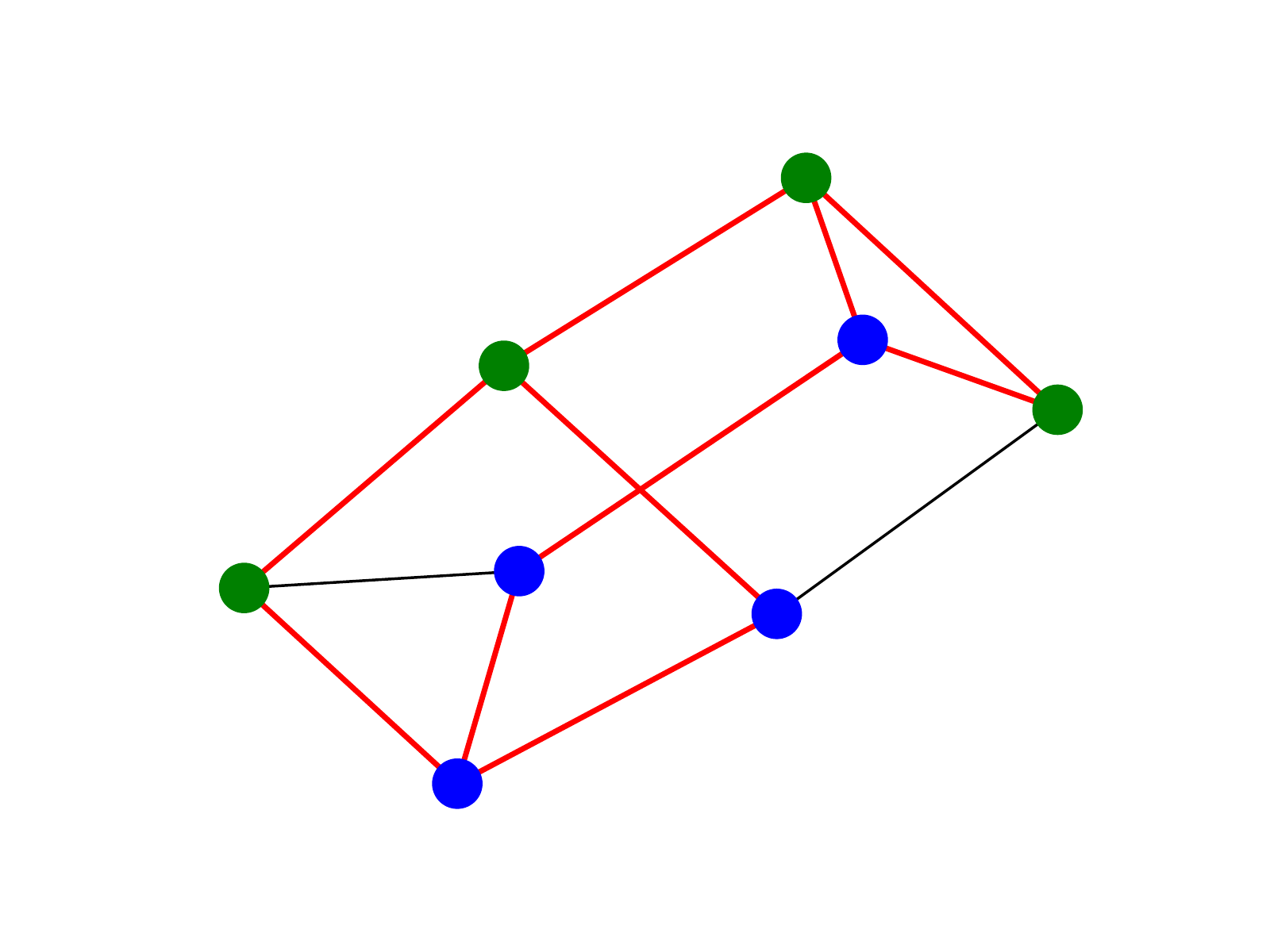}
    \includegraphics[width=0.13\textwidth]{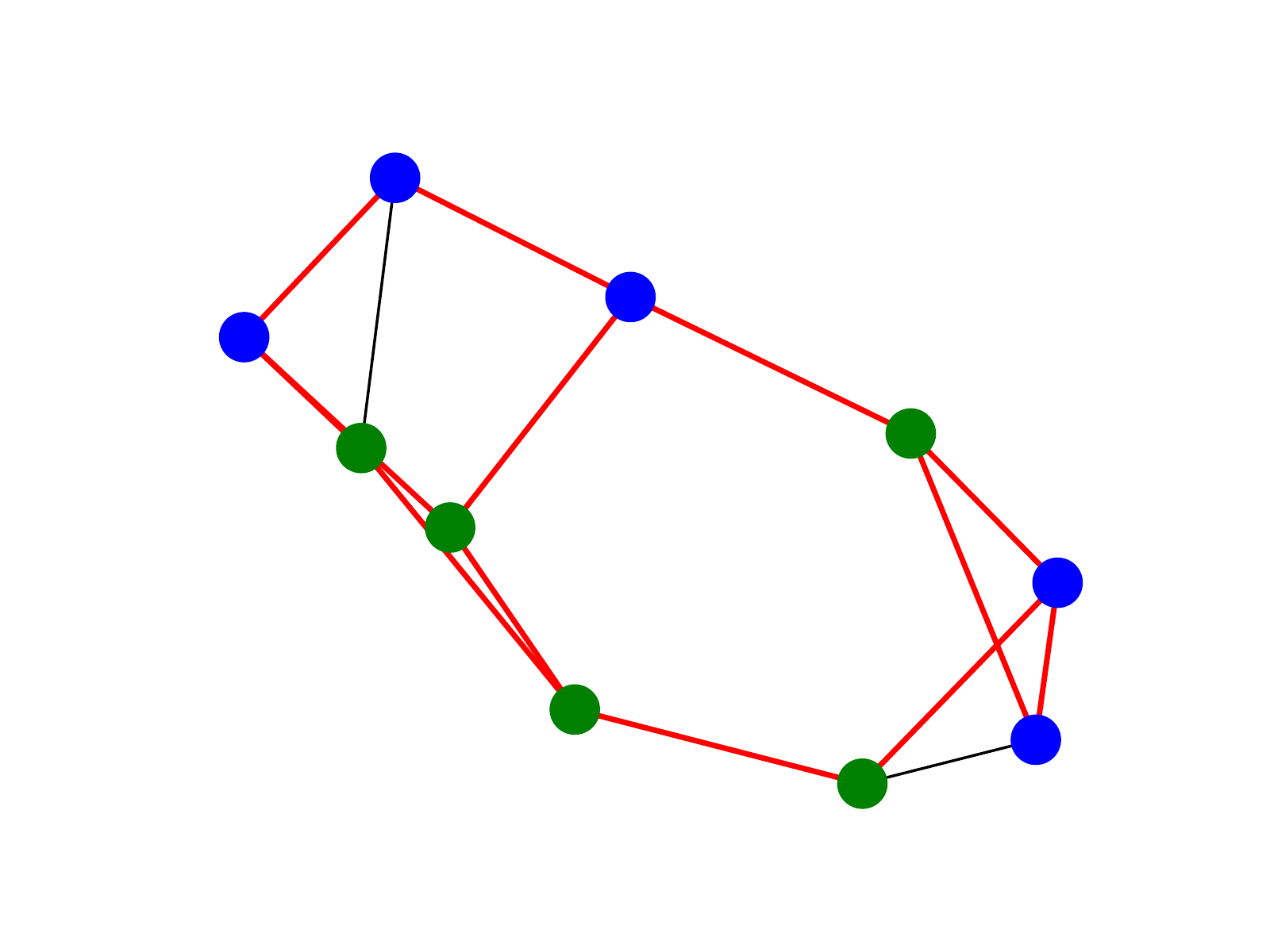}
    \includegraphics[width=0.14\textwidth]{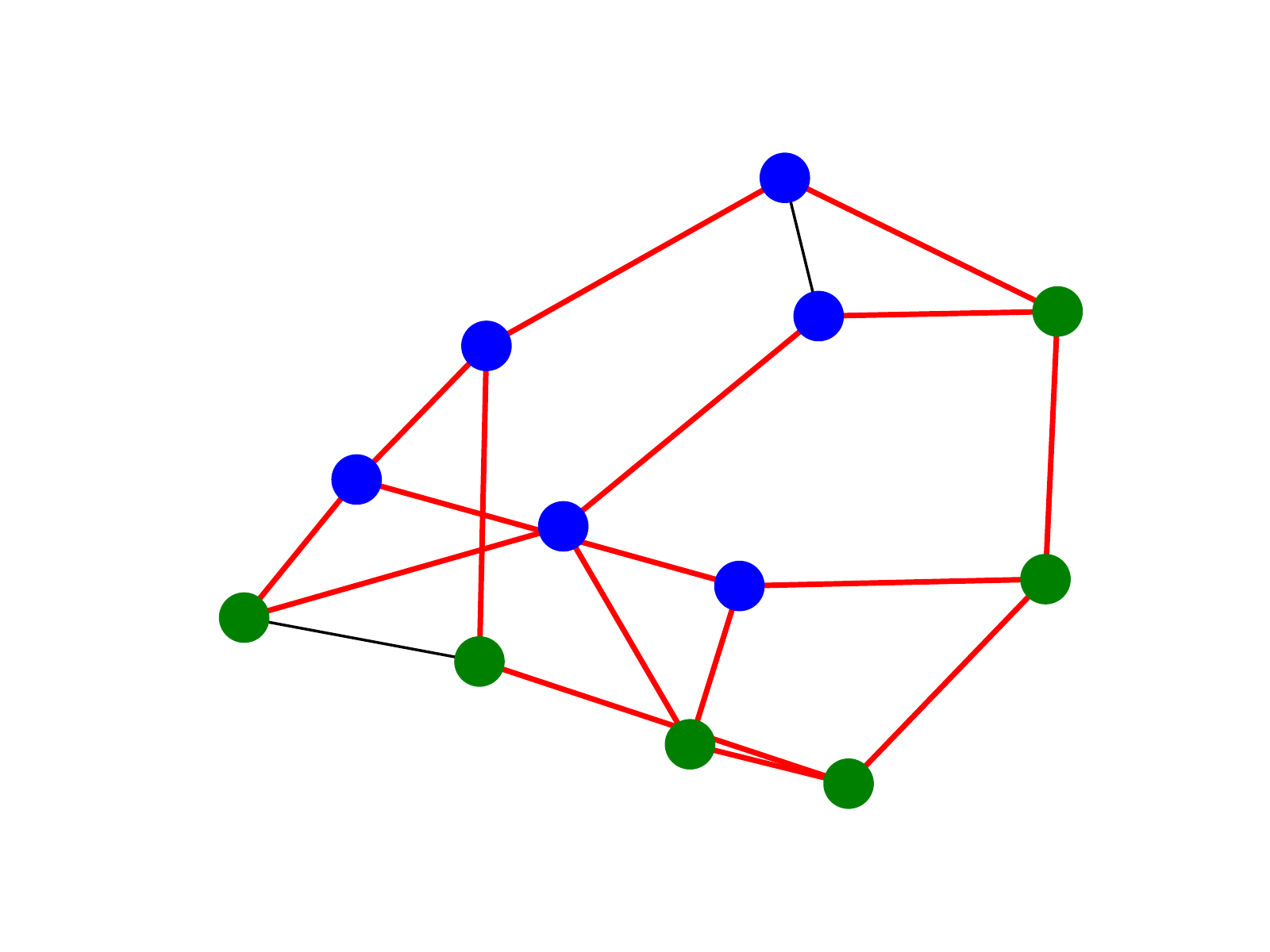}
    \includegraphics[width=0.14\textwidth]{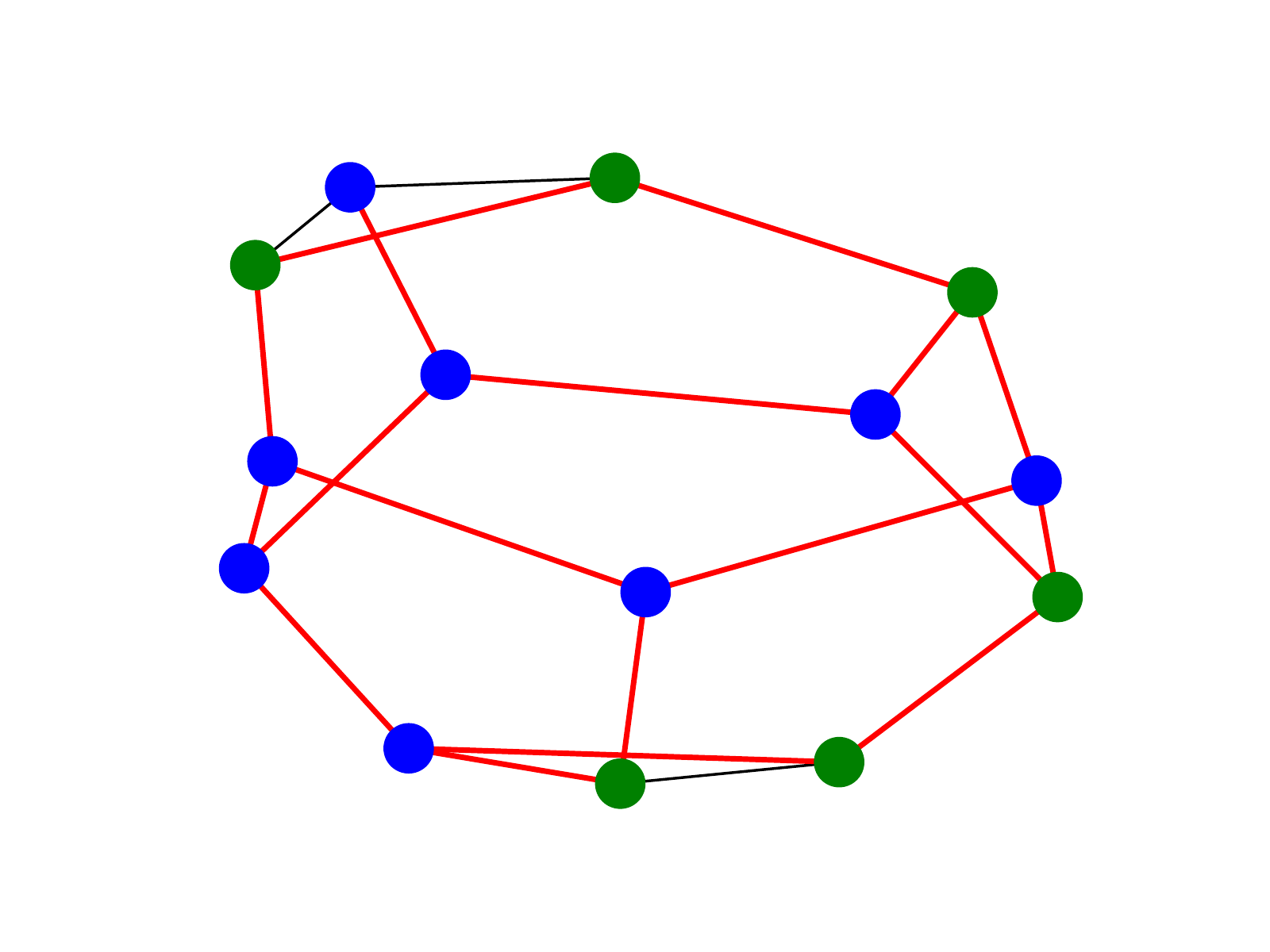}
    \includegraphics[width=0.14\textwidth]{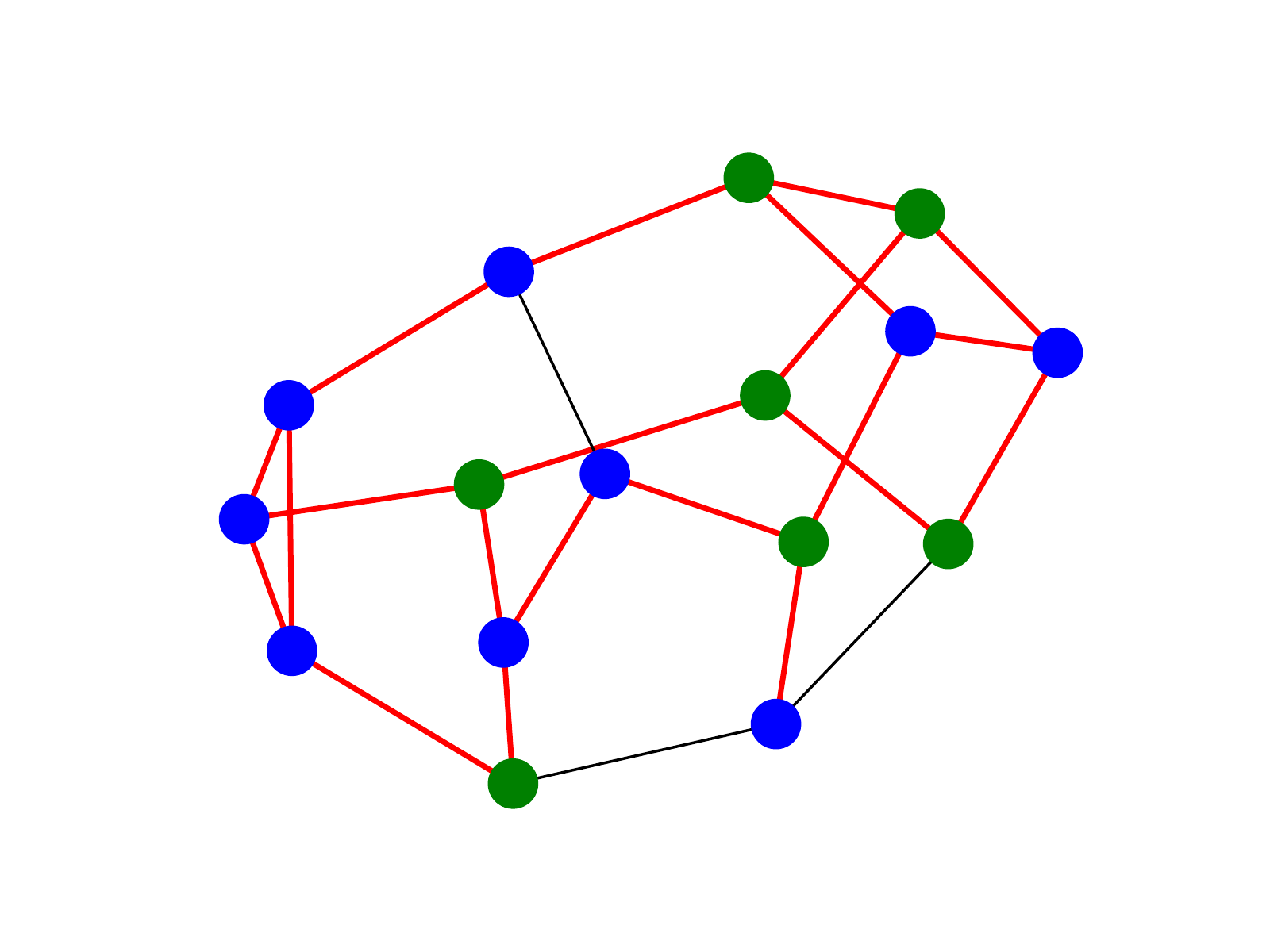}
    \caption{\textbf{Graph Instances Chosen for the Benchmark Implementations.} For each problem size ranging from 4 to 16 in increments of 2, we used both QAOA and QA to solve the MaxCut problem for one 3-regular graph with that number of nodes. (For QA, larger graphs up to 320 nodes were also generated). These graphs show one solution to the MaxCut problem using colored nodes and edges. Nodes with different colors belong to the two sets of the solution cut. The number of red edges connecting nodes from different sets is the MaxCut for that graph.}
    \label{fig:graph_diagrams_with_maxcut}
\end{figure*}

\begin{figure*}[t!]
    \includegraphics[width=0.78\textwidth]{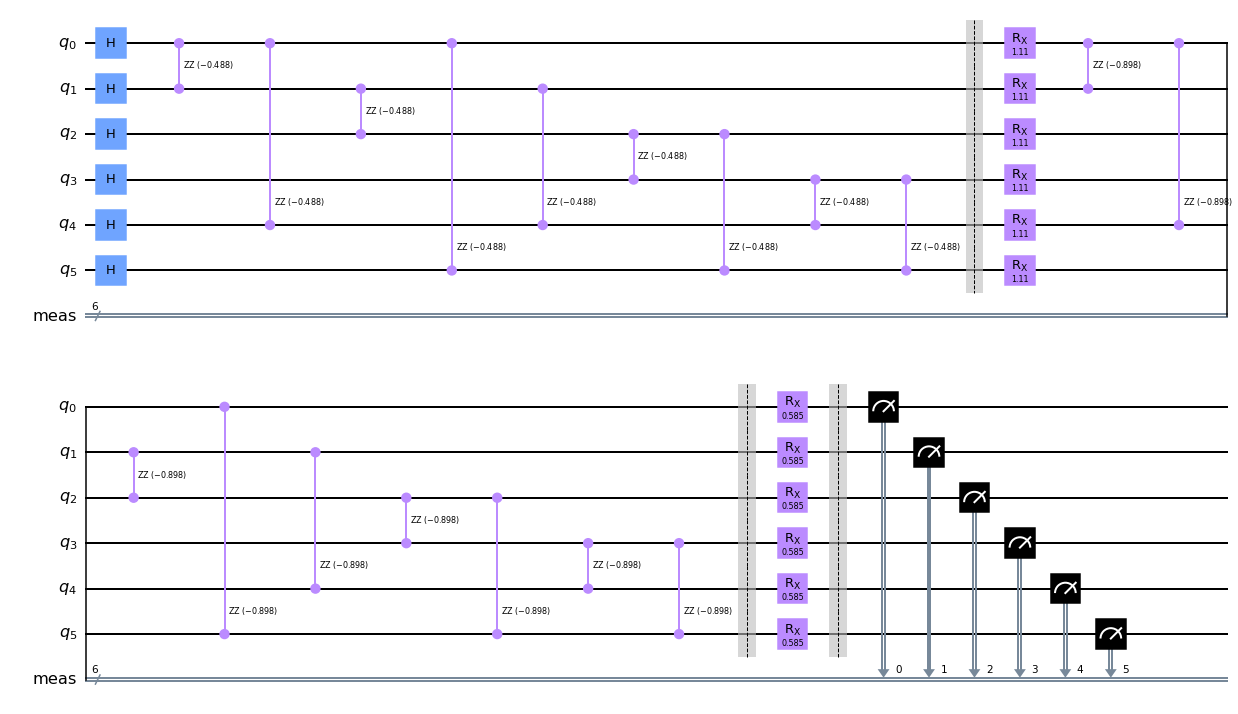}
    \caption{{\bf Sample QAOA Ansatz Circuit Diagram.} The quantum circuit shown here is the ansatz created for the MaxCut problem with 6 variables shown above, implemented using 2 rounds on 6 qubits. 
    Each of the two sets of 9 parameterized RZZ gates maps the edges within the graph to the circuit and represents the problem Hamiltonian. The two sets of parameterized RX gates represent the mixer Hamiltonian. 
    This circuit implements what is shown in \autoref{fig:qaoa_circuit} for the specific graph used in the benchmark.}
    \label{fig:circuit_diagram}
\end{figure*}

The Quantum Approximate Optimization Algorithm~\cite{farhiQuantumApproximateOptimization2014} is arguably the leading candidate for solving combinatorial optimization problems using gate model quantum processors.
QAOA belongs to the class of Variational Quantum Algorithms (VQA)~\cite{Cerezo_2021} and is usually implemented iteratively wherein a classical optimizer `trains' a parameterized quantum circuit.
QAOA is a heuristic that attempts to solve combinatorial optimization problems such as QUBO problems. Specifically, the problem is encoded in the form of a specified quadratic function of binary variables, and the objective is to find an assignment for those variables that minimizes the function. 

At the core of QAOA is an `ansatz circuit', a parameterized quantum circuit. 
Measurements in the computational basis at the end of the circuit correspond to sampling from a probability distribution over possible answers to the problem. 
A classical optimizer obtains parameter values with a significant probability of producing optimal or near-optimal solutions.
Finally, repeatedly measuring the circuit with the parameter values the optimizer determines provides approximate solutions to the problem.

The problem is first codified as a Hamiltonian $H_P$, such that an optimal problem solution corresponds to a ground state(s) or lowest energy eigenstate(s). For a given choice of the number of `rounds' denoted by $p$, the QAOA ansatz is given by
\begin{align}
    \ket{\vv \beta, \vv \gamma} = e^{-i\beta_p H_M} e^{-i \gamma_p H_P} \dotsc e^{-i\beta_1 H_M} e^{-i \gamma_1 H_P} \ket{+}, \label{eq:qaoa_ansatz}
\end{align}
where $H_M = \sum_i X_i$, is the so-called mixer Hamiltonian, and $\ket{+}= \otimes_i H \ket{0}$ is the equal superposition state. Here, $X$ is the Pauli X matrix, defined by $X\ket{0}=\ket{1}$ and $X\ket{1}=\ket{0}$. The ansatz state is thus obtained by implementing repeating and alternating rotations about $H_P$ and $H_M$ as shown in~\autoref{fig:qaoa_circuit}.

The Max-Cut problem can be framed in terms of obtaining the ground state of the Hamiltonian
\begin{align}
    H_P &= \frac{-1}{2}\sum_{\langle i,j\rangle\in E}(1 - Z_i Z_j), \label{eq:maxcut_hamiltonian}
\end{align}
where $E$ denotes the set of (undirected) edges of the graph, and $Z$ is the Pauli-Z matrix satisfying $Z\ket{0}=\ket{0}$ and $Z\ket{1}=-\ket{1}$. 
Each computational basis vector corresponds to a possible cut, and its energy represents the negative of the cut size. 
Note that the eigenvalues of $H_P$ are all nonnegative integers and that \autoref{eq:maxcut_hamiltonian} corresponds to an Ising model with all the coupling constants set to $1/2$.

A quantity called the approximation ratio is usually computed to characterize the quality of solutions. The approximation ratio $r$ is defined as the ratio of the energy expectation value $F_{\vv \beta, \vv \gamma}\coloneqq \bra{\vv \beta, \vv \gamma} H_P \ket{\vv \beta, \vv \gamma}$, and the ground state energy value $\emin$: 
\begin{align}
    \label{eq:approximation_ratio}
    r_{\vv \beta, \vv \gamma} = \frac{F_{\vv \beta, \vv \gamma}}{\emin} = \frac{\bra{\vv \beta, \vv \gamma} H_P \ket{\vv \beta, \vv \gamma}}{\emin}.
\end{align}
Note that the numerator is less than or equal to zero, whereas the denominator, which is the negative of the largest cut size, is strictly negative. Consequently, $0\leq r\leq 1$.
The classical optimizer routine aims to obtain optimal values of the angles $\vv \beta$ and $\vv \gamma$, i.e., values corresponding to the highest approximation ratio value.
$F_{\vv \beta, \vv \gamma}$, and hence $r_{\vv \beta, \vv \gamma}$ cannot be computed exactly and are instead approximated by measuring $\ket{\vv \beta, \vv \gamma}$ many (say M) times, or \emph{shots}, in the computational basis at the end of the circuit (see~\autoref{fig:qaoa_circuit}). 
Specifically, $F_{\vv \beta,  \vv \gamma}$ are approximated by the empirical average of energy.


\subsection{Quantum Annealing}
\label{subsec:QA}


With quantum annealing, an optimization problem is encoded into the machine, after which the solution is determined through quantum adiabatic evolution to arrive at a near-optimal final state.
The algorithmic approach of quantum annealing is to take advantage of the dynamic evolution of a quantum system to transform an {\em initial} ground state (which is easy to prepare) into the ground state of a {\em target} Hamiltonian, which is unknown and difficult to compute by other means.
At a high level, the protocol strives to identify the low-energy states of a user-specified $H_{\text{Target}}$ model by conducting an analog interpolation process of the following Hamiltonian:
\begin{equation}
    H(s) = (1-s)H_{\text{Init}} + (s) H_{\text{Target}}.
    \label{eq:qah}
\end{equation}
The interpolation process starts with $s = 0$ and in the ground state of $H_{\text{Init}}$. The annealing process involves a smooth interpolation of $s$ from $0$ to $1$. 
For a sufficiently long annealing time, the adiabatic theorem demonstrates that a quantum system remains at the minimal eigenvector of the interpolating Hamiltonian, $H(s)$ \citep{Bor1928, Kat1950, Jan2007}, and therefore arrives at minimum energy states of $H_{\text{Target}}$ at the end of the evolution.

Currently, available quantum annealing hardware focuses on a particular case of \autoref{eq:qah} that is limited to the Transverse Field Ising model,
\begin{equation}
    H(s) = A(s) \left( \sum_i X_i \right) + B(s) \left( \sum_i h_i Z_i + \sum_{i,j} J_{ij} Z_i Z_j \right). \label{eq:qatising}
\end{equation}
Where $X_i$ denotes the Pauli X operator applied to qubit $i$,   $Z_i$ denotes the Pauli Z operator applied to qubit $i$, and $Z_i Z_j$ is the tensor product of Z operators on qubits $i$ and $j$.
The two interpolation functions $A(s)$ and $B(s)$ control a transition from a strong $H_{\text{Init}}$ and weak $H_{target}$ to a weak $H_{\text{Init}}$ and strong $H_{target}$; that is, $A(0) \gg B(0)$ and $A(1) \ll B(1)$.
The hardware implements a default annealing ``path'' through these functions, which user parameters can modify.
This way, this hardware, and the QA algorithm can find ground and low-energy states of a user-specified classical Ising model specified on the $Z$ basis via the parameters $h$ and $J$, which encode the local fields and coupling strengths, respectively.
Note that the Max-Cut problem considered in this work is encoded in this model by setting $h=0$ and $J_{ij} = +1$ for each edge $(i,j) \in E$ that appears in the given Max-Cut graph.
If the Max-Cut graph cannot be encoded naively in the quantum annealing hardware (i.e., the edge set of the Max-Cut problem is not a subgraph of the $Z_i Z_j$ terms in the quantum annealing hardware), then a process known as {\em minor embedding}~\cite{choi2008minor} is used to map the Max-Cut problem into a mathematically equivalent hardware-native problem. 

It is interesting to note that the QAOA algorithm outlined in~\autoref{subsec:QAOA} can be interpreted as a Trotterized version of \autoref{eq:qatising} where the number of rounds $p$ determines the Trotter order.
That is, the limit of a QAOA circuit can model the smooth analog QA transition as $p \rightarrow \infty$.
The approximation ratio is computed for QA in the same way as QAOA by transforming the samples obtained after annealing to an equivalent distribution.


\section{Problem and Implementation Details}
\label{apdx:problem_details}

To generate the results presented in this paper, we execute the MaxCut benchmark on a single problem instance at each problem size (or number of qubits).
The instance is a randomly chosen 3-regular graph at each size.
A data set defining each instance is contained in the QED-C benchmark repository at
\href{https://github.com/SRI-International/QC-App-Oriented-Benchmarks}{https://github.com/SRI-International/QC-App-Oriented-Benchmarks}.
The benchmark can be modified to use other graphs if desired.

For reference, in \autoref{fig:graph_diagrams_with_maxcut}, we present some of the graphs used for different size problems.
Each graph shows one solution to the MaxCut problem using colored nodes and edges, as described in the caption.

In \autoref{fig:circuit_diagram}, we show the QAOA ansatz circuit generated for the problem of size 6 (number of variables/qubits) used in this benchmark.
The caption describes how the components of the circuit represent the Hamiltonian that defines the problem.

Below, we show the $J$ matrix required to specify the 6-variable Max-Cut problem for the quantum annealing hardware according to \autoref{eq:qatising}.

\begin{equation}
    \label{eq:J_6}
    J=
    \begin{bmatrix}
    0 &  -1 &  0 &  0 &  -1 &  -1 \\
    -1 &  0 &  -1 &  0 &  -1 &  0 \\
    0 &  -1 &  0 &  -1 &  0 &  -1 \\
    0 &  0 &  -1 &  0 &  -1 &  -1 \\
    -1 &  -1 &  0 &  -1 &  0 &  0 \\
    -1 &  0  &  0 &  0 \end{bmatrix}
\end{equation}


\section{Analysis of Execution Time}
\label{apdx:exec_time_analysis}

\paragraph{\textbf{Execution Time in QAOA}}
The total time the QAOA algorithm consumes includes both quantum and classical components. 
There is a time associated with executing the quantum ansatz circuit on the quantum processor.
Additionally, time is spent on a classical processor to perform the minimization function that computes new parameters from measurements obtained after each execution of the ansatz.

The time to execute the quantum portion of the algorithm itself is broken down into several components.
One is the `Quantum Execution Time', defined as the time to execute $N$ shots of a quantum circuit within a quantum processor. 
Quantum computer hardware providers typically report this time in a result record and include the time required to initialize the quantum system before execution and the delay between shots \cite{wack_clops_2021} as shown in \autoref{eq:quantum_execution_time}.
\begin{equation}
    \label{eq:quantum_execution_time}
    t_{\rm quantum} = t_{\rm init} + N_{\rm shots} \cdot (t_{\rm shot} + t_{\rm delay})
\end{equation}

Using QAOA, a quantum circuit is executed repeatedly but with varying parameters.
The total time to execute the circuit, the `Elapsed Quantum Execution Time', includes the time required to either compile the circuit or to apply parameters before execution, and to validate and load the compiled circuit for execution.
Another highly variable component is the time spent in a queue awaiting execution.
The elapsed quantum execution time is defined in \autoref{eq:elapsed_quantum_execution_time} and must be collected as part of the benchmarking algorithm, as we did not find this metric directly available in most systems.
\begin{equation}
    \label{eq:elapsed_quantum_execution_time}
    t_{\rm elapsed\_quantum} = t_{\rm queue} + t_{\rm compile} + t_{\rm load} + t_{\rm quantum}
\end{equation}

The sum of the quantum and elapsed quantum times for all iterations of the QAOA algorithm, the `Cumulative Quantum Execution Time' and `Cumulative Elapsed Quantum Execution Time' respectively, are defined in \autoref{eq:cum_quantum_execution_time} and \ref{eq:cum_elapsed_quantum_execution_time}.
The financial cost of quantum computation is often tied to the cumulative quantum execution time in many hardware systems.
The elapsed time for each iteration depends on the execution parameters and may be influenced by the system's ability to support parameterized execution or the inclusion of hidden classical post-processing time, such as error mitigation.
\begin{equation}
    \label{eq:cum_quantum_execution_time}
    t_{\rm cum\_quantum} = \sum_{iter=1}^{N_{\rm iter}} t_{\rm quantum(iter)}
\end{equation}
\begin{equation}
    \label{eq:cum_elapsed_quantum_execution_time}
    t_{\rm cum\_elapsed\_quantum} = \sum_{iter=1}^{N_{\rm iter}} t_{\rm elapsed\_quantum(iter)}
\end{equation}

`Classical Execution Time' for QAOA is the sum of the time needed to create the ansatz with specific parameters and the time used by the minimizer to process measurement results and generate new parameters during a particular iteration $iter$, as in \autoref{eq:classical_execution_time}.
The `Cumulative Classical Execution Time' is the sum of the classical execution times for all iterations as in \autoref{eq:cum_classical_execution_time}.
For small problems, this time is typically insignificant but can increase with problem size and rounds.
It may also be impacted by the system's ability to support parameterized execution and reduce creation time.
\begin{equation}
    \label{eq:classical_execution_time}
    t_{\rm classical} = t_{\rm create} + t_{\rm optimize}
\end{equation}
\begin{equation}
    \label{eq:cum_classical_execution_time}
    t_{\rm cum\_classical} = \sum_{iter=1}^{N_{\rm iter}} t_{\rm classical(iter)}
\end{equation}

The total execution time for QAOA is the sum of the cumulative elapsed quantum time $t_{\rm cum\_elapsed\_quantum}$, and classical compute time $t_{\rm cum\_classical}$.
Variability due to different processing options or choice of classical optimizer can result in widely differing result quality and execution times.
\vspace{0.3cm}

\paragraph{\textbf{Execution time in QA}}
For QA, the `Quantum Execution Time' is defined as the time the quantum processing unit (QPU) takes to execute $N$ reads (samples) using a specified anneal time. 
This time is reported by the hardware as `qpu\_access\_time' and includes `qpu\_programming\_time' of \textasciitilde~$16 ms$, and `qpu\_sampling\_time',  which is  `anneal\_time' plus `readout\_time' 
(\textasciitilde~$0.25 ms$), multiplied by the number of reads. 
Quantum execution time is defined in~\autoref{eq:qa_quantum_execution_time}.
\begin{equation}
    \label{eq:qa_quantum_execution_time}
    t_{\rm quantum} = t_{\rm qpu\_access} = t_{\rm qpu\_programming} + t_{\rm qpu\_sampling}
\end{equation}

The `Elapsed Quantum Execution Time' includes the time required to issue the sample command to the (remote) backend hardware system, wait for it to complete, and receive a fully resolved sample set.
It is defined in \autoref{eq:qa_elapsed_quantum_execution_time}. It includes the quantum execution time, along with the time for computing a minor embedding of the input to match the specific qubit connection structure inside the QPU, and the time to resolve solutions by mapping them back to the original (unembedded) problem. 
The embedding cost is measured once for each annealing time we test, which is not always necessary in practice because embeddings can be reused. 
\begin{equation}
    \label{eq:qa_elapsed_quantum_execution_time}
    t_{\rm elapsed\_quantum} = t_{\rm queue} + t_{\rm embed} + t_{\rm sample} + t_{\rm quantum} + t_{\rm resolve}
\end{equation}

For QA, the cumulative times reported in~\autoref{fig:opt_intro_examples_2} reflect measurements of  $t_{\rm quantum}$ for increasing anneal times, as specified on line 6 of the benchmarking code.
The $t_{\rm elapsed\_quantum}$ metric includes all the time needed to perform the annealing operation and obtain a final sample set.
This is comparable to the cumulative times in QAOA.
To illustrate the difference between the two, we visualize the data in a slightly different style. Each bar represents a different anneal time and has a slight vertical offset from the one before it to convey that it represents the time starting at 0.
See~\autoref{sec:annealing_hardware_execution} for a presentation of these metrics collected from execution on quantum annealing hardware.


\section{Result Quality and Hyper-Parameters}
\label{apdx:result_quality_assessment}

The performance of an optimization algorithm is often studied in terms of the trade-off between the quality of the obtained result and the resources required to achieve it.
In many real-world applications, a high-quality result is required in a limited time.
It is desirable to predict whether obtaining a solution with acceptable quality within the available time budget and to determine the parameter values that result in high-quality outputs is possible.

However, many options (or ``hyper-parameters'') can be used to control the execution of QAOA, and conclusions must not be drawn from just one set of results obtained with limited exploration. 
While various hyper-parameters, such as the number of shots, choice of the classical optimizer, number of iterations, and rounds, affect result quality, we focus on the effects of values of initial angles on the performance in this section. We end this section by discussing a parameter tuning strategy to help identify good hyper-parameters for QAOA execution. (Similar strategies can be developed for QA but are not discussed here.) 

While this section refers to QAOA in the context of the Max-Cut problem, most of the conclusions hold more generally for QAOA. Throughout this section, we use the terms cut size and energy interchangeably, where energy refers to the eigenvalues of the Hamiltonian in \autoref{eq:maxcut_hamiltonian}.

\begin{figure*}[t!]
    \centering
    \begin{subfigure}{0.32\textwidth}
        \includegraphics[width=\textwidth]{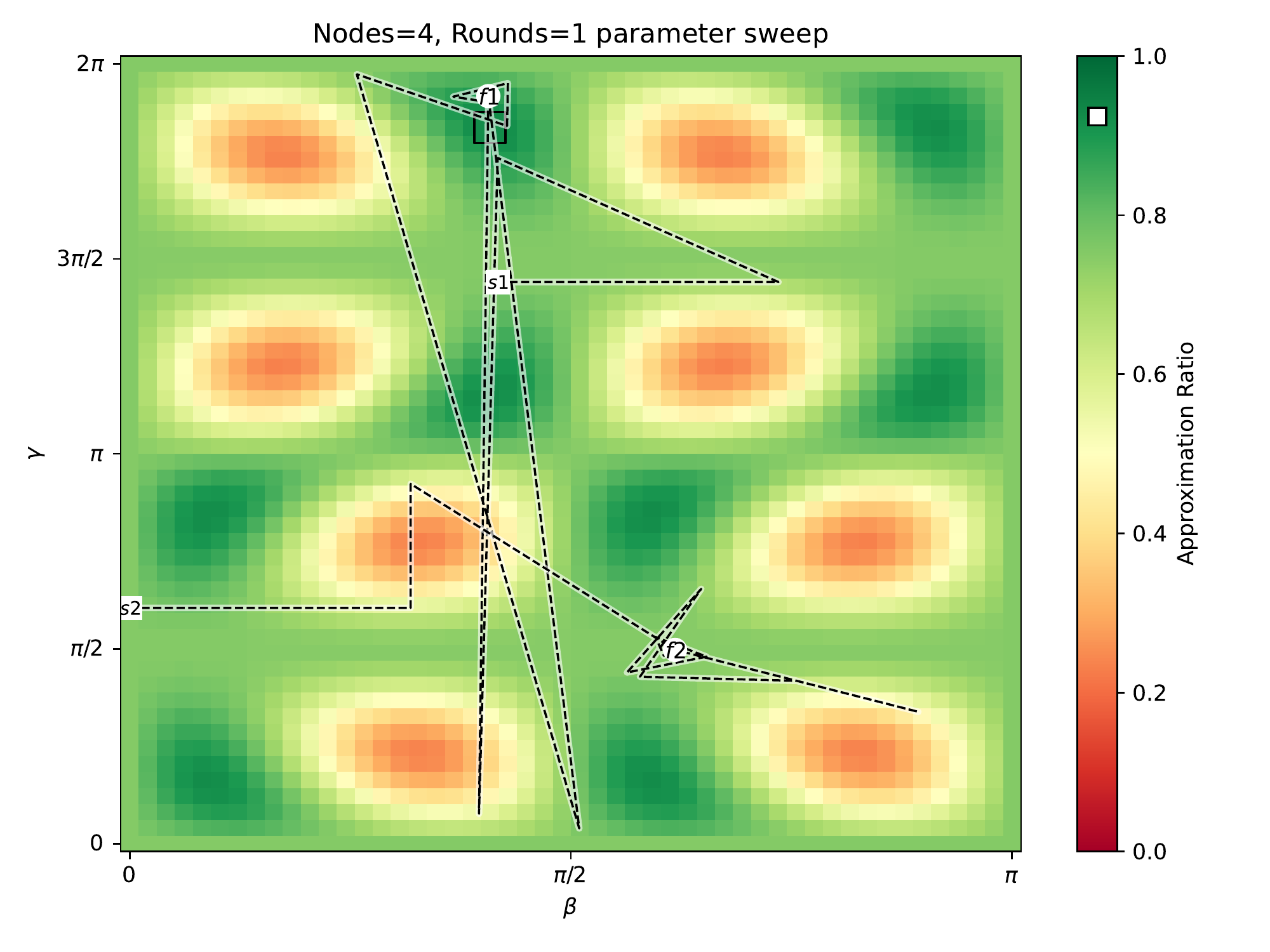}%
        \caption{}%
        \label{fig:landscape}%
    \end{subfigure} \hfill
    \begin{subfigure}{0.32\textwidth}
        \includegraphics[width=\textwidth]{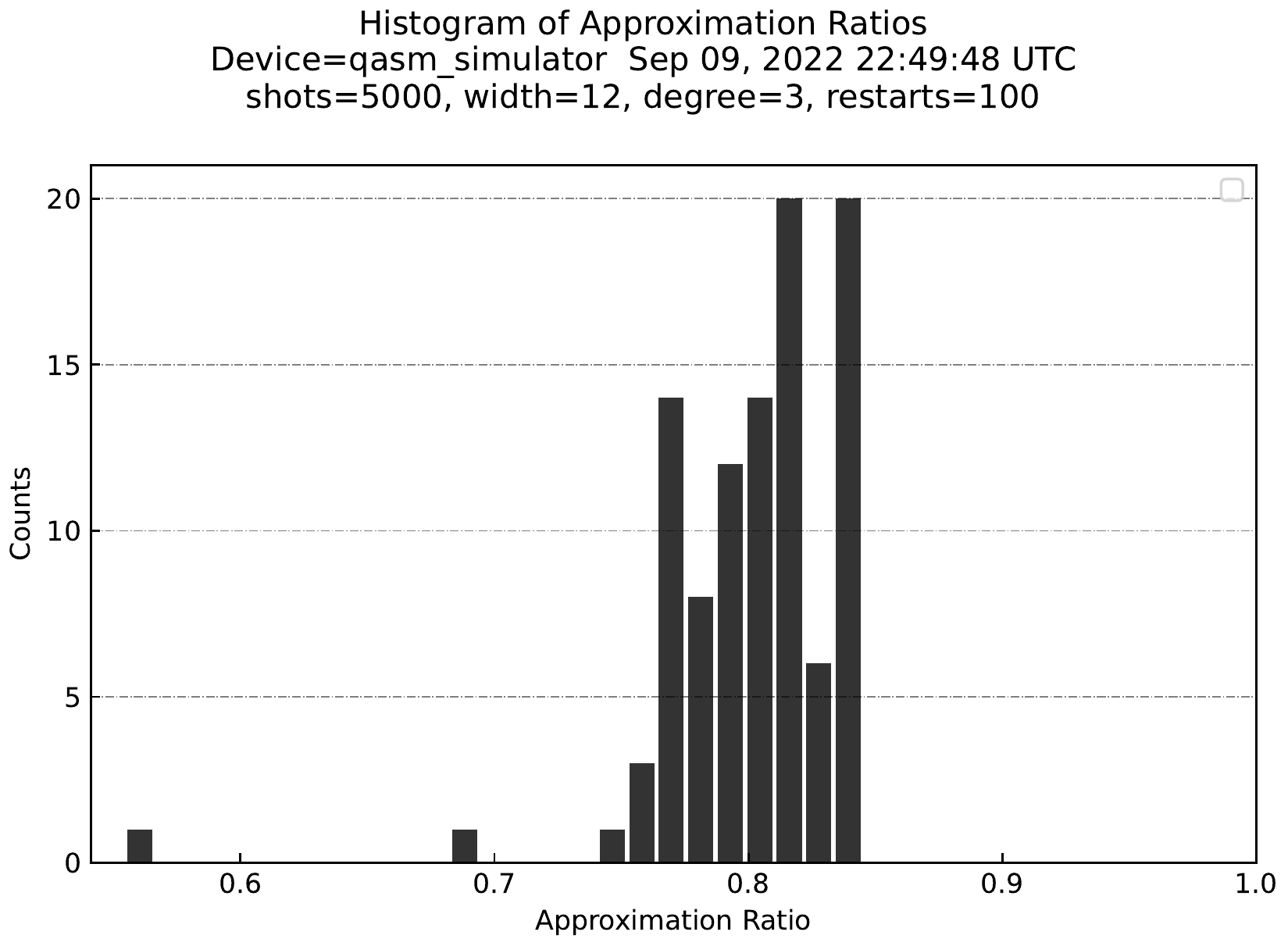}%
        \caption{}%
        \label{fig:histogram_of_ARs}%
    \end{subfigure}\hfill
    \begin{subfigure}{0.32\textwidth}
        \includegraphics[width=\textwidth]{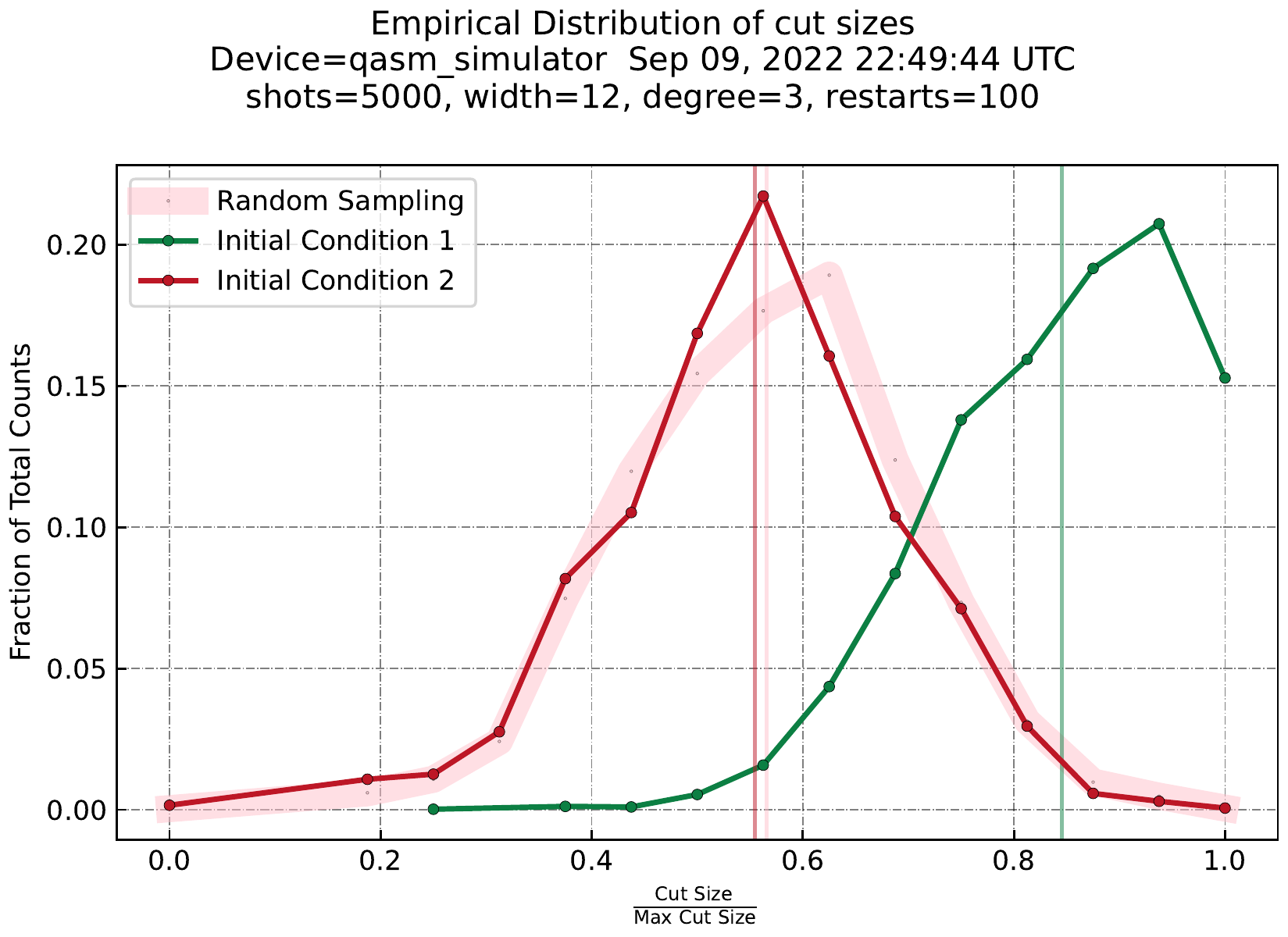}
        \caption{}%
        \label{fig:good_vs_bad}%
    \end{subfigure}
    \caption{{\bf Values of Initial Angles Affect Result Quality}. (a) {\it Parameter trajectories and approximation ratio landscape:} The COBYLA optimizer navigates the parameter space differently depending on the initial parameter values.
    The trajectory taken for two randomly chosen initial angle values (labeled $s1$ and $s2$) is shown in the background of the approximation ratio landscape, obtained from a state-vector simulation.
    The parameters at the end of the 30 iterations are labeled $f1$ and $f2$, respectively. (b) {\it Histogram of approximation ratios} obtained at the end of 30 COBYLA iterations from QAOA simulations for 100 restarts succinctly shows the variability associated with initial conditions. Although most restarts result in an approximation ratio between 0.75 and 0.85, some result in a substantially lower approximation ratio. (c) The distribution of cut sizes at the end of 30 iterations for two initial conditions is substantially different, with initial condition 2 almost overlapping with a random sampling of bit-strings, while initial condition 1 results in a relatively high approximation ratio of $\approx$ 0.83.}
    \label{fig:initial_conditions_three_subfigs}
\end{figure*}


\subsection{Initial Angles and Restarts}
\label{subsec:initial_conditions_effects}

While several hyper-parameters, such as rounds, shots, number of optimizer iterations, etc., affect the ability of the iterative QAOA execution to obtain a high-quality output, perhaps the most critical and non-trivial choice is that of the initial values of the angles.

The classical optimization routine faces several challenges. 
Finding the optimal angles for QAOA has been shown to be an \(\text{NP-HARD}\) problem~\cite{Bittel2021b}. Additionally, the landscape of the objective function suffers from `barren plateaus', a condition where the gradient of the objective function is close to zero, hindering training of the angles~\cite{McClean2018}. Barren plateaus can also be exacerbated by the choice of objective function~\cite{Cerezo2021}, noise in quantum hardware~\cite{Wang2021}, or large entanglement in the ansatz~\cite{Marrero2021}.

A consequence of these challenges is that the choice of the initial angles (i.e., $\vv \beta$ and $\vv \gamma$) can substantially affect the optimizer's ability to reach the optimal value of the objective function. For example, \autoref{fig:landscape} shows the trajectories of the angles probed by the optimizer for two randomly chosen initial angles. The distribution of the cut sizes obtained at the end of 30 optimizer iterations is substantially different, as shown in \autoref{fig:good_vs_bad}. While one choice results in an output practically indistinguishable from a random sampling of bitstrings, the other results in a high-quality distribution of cuts. We also plot a histogram of the approximation ratios in \autoref{fig:histogram_of_ARs} from 100 random initializations.

These issues have spurred substantial research to address and overcome these challenges.
Although many proposals have been put forth, keeping in mind our objective of benchmarking the performance of quantum solutions available to end users, we focus on implementing the most basic approach for mitigating some of these effects. Specifically, we implement multiple `restarts', i.e., we run QAOA multiple times using random angles as initial angles for the optimizer and report the output corresponding to the best restart.

\begin{figure}[t!]
\includegraphics[width=0.80\columnwidth]{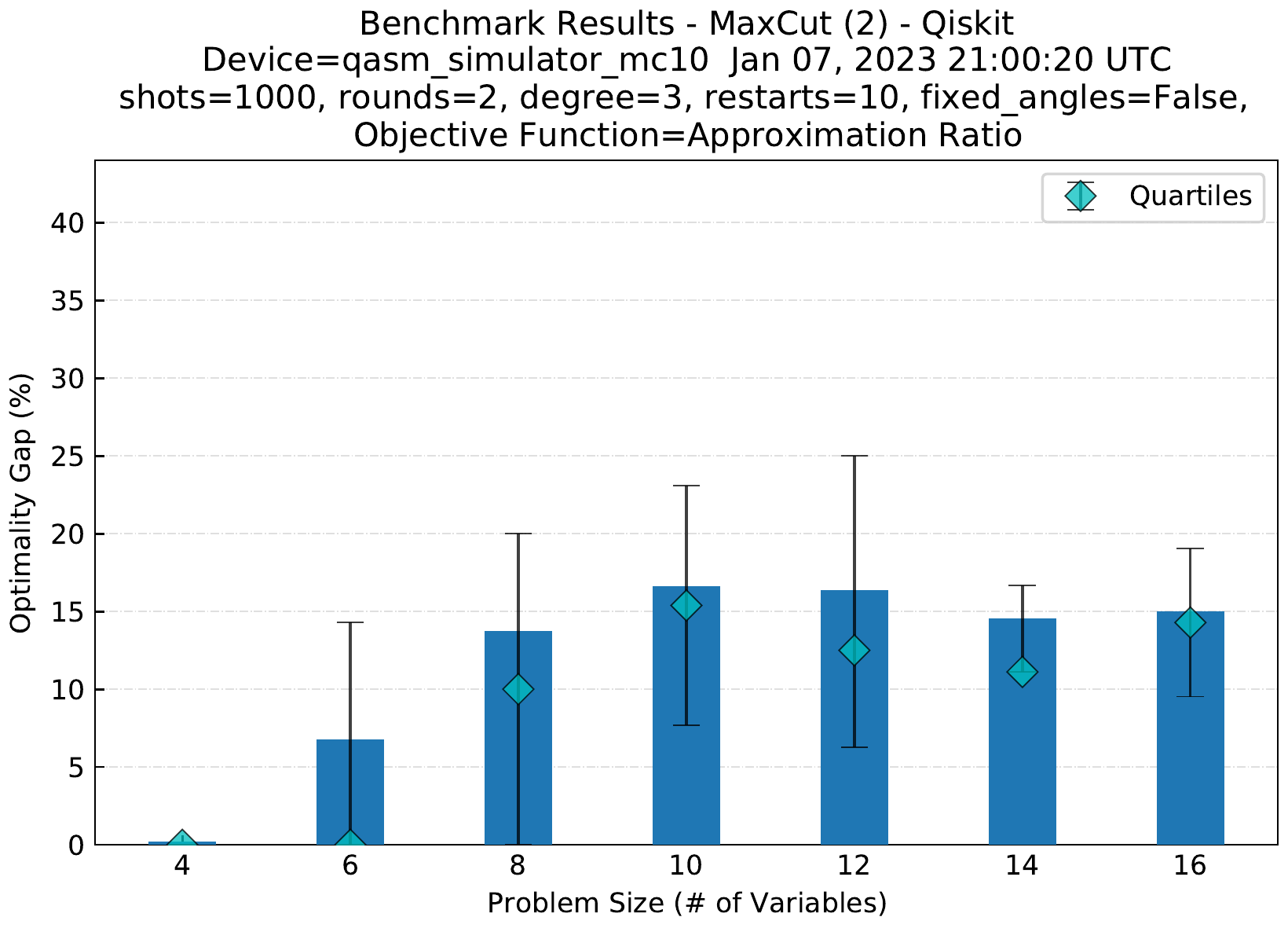}
\caption{\textbf{Optimality Gap with Multiple Restarts.} We show the results obtained when executing the Max-Cut benchmark 10 times at each problem size. The result shown for each problem size represents the `best' result obtained for that problem size across all 10 restarts, defined as the result showing the highest approximation ratio.}
\label{fig:optgap_ten_restarts}
\end{figure}

To this end, our benchmarking framework allows users to specify the number of restarts through a parameter called \verb|max_circuits|. This parameter is set to 1, and all the initial $\beta$ and $\gamma$ angles are set to 1. Thus, all the results in~\autoref{sec:execution_on_hardware} use these starting angles. For restarts~$>1$, for each problem size, the output corresponding to the best restart is displayed in the plots. \autoref{fig:optgap_ten_restarts} shows the output corresponding to $10$ restarts for the same parameters as \autoref{fig:optgaps_barchart}. The quality of the results is noticeably better for smaller problem sizes. The user can also specify initial angles manually using the \verb|thetas_array| parameter.


\subsection{Fixed Angle Conjecture}
\label{sec:fixed_angle_conjecture}

Although multiple initializations or restarts mitigate some of the difficulties faced by the optimizer, the cost of implementing the optimizer routine multiple times can be substantial, requiring many-fold quantum) processing unit access time.

\begin{figure}[t!]
    \includegraphics[width=0.80\columnwidth]{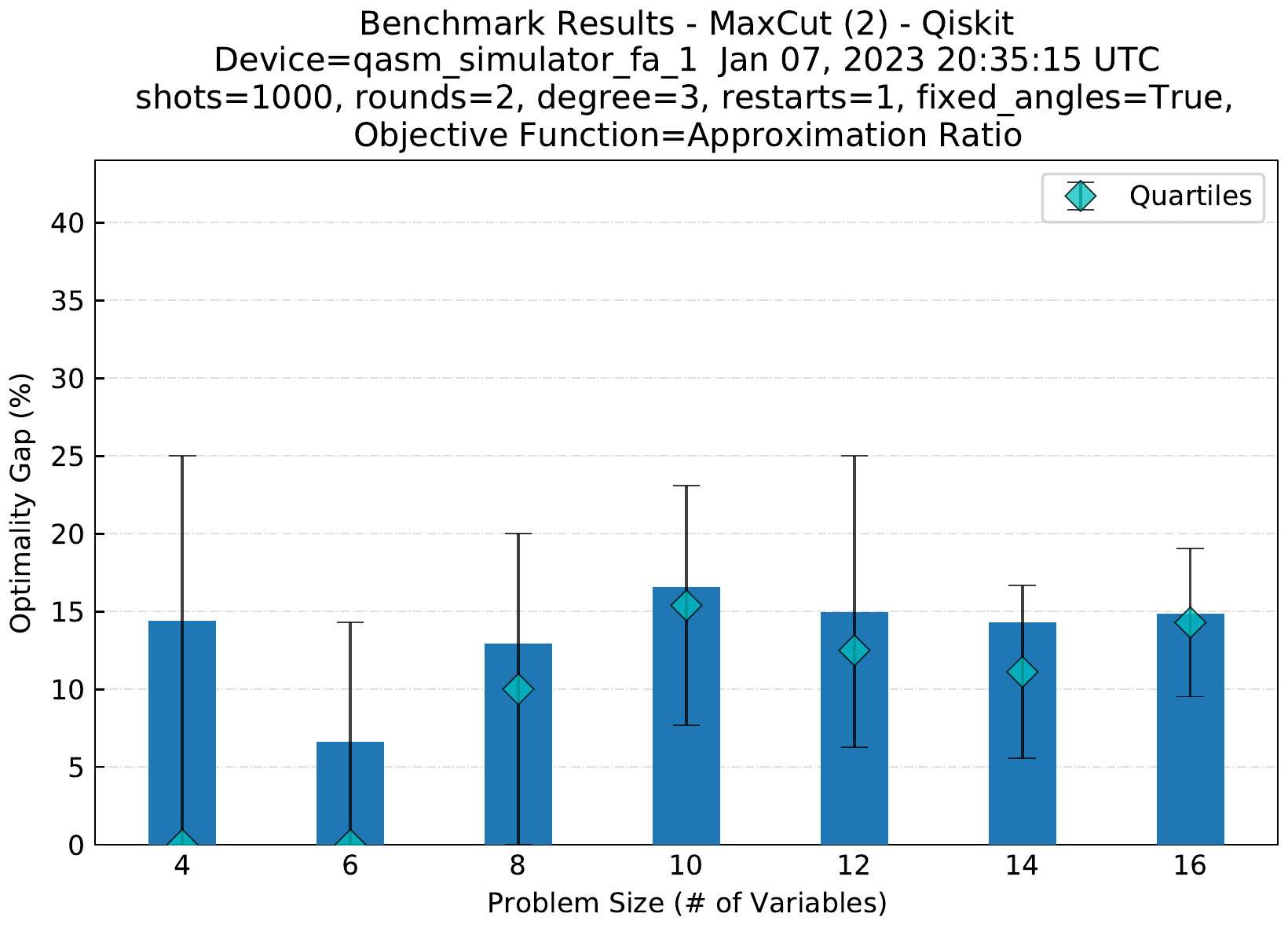}
    \caption{{\bf Optimization-free QAOA Implementation using Fixed Angles.} To avoid restarts and the costly optimization loop, one may use the `fixed angles` guaranteed to produce a good quality output~\cite{WurtzFixedangle2021}. Here is the optimality gap using fixed angles for rounds=2 without implementing the minimizer routine.}
    \label{fig:optgap_fixed_angle_one_iteration}
\end{figure}

A recent study~\cite{WurtzFixedangle2021} proposes an optimization-free QAOA implementation, executing the ansatz for each problem only once using the so-called `fixed angles'. The authors show that at these angles, the approximation ratio is higher than the threshold for every problem instance for 3-regular graphs. Although these angles are not the global maxima of the approximation ratio landscape, they guarantee close to optimal performance without performing the costly optimizer loop.
For example, \autoref{fig:optgap_fixed_angle_one_iteration} shows that the optimality gap for all problem sizes (except 4) with fixed angles is practically the same as in \autoref{fig:optgap_ten_restarts}, which required 10 restarts with 30 optimizer iterations each. This corresponds to a reduction in QPU access time by a factor of $\approx$300 while yielding similar quality results. 

Hence, the benchmark framework includes a provision for choosing the initial angles to be the fixed angles by setting the \verb|use_fixed_angles| flag to \verb|True|. The optimizer iterations can be set simultaneously to 1 to avoid using the optimizer routine.

\begin{figure}[t!]
    \includegraphics[width=0.80\columnwidth]{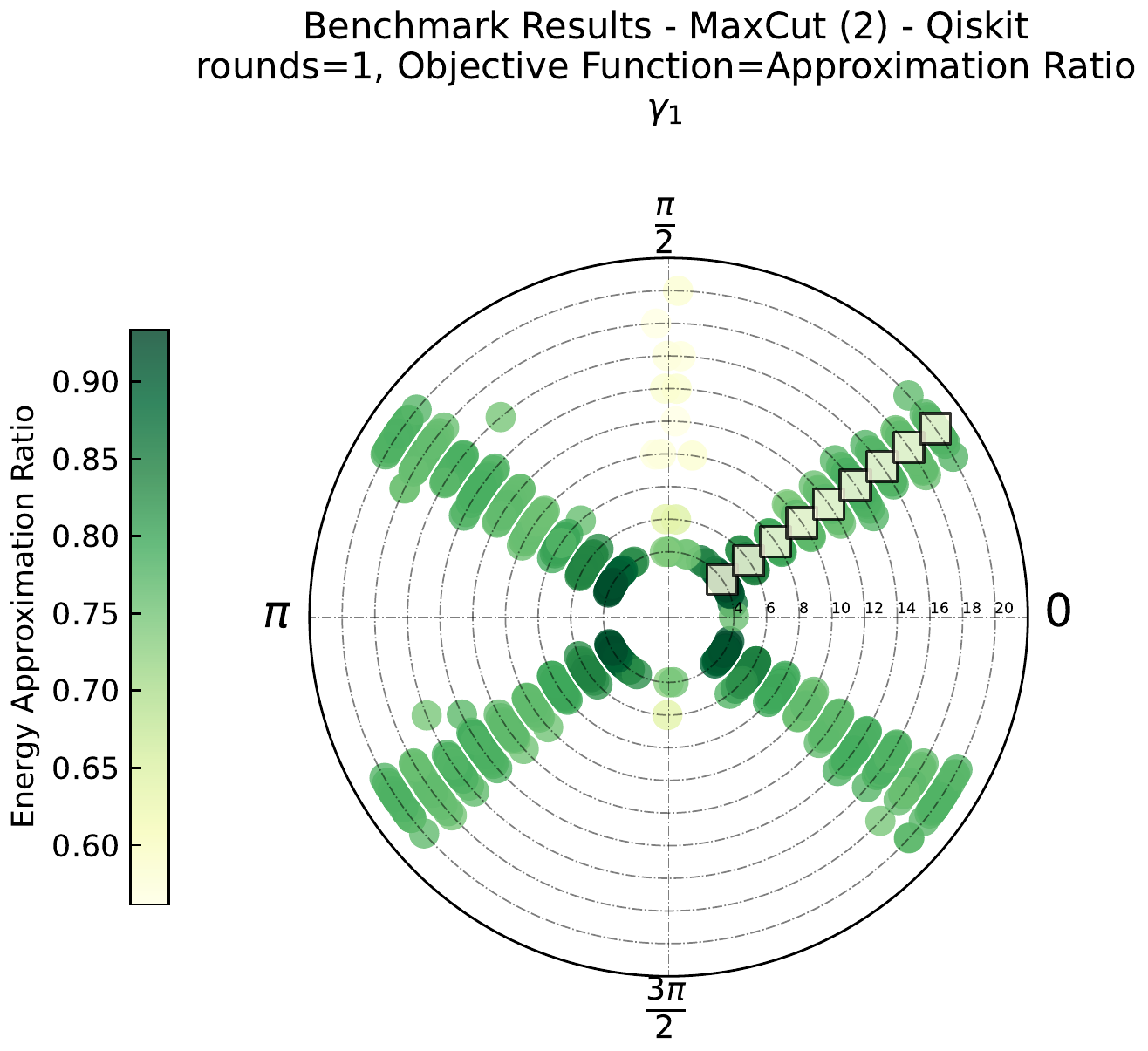}
    \caption{{\bf Angles cluster around certain values.} For 3-regular graphs with sizes ranging from 4 qubits to 20 qubits, we choose 100 random initial angles and run 30 (COBYLA) optimizer iterations each. The final angles obtained by the optimizer cluster around four values. The $\gamma$ values are shown here, along with the corresponding approximation ratios.}
    \label{fig:gamma-1}
\end{figure}

In \autoref{fig:gamma-1}, we present results from a test run using this benchmark feature to explore `parameter concentration'~\cite{AkshayParameter2021}. For problem sizes ranging from 4 to 20 qubits on 3-regular graphs, 100 random initial angles were tested using the Max-Cut benchmark, with 30 optimizer iterations each.  This plot shows the $\gamma$ values obtained as final values by the optimizer and the corresponding approximation ratios.
The angles obtained by the optimizer are shown to cluster around four values, most of which match the values proposed in~\cite{WurtzFixedangle2021}.
The choice of initial angles influences the algorithm's outcome, and a strategy for selecting these angles is critical for optimal performance.

\subsection{Parameter Selection Strategy}
\label{sec:hyper_parameter_selection}

Previously, we showed how the choice of initial angles and the number of rounds, shots, and restarts could affect the performance of the QAOA implementation.
In addition, the performance could vary from one problem instance to another.
These considerations raise the following question: For previously unseen problem instances, can we predict parameter values that are likely to result in the best performance?
Specifically, given a notion of resource (e.g., QPU access time) and a metric for result quality (e.g., approximation ratio), what parameter values should be used to get optimal performance given a resource budget?

With that goal in mind, a benchmarking framework is being developed for parameterized stochastic optimization algorithms such as QAOA and quantum annealing in a parallel effort~\cite{bernal2022benchmarking}.
Although this framework~\cite{WS_github} applies to other algorithms, we apply it to the QAOA simulations using results obtained from the QED-C benchmarking framework.
This framework generates parameter recommendations over a grid of resource values and also plots the corresponding performance compared to the best performance seen in the data.  

The input to the framework consists of performance data obtained empirically by implementing an algorithm on various problem instances.
The data includes the quality of the result, which we call the performance metric, corresponding to many algorithm executions over a range of parameter value settings.
A function is provided to compute the resources expended for each execution.

The framework splits the problem instances into testing and training sets.
A statistical analysis of the training set data is then used to identify the parameter values likely to lead to high performance when applied to the test set.
On the other hand, for each instance in the test set, parameter values as a function of the resource corresponding to the highest result quality found so far are identified from the available data for all resource grid values.
These are summarized in a curve denoted as `virtual best'.

\begin{figure}[t!]
    \centering
    \includegraphics[width=0.80\columnwidth]{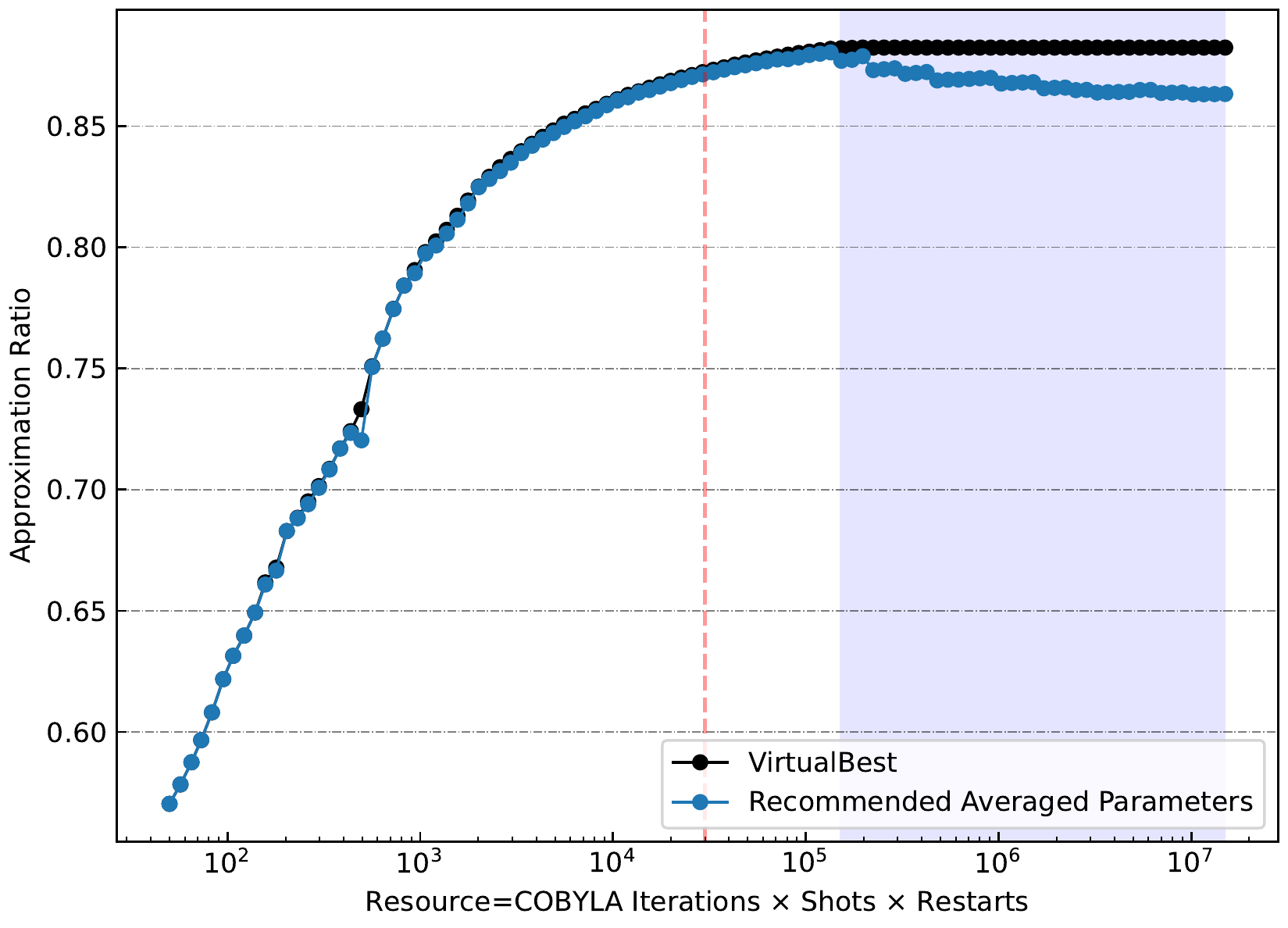}
    \caption{{\bf Solution Quality vs. Total Resource Utilization:} The virtual best provides a bound on the best performance attainable by any parameter strategy. Here are the performance profiles of the virtual best, along with the performance obtained from the parameters recommended by the stochastic-benchmarking framework~\cite{WS_github}. The red dashed vertical line corresponds to the resource value used throughout the hardware section (30 iterations, 1000 shots, 1 restart). The shaded area highlights the regime, after which the approximation ratio drops with increasing resources.}
    \label{fig:performance}
\end{figure}

Thus, the parameter values corresponding to the virtual best simulate knowing ahead of time for each instance what the best parameters would be for any resource value.
The virtual best provides a bound on the performance that any parameter-setting strategy using the data provided for the analysis can provide.
Thus, the recommended and virtual best parameters are plotted together in a `strategy plot'.
The virtual best performance is plotted in a separate plot along with the performance obtained on the test set using the recommended parameters.

\begin{figure}[t!]
    \centering
    \includegraphics[width=0.84\columnwidth]{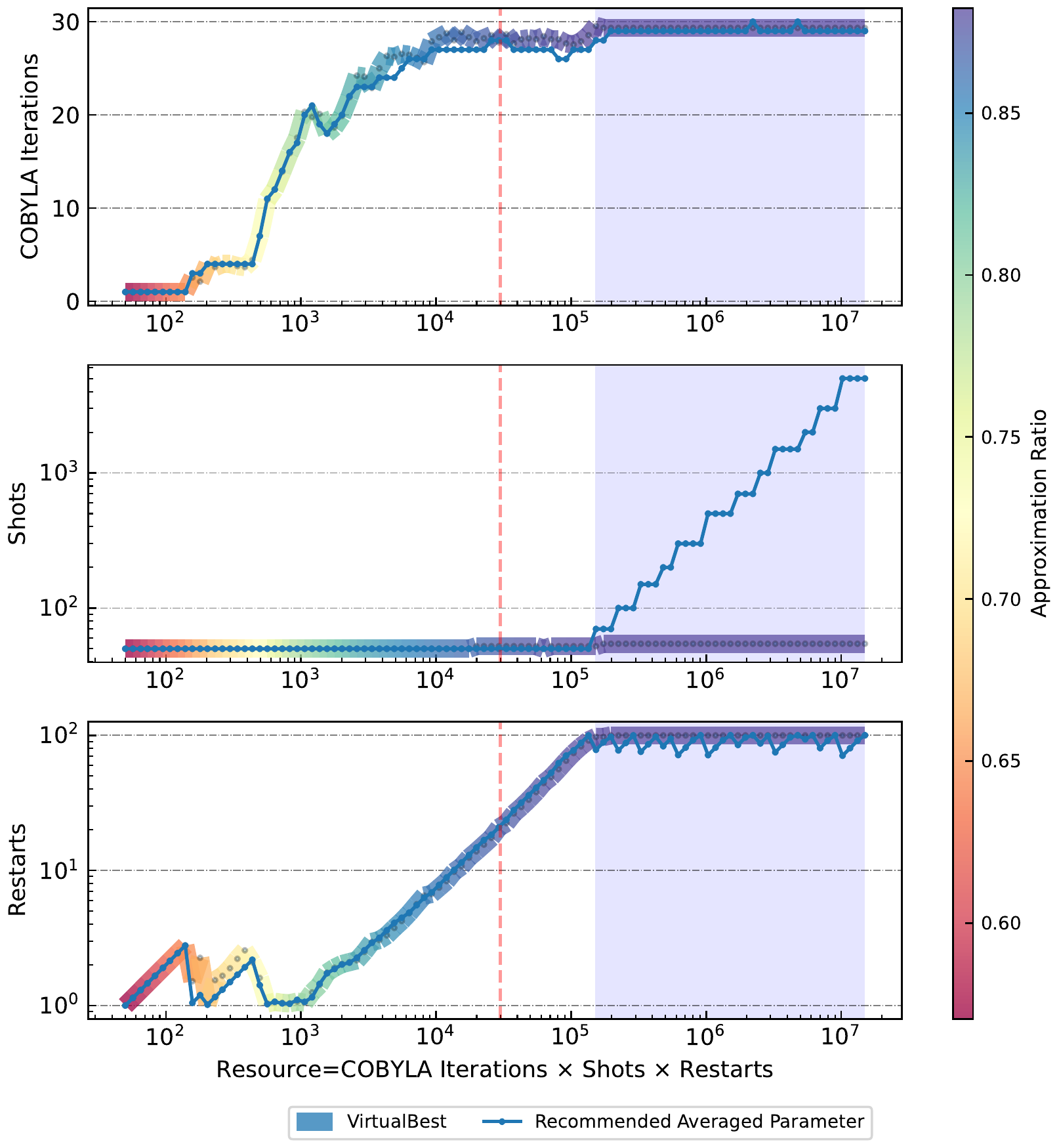}
    \caption{{\bf Strategy Plots:} For each resource value, the framework recommends values for minimizer iterations, shots, and the number of restarts likely to lead to the best performance. For comparison, the virtual best parameters are also plotted as colored curves, with the colors indicating the corresponding approximation ratio. The red dashed vertical line corresponds to the resource value used throughout the hardware section (30 iterations, 1000 shots, 1 restart). The shaded area highlights the regime, after which the approximation ratio drops with increasing resources.}
    \label{fig:all_params}
\end{figure}

We now present an analysis of QAOA using this framework. \autoref{fig:performance} shows the obtained performance profile, while \autoref{fig:all_params} shows the strategy plots generated by the framework, using an 80\%-20\% train/test instances split.
The QAOA algorithm uses noiseless simulations with two rounds for 50 distinct 3-regular graphs of size 12.
We implemented runs corresponding to a range of values for restarts [1,\dots,100], number of classical optimizer (COBYLA) iterations [1,\dots,30], and shots [50,\dots,5000].
We capture the number of times the processing unit was accessed by defining the resource as the product of these parameters. 

\autoref{fig:performance} can be used to determine the relative performance of the recommended parameter values with respect to the optimistic bound given by the virtual best.
In this example, we observe that both lines almost overlap, showing that good parameter values are shared across the training (recommended parameter values) and testing (virtual best parameter values) instances.
In particular, these results show that, with increasing access to the processing unit, the quality of the response increases, as measured by the approximation ratio up to a certain point.
The shaded area in this figure shows a regime of the resource quantity in which the performance metric decreases with increasing resources.
This is counterintuitive and reveals that, given the data used to generate this analysis, the best parameter values are given with the COBYLA iterations set to 30 and 100 restarts and only 50 shots.
Combinations of parameter values that yield larger resource usage can diminish the approximation ratio.
This observation suggests that increasing the number of shots deteriorates the performance if allowed more processing unit access.

Moreover, it highlights that the number of classical minimizer iterations should be increased before the number of restarts when more resources become available, always aiming to keep the number of shots small.
The dashed red line shows the equivalent resource usage of the simulations in the remaining of~\autoref{apdx:result_quality_assessment}.
Notice how the recommended parameter values, as seen in \autoref{fig:all_params}, i.e., 27 COBYLA iterations, 20 restarts, and 50 shots, are not the same as the ones used in the other hardware demonstrations, i.e., 30 COBYLA iterations, 1 restart, and 1000 shots.
Using this analysis and specifying a performance metric and a resource function, empirical data can be used to inform parameter setting values.
Moreover, these results can also inform about the instances themselves.
In this example, the problem instances are relatively small, i.e., 12 node graphs with a solution space of size $2^{12} = 4096$.
When solving the problem with QAOA, sampling the output distributions extensively with many shots does not improve the approximation ratio as much as reinitializing the problem (restarts) or allowing more classical optimization iterations.

These parameter-setting strategy analyses provide practical recommendations for using algorithms like the one discussed in this manuscript.





\end{document}